\documentclass[amsmath,amssymb,reprint,superscriptaddress,floatfix, nofootinbib]{revtex4-2}
\usepackage[dvipsnames]{xcolor}
\usepackage{graphicx}
\usepackage{changepage} 
\usepackage{multirow} 
\usepackage{hyperref}
\usepackage{float}
\usepackage{soul} 
\usepackage[maxfloats=256]{morefloats}
\usepackage{ulem}
\usepackage{url}
\usepackage{subcaption}
\usepackage[percent]{overpic} 
\usepackage{array} 
\usepackage{booktabs}
\usepackage{enumitem}
\usepackage[justification=raggedright,singlelinecheck=false]{caption}
\usepackage{colortbl}

\usepackage[capitalise,noabbrev,nameinlink]{cleveref}
\hypersetup{
citecolor=blue,
 colorlinks=true,
 linkcolor=blue,
 filecolor=magenta,
 urlcolor=blue,
}
\usepackage{orcidlink}

\graphicspath{{./images/}}
\Crefname{equation}{Eq.}{Eqs.}

\newcommand{\pcacellhex}[2]{%
  \cellcolor[HTML]{#1}\makebox[1.4cm][c]{$\displaystyle #2$}%
}
\newcommand{\pcacellhexn}[2]{%
  \cellcolor[HTML]{#1}\makebox[1.4cm][c]{\textcolor{magenta}{$\displaystyle #2$}}%
}
\newcommand{\pcavsep}{%
  \makebox[0pt][l]{%
    \raisebox{-1.5ex}{\rule{1.25pt}{3.6ex}}%
  }%
}
\newcommand{\pcahsep}{%
  \makebox[0pt][l]{%
    \raisebox{-1.5ex}{\rule{1.4cm}{1pt}}%
  }%
}
\newcommand{\pcacellhextable}[2]{%
  \cellcolor[HTML]{#1}\makebox[1.5cm][c]{$\displaystyle #2$}%
}
\newcommand{\pcacellhextablen}[2]{%
  \cellcolor[HTML]{#1}\makebox[1.5cm][c]{\textcolor{magenta}{$\displaystyle #2$}}%
}
\newcommand{\pcacolorcell}[1]{%
  \cellcolor[HTML]{#1}%
  \makebox[1.4cm][c]{%
    \rule{0pt}{0.45cm}%
    \raisebox{0.04cm}[0pt][0pt]{\large$\textcolor{black}{+}\;\textcolor{magenta}{-}$}%
  }%
}
\newcommand{\pcacolorbar}{%
\begin{tabular}{@{}c@{}c@{}c@{}c@{}c@{}c@{}}
\pcacolorcell{EAF2FB} &
\pcacolorcell{D4E6F8} &
\pcacolorcell{A9CCE3} &
\pcacolorcell{7FB3D5} &
\pcacolorcell{5499C7} &
\pcacolorcell{2E86C1} \\
{\footnotesize $[0,0.1)$} &
{\footnotesize $[0.1,0.2)$} &
{\footnotesize $[0.2,0.3)$} &
{\footnotesize $[0.3,0.4)$} &
{\footnotesize $[0.4,0.5)$} &
{\footnotesize $[0.5,1.0]$}
\end{tabular}%
}
\makeatletter
\renewcommand\paragraph{%
  \@startsection{paragraph}{4}{2.5em}
    {0.8ex \@plus .2ex \@minus .2ex}
    {-1em}
    {\normalfont\normalsize\itshape}
}
\makeatother

\begin{document}
\title{Sensitivity of neutron star observables to \\ microscopic nuclear parameters of realistic equations of state}
\author{Nikolas Cruz-Camacho\,\orcidlink{0009-0004-7870-0039}}
\affiliation{The Grainger College of Engineering, Illinois Center for Advanced Studies of the Universe, Department of Physics, University of Illinois at Urbana-Champaign, Urbana, IL 61801, USA} 
\author{Carlos Conde-Ocazionez\,\orcidlink{0000-0002-5393-0565}}
\affiliation{The Grainger College of Engineering, Illinois Center for Advanced Studies of the Universe, Department of Physics, University of Illinois at Urbana-Champaign, Urbana, IL 61801, USA}
\author{Veronica Dexheimer\,\orcidlink{0000-0001-5578-2626}}
\affiliation{Center for Nuclear Research, Department of Physics, Kent State University, Kent, OH 44243, USA}
\author{Jacquelyn Noronha-Hostler\,\orcidlink{0000-0003-3229-4958}}
\affiliation{The Grainger College of Engineering, Illinois Center for Advanced Studies of the Universe, Department of Physics, University of Illinois at Urbana-Champaign, Urbana, IL 61801, USA}
\author{Nicol\'as Yunes\,\orcidlink{0000-0001-6147-1736}}
\affiliation{The Grainger College of Engineering, Illinois Center for Advanced Studies of the Universe, Department of Physics, University of Illinois at Urbana-Champaign, Urbana, IL 61801, USA}
\noaffiliation
\date{\today}
\begin{abstract}
{
The equation of state of matter at supranuclear densities governs the astrophysical observables of neutron stars. 
A realistic, though complex, description is provided by the Chiral-Mean-Field model, which depends on many microscopic nuclear-physics parameters. 
We present a Fisher-information-inspired analysis of the sensitivity of neutron-star observables to the parameters of the Chiral-Mean-Field model at $\beta$-equilibrium using SLy as a crust.  
We then  compute neutron-star sequences and extract masses, radii, compactnesses, and tidal deformabilities. 
From the logarithmic derivatives of these observables with respect to each nuclear parameter, we construct a dimensionless, Fisher-inspired sensitivity matrix and perform a principal-component analysis to identify the effective combinations of nuclear parameters that most strongly affect neutron-star observables. 
Although the ranking depends mildly on the observable, the three most important nuclear parameters are the vacuum value of the dilaton field $\chi_0$ (which sets the overall scale of the scalar potential and trace-anomaly contribution), the scalar singlet strength $g_{1}^X$ (which controls the overall scalar attraction through the baryon effective masses), and the $k_0$ quadratic scalar term (which governs the curvature of the scalar potential). 
This framework provides a reproducible, data-driven approach to quantify parameter sensitivities in dense-matter models and to guide future Bayesian inference of nuclear information from multi-messenger astrophysical observations.
}
\end{abstract}
\maketitle

\section{Introduction}
Neutron stars are unique laboratories for quantum chromodynamics (QCD) at supranuclear densities \cite{Lattimer:2012nd,Oertel:2016bki,Baym:2017whm}. Their macroscopic properties, gravitational mass $M$, radius $R$, compactness $C$, and (dimensionless) tidal deformability $\Lambda$,
encode the microphysics of dense matter via the cold, $\beta$-equilibrated equation of state (EoS). Recent multi-messenger observations, combining gravitational waves from binary mergers \cite{LIGOScientific:2017vwq} with X-ray pulse-profile modeling \cite{Riley:2019yda,Miller:2019cac,Raaijmakers:2021uju,Miller:2021qha}, have transformed these objects into precision probes of nuclear interactions at densities inaccessible in terrestrial experiments (for a summary of constraints see \cite{MUSES:2023hyz}). 

Yet, despite this progress, the mapping from a given nuclear model to astrophysical observables remains highly non-unique \cite{Mroczek:2023zxo}: different microscopic assumptions and parameterizations can reproduce similar macroscopic trends. Quantifying which model ingredients actually drive the observable response is therefore as important as fitting the data themselves. Linking nuclear-physics parameters to stellar observables is essential for two reasons. First, it identifies the effective combinations of microphysical inputs that control the stiffness of the EoS, and thus, the stellar mass, radius, compactness, and tidal deformability measured by NICER and LIGO/Virgo/KAGRA (for example, \cite{Essick:2021kjb}). Second, it guides principled priors and low-dimensional embeddings for future Bayesian inference \cite{Landry:2020vaw}, reducing wasted exploration in directions in which the observables are only weakly sensitive.

Not all frameworks constructed to model dense matter and neutron-star interiors depend on physically-motivated, microscopic, nuclear physics parameters. 
Some models, such as speed-of-sound and spectral parameterizations \cite{Lindblom:2010bb,Greif:2018njt} or non-parameteric, Gaussian processes \cite{Landry:2020vaw,Landry:2018prl}, are purely phenomenological or attempt to be agnostic. The parameters that enter these models, therefore, do not have a direct link to the high-energy, particle or nuclear physics of the Standard Model. Other models, such as non-relativistic energy density functionals \cite{Chabanat:1997un}, relativistic mean-field models \cite{Serot:1997xg,Typel:1999yq}, and chiral effective field theory extrapolations \cite{Drischler:2021bup, Hebeler:2013nza,Tews:2018iwm}, are constructed to serve as a bridge between Standard Model physics and neutron stars. 
An example of the latter is the Chiral Mean Field (CMF) model \cite{Dexheimer:2008ax,Dexheimer:2009hi}, which provides a unified, relativistic-mean-field description in which baryon effective masses and interactions are dynamically generated by mesonic condensates, with symmetry structure rooted in SU(3), non-linear, chiral realizations and spontaneous symmetry breaking. CMF's modern C++ implementation within the Modular
Unified Solver of the Equation of State  (MUSES) ecosystem \cite{ReinkePelicer:2025vuh}, hereafter \texttt{CMF++} \cite{Cruz-Camacho:2024odu}, allows for transparent studies of the effect that physically-motivated, nuclear-physics parameters across scalar, vector, and explicit-symmetry-breaking sectors have on neutron-star observables. 

An exhaustive, fully agnostic survey of all dense-matter possibilities in any realistic microscopic model is currently infeasible due to the large dimensionality of the nuclear physics parameter space. Here we take first steps in this direction by adopting a focused, end-to-end workflow for cold neutron-proton-electron (\textit{npe}) matter, building upon the open-source MUSES framework \cite{ReinkePelicer:2025vuh}. Schematically, \texttt{CMF++} generates a neutron star core EoS, which is then leptonized to enforce neutron-star conditions, and finally merged with a SLy crust~\cite{Douchin:2001sv}; with the EoS constructed, we then compute neutron-star observables through the QLIMR MUSES module, which solves the perturbed Einstein equations for non-rotating neutron stars. We first define and validate this workflow, ensuring thermodynamic continuity at the crust-core interface, TOV integration quality, and nuclear saturation sanity checks, and then justify it as a controlled baseline with which to perform a systematic sensitivity analysis.

Building on this framework, the central goal of this paper is to identify which \texttt{CMF++} coupling and vacuum parameters most strongly affect neutron-star observables. We focus on the neutron-star mass $M$, radius $R$, tidal deformability $\Lambda$ (due to perturbations from a companion), and compactness $C := G M/(c^2 R)$, and study their variability as \texttt{CMF++} parameters are varied about a fiducial set. We further examine how the parameter sensitivity of observables compares in magnitude across the parameter set, whether these sensitivities collectively follow a consistent hierarchy as revealed by a Fisher-like matrix analysis, and how robust these trends remain under different physical and numerical conditions. We finally carry out a principal component analysis (PCA) on the parameter space to determine the (eigenvector) combination of parameters that leads to the most variability for different observables. 

Our analysis reveals a coherent and low-dimensional structure in the response of neutron-star observables to variations of the \texttt{CMF++} parameters. In the original CMF parameter basis, the most influential individual parameters are typically the vacuum value of the dilaton $\chi_0$ (which controls the pace of chiral restoration and the overall scalar scale), the scalar singlet strength $g_{1}^X$ (which regulates the overall scalar attraction through the baryon effective masses), and the quadratic scalar coefficient $k_0$. These parameters consistently appear among the largest individual contributions to the sensitivity matrices and set the dominant stiffness trends of the EoS, and therefore of the macroscopic neutron-star observables. At the same time, the PCA shows that the observable response is not controlled by only three original parameters. Rather, it is organized into two dominant, and at most three, principal directions in parameter space. These rotated directions are low-dimensional in number, but each is generally supported by several original CMF parameters. In practice, the leading eigenvectors typically contain $\sim 10$--$11$ non-negligible components, with recurrent contributions from $\{\chi_0, g_{1}^X, k_0, k_1, k_2, k_3, f_\pi, m_N, m_K\}$ and, at subleading level, from isoscalar-vector quantities such as $g_{N\omega}$ and $m_\omega$. In this sense, the CMF parameter space around the fiducial point behaves as a low-rank response manifold, even though the corresponding principal directions are distributed over a broader subset of microscopic parameters.

Interestingly, the scalar sector is the least explored one in the CMF model. It is usually held fixed, as in many chiral approaches, to reproduce vacuum physics, while the vector sector has been more systematically varied in connection with saturation properties and neutron-star observables.~\cite{Dexheimer:2015qha,Dexheimer:2018dhb} In this work, we adopt a different strategy and vary all relevant parameters simultaneously, allowing correlated scalar and vector contributions to emerge directly from the sensitivity analysis.

The remainder of this paper explains the details that led to the conclusions summarized above and it is organized as follows. 
\Cref{sec:workflow} summarizes the \texttt{CMF++} \textit{npe} workflow and the crust-core merging. \Cref{sec:nsobs} reviews the stellar-structure and tidal calculations using QLIMR. \Cref{sec:Fisher_theory} formulates the Fisher-inspired, scale-invariant sensitivity and PCA. \Cref{sec:results} presents our main results, beginning with relative responses and culminating in the principal directions across observables and composites. We discuss implications in \cref{sec:discussion}, and provide numerical-derivative diagnostics and reproducibility notes in the appendices. 

Natural units $c=\hbar=1$ are used for the EoS and geometric units $c=G=1$ for the calculation of the $M(R)$ curves and gravitational-wave observables; we follow the sign and metric conventions stated in \Cref{sec:nsobs}.

\section{Crust to core EoS}
\label{sec:workflow}
In this section, we review the \texttt{CMF++} workflow and describe the matching between the EoS we use to describe neutron star cores and the SLy EoS we use to describe crusts. 

\subsection{Core EoS: CMF}
The primary model employed in this work to describe strongly-interacting nuclear matter is the CMF model, which has been recently rewritten in modern C++ by the MUSES collaboration~\cite{Cruz-Camacho:2024odu}. The theoretical framework summarized in this section follows Sec.~II of~\cite{Cruz-Camacho:2024odu}, focusing on \textit{npe} cold matter (in the MeV scale), where it is reasonable to assume $T=0$.

The CMF model is a relativistic mean-field framework for dense hadronic matter~\cite{Dexheimer:2008ax}, built upon a non-linear realization of the SU(3) sigma model~\cite{Papazoglou:1998vr}. In this approach, baryons interact (and have effective masses) through the exchange of scalar and vector mesons: the scalar-isoscalar $\sigma$ (associated with the $f_0(500)$ meson), the strange scalar-isoscalar $\zeta$ ($f_0(980)$), the scalar-isovector $\delta$ ($a_0(980)$), and the vector-isoscalar $\omega$ and $\phi$, together with the vector-isovector $\rho$. Attractive medium-range interactions arise from the scalar fields, while short-range repulsion is mediated by the vector fields. In addition, a scalar dilaton field $\chi$ mimics the QCD trace anomaly, ensuring approximate scale invariance.
The non-linear realization of chiral symmetry is important for enabling the introduction of explicit symmetry-breaking terms without violating the Partially Conserved Axial-Vector Current (PCAC) relations by treating pseudoscalar mesons as angular parameters of the chiral transformation. This avoids the inconsistencies of the linear sigma model and yields realistic hyperon potentials and baryon interactions~\cite{Papazoglou:1998vr}.

\subsubsection{Lagrangian}
The CMF Lagrangian density can be organized as
\begin{equation}
\mathcal{L}_{\rm CMF} = \mathcal{L}_{\rm kin} + \mathcal{L}_{\rm int} + 
\mathcal{L}_{\rm scal} + \mathcal{L}_{\rm vec} + \mathcal{L}_{\rm esb} + \mathcal{L}_{m_0},
\end{equation}
where $\mathcal{L}_{\rm kin}$ represents the kinetic-energy term, $\mathcal{L}_{\rm int}$ the baryon-meson interaction term, $\mathcal{L}_{\rm scal}$ and $\mathcal{L}_{\rm vec}$ the scalar and vector meson self-interaction terms respectively, $\mathcal{L}_{\rm esb}$ the explicit chiral symmetry-breaking term, and, for the C4 parametrization, $\mathcal{L}_{m_0}$ is an additional bare-mass contribution. We now describe each of these components in turn. 

\paragraph{Scalar sector}
In relativistic mean field models, scalar interactions mimic attractive interactions between baryons, as well as bosonic self-interactions. 
After applying the mean-field approximation, the scalar-sector Lagrangian reads
\begin{align}
\mathcal{L}_{\rm scal} =& -\frac{1}{2}k_0\chi_0^2(\sigma^2+\zeta^2+\delta^2)
+ k_1(\sigma^2+\zeta^2+\delta^2)^2 \nonumber \\
&+ k_2\!\left[\frac{\sigma^4+\delta^4}{2}+\zeta^4+3(\sigma\delta)^2\right]
+ k_3\chi_0(\sigma^2-\delta^2)\zeta \nonumber\\
&+ \frac{\epsilon}{3}\chi_0^4\ln\!\left[\frac{(\sigma^2-\delta^2)\zeta}{\sigma_0^2\zeta_0}\right],
\end{align}
where $\sigma_0=-f_\pi$ and $\zeta_0={f_\pi}/{\sqrt{2}} - \sqrt{2}f_K$ are the vacuum expectation values of the $\sigma$ and $\zeta$ fields, respectively. The scalar self-interactions are governed by the parameters $\{\chi_0,k_0,k_1,k_2,k_3,\epsilon,f_\pi,f_K\}$, which control the restoration of chiral symmetry related to baryon masses. 

\paragraph{Vector sector}
The vector mesons $\{\omega,\rho,\phi\}$ provide repulsive interactions between baryons, required to stabilize nuclear matter. Their self-interactions are described by coupling terms parameterized by $g_4$, with several chiral-invariant structures (C1-C4) that contribute to the Lagrangian via $ \mathcal{L}_{\rm vec}^{\rm SI}$~\cite{Dexheimer:2015qha}. The specific choice of structure influences the stiffness of the EoS and the nuclear incompressibility. In particular, the C4 parametrization, used in this work as it provides the best agreement with neutron star observables, requires the introduction of a bare baryon mass $m_0$ ~\cite{Dexheimer:2008ax}.
After applying the mean-field approximation, the vector-meson Lagrangian becomes
\begin{align}
\mathcal{L}_{\rm vec} = 
\frac{1}{2}\!\left(m_\omega^2\omega^2+m_\phi^2\phi^2+m_\rho^2\rho^2\right)
+ \mathcal{L}_{\rm vec}^{\rm SI},
\end{align}
and, for the C4 case, the last term takes the explicit form
\begin{align}
\mathcal{L}_{\rm vec}^{\rm SI,C4} =&  g_4 \!\left(\omega^4 + \frac{\phi^4}{4} + 3\omega^2\phi^2 
+ 2\sqrt{2}\omega^3\phi + \sqrt{2}\omega\phi^3\right).
\label{eq:L_vec}
\end{align}
The parameters $\{m_\omega, m_\rho, m_\phi, g_4\}$ govern the repulsive self-interaction contributions to the EoS.

\paragraph{Explicit symmetry breaking}
Masses of pseudoscalar mesons and hyperon potentials are reproduced by explicit chiral symmetry-breaking terms. Under the mean-field approximation, the leading term simplifies to
\begin{equation}
\mathcal{L}^u_{\rm esb} = 
-\!\left[m_\pi^2 f_\pi \sigma 
+ \!\left(\sqrt{2}m_K^2 f_K - \frac{1}{\sqrt{2}}m_\pi^2 f_\pi\right)\!\zeta \right].
\label{eq:L_esb}
\end{equation}
This sector is characterized by the parameters $\{m_\pi, m_K, f_\pi, f_K\}$, which we will vary in this work. Although hyperons are not included, the kaon parameters $m_K$ and  $f_K$ must still be retained to preserve the SU(3) chiral symmetry of the model, as they determine the strange condensate $\zeta_0$ and influence the baryon-meson couplings.

\paragraph{Baryon-meson couplings and bare-mass term}
Baryon-meson couplings are determined from SU(6) symmetry relations and adjusted to reproduce nuclear saturation properties.   Non-strange baryons decouple from the $\phi=s\bar{s}$ meson by construction, while the $\delta$ and $\rho$ mesons govern isospin asymmetry.
The interaction Lagrangian can be written as a sum over baryons $B$,
\begin{equation}
\begin{split}
\mathcal{L}_{\rm int} = -\!\sum_{i \in B} 
\bar{\psi}_i \big[ &\,\gamma_0 ( g_{i\omega}\omega + g_{i\rho}\rho + g_{i\phi}\phi ) \\
&+ g_{i\sigma}\sigma + g_{i\zeta}\zeta + g_{i\delta}\delta \big]\psi_i ,
\end{split}
\label{eq:L_int}
\end{equation}
where the couplings $g_{i\,\rm{meson}}$ with $\rm{mesons}=\{\sigma,\zeta,\delta,\omega,\phi,\rho\}$ are expressed in terms of $\{g_1^X, g_8^X, m_N\}$~\cite{Cruz-Camacho:2024odu}, with $m_N$ being the nucleon vacuum mass, for the scalar sector and $\{g_{N\omega}, g_{N\rho}, g_{N\phi}\}$ for the vector sector.  These parameters affect the particle population densities and, consequently, the thermodynamic observables.

For the C4 parametrization, an additional bare-mass term is required to reproduce the empirical nuclear incompressibility at saturation~\cite{Dexheimer:2008ax}. After the mean-field approximation, it takes the form
\begin{equation}
\mathcal{L}_{m_0} = -\!\sum_{i \in B} \bar{\psi}_i m_0 \psi_i .
\label{eq:m0_lagrangian}
\end{equation}
Combining \Cref{eq:L_int} and \Cref{eq:m0_lagrangian}, the corresponding effective baryon masses become
\begin{equation}
m_i^* = g_{i\sigma}\sigma + g_{i\zeta}\zeta + g_{i\delta}\delta + \Delta m_i,
\label{eq:eff_mass_nopol}
\end{equation}
which encode the additional parameter $\{m_0\}$.
%

\subsubsection{Core EOS from a Lagrangian}
In the mean-field and no-sea approximations, only the temporal components of the vector fields survive. Solving the Euler-Lagrange equations for $\{\sigma,\zeta,\delta,\omega,\rho,\phi\}$ at a given set of chemical potentials yields the effective baryon masses and chemical potentials. In this work, we consider finite baryon chemical potential $\mu_B$ and charge chemical potential $\mu_Q$, while the strangeness chemical potential is fixed to $\mu_S=0$. These effective quantities fully determine the thermodynamic observables. The pressure and energy density are then obtained from the energy--momentum tensor using the effective masses $m_i^*$ and chemical potentials $\mu_i^*$~\cite{Cruz-Camacho:2024odu}. At zero temperature, the effective chemical potential of the $i$-th particle is given by $\mu_i^* = B_i\mu_B + S_i\mu_S + Q_i\mu_Q - g_{i\omega}\omega - g_{i\rho}\rho - g_{i\phi}\phi$, where $B_i$, $Q_i$, and $S_i$ denote the baryon number, electric charge, and strangeness of species $i$, respectively.

To model cold, non-strange neutron-star matter, we impose $\beta$ equilibrium and charge neutrality on the 2D CMF surface ($\mu_B$ and $\mu_Q$, with $\mu_S=0$). With lepton contributions provided by the \texttt{Lepton} module of MUSES, the equilibrium conditions select a unique trajectory:
\begin{align}
\beta\text{ equilibrium:} \quad 
& \mu_n - \mu_p \;=\; \mu_e , \label{eq:beta-eq} \\
\text{charge neutrality:} \quad 
& n_p \;=\; n_e, \label{eq:neutrality} \\
\text{baryon density:} \quad 
& n_B \;=\; n_n + n_p, \label{eq:nB}\\
\text{pressure:} \quad 
& p \;=\; p_{\rm{CMF}} + p_{e},\\
\text{energy density:} \quad 
& \varepsilon \;=\; \varepsilon_{\rm{CMF}} + \varepsilon_{e}. \label{eq:energy_density}
\end{align}
Solving Eqs.\ \eqref{eq:beta-eq}-\eqref{eq:energy_density} with an ideal Fermi gas of electrons provided by the MUSES \texttt{Lepton} module produces a one-dimensional, barotropic core EoS $p(\varepsilon)$ along the equilibrium path. This CMF$\to$Lepton barotrope is what we match to the SLy EoS crust, as we explain in ~\Cref{sec:crust_core_match} after discussing the core EoS parameters.

\subsection{Core EoS parameters}
%

\subsubsection{Parameter set}
\label{sec:parameter_set}
The complete parameter set entering the mean-field Lagrangian, together with their associated sectors and physical interpretations, is summarized in \Cref{tab:cmf_params}. These parameters encompass couplings and vacuum physical properties, including the scalar self-interaction $\{\chi_0,k_0,k_1,k_2,k_3,\epsilon,f_\pi,f_K\}$, vector self-interaction  $\{m_\omega, m_\rho, m_\phi, g_4\}$, explicit symmetry-breaking $\{m_\pi, m_K, f_\pi, f_K\}$, and baryon-meson interaction  $\{g_1^X, g_8^X, m_N,g_{N\omega}, g_{N\rho}, g_{N\phi},m_0\}$, sectors introduced in the preceding subsections. When combined and simplified (excluding duplicate parameters), they result in the following set of $21$ parameters:
\begin{eqnarray}
\boldsymbol{\lambda} \equiv \{\chi_0, k_0, k_1, k_2, k_3, \epsilon, m_\omega, m_\rho, m_\phi, g_4, m_\pi, f_\pi, \nonumber\\ m_K, f_K, g_1^X, g_8^X, m_N,g_{N\omega}, g_{N\rho}, g_{N\phi},m_0\} ,
\label{eq:param-vector}
\end{eqnarray}
which define the specific C4 \textit{npe} configuration studied in this work.
\begin{table*}[t]
\centering
\small
\begin{tabular}{l l l l}
\toprule
\textbf{Parameter} &
\textbf{Sector} &
\parbox[t]{6.6cm}{\raggedright\textbf{Lagrangian terms}} &
\parbox[t]{6.6cm}{\raggedright\textbf{Description}} \\
\midrule

$\chi_0$ & scalar &
\parbox[t]{6.6cm}{
\(
\begin{aligned}
&-\tfrac{1}{2}k_0\chi_0^2(\sigma^2+\zeta^2+\delta^2)+\,k_3\chi_0(\sigma^2-\delta^2)\zeta\\
&\quad +\,\tfrac{\epsilon}{3}\chi_0^4
\ln\!\Big[\tfrac{(\sigma^2-\delta^2)\zeta}{\sigma_0^2\zeta_0}\Big]
\end{aligned}
\)
} &
\parbox[t]{6.6cm}{\raggedright
Vacuum expectation value of the scale-breaking (dilaton) field; sets the overall scale of the scalar potential and couples to the logarithmic trace-anomaly term.
} \\

$k_0$ & scalar &
\parbox[t]{6.6cm}{$-\tfrac{1}{2}k_0\chi_0^2(\sigma^2+\zeta^2+\delta^2)$} &
\parbox[t]{6.6cm}{\raggedright
Quadratic scalar-potential coefficient controls location/curvature near vacuum.
} \\

$k_1$ & scalar &
\parbox[t]{6.6cm}{$k_1(\sigma^2+\zeta^2+\delta^2)^2$} &
\parbox[t]{6.6cm}{\raggedright
Quartic scalar self-interaction coefficient controlling large-field stabilization of the scalar potential.
} \\

$k_2$ & scalar &
\parbox[t]{6.6cm}{$k_2\!\left[\tfrac{\sigma^4+\delta^4}{2}+3\sigma^2\delta^2+\zeta^4\right]$} &
\parbox[t]{6.6cm}{\raggedright
Quartic SU(3) scalar interaction coefficient with $\sigma-\delta$ structure; controls isovector-scalar nonlinearities and the density dependence of $\delta$ in asymmetric matter.
} \\

$k_3$ & scalar &
\parbox[t]{6.6cm}{$k_3\chi_0(\sigma^2-\delta^2)\zeta$} &
\parbox[t]{6.6cm}{\raggedright
Cubic scalar mixing term coupling the strange condensate $\zeta$ to $\sigma^2-\delta^2$; affects the coupled evolution of $\sigma, \zeta, \delta$ with density.
} \\

$\epsilon$ & scalar &
\parbox[t]{6.6cm}{$\tfrac{\epsilon}{3}\chi_0^4\ln\!\big[\tfrac{(\sigma^2-\delta^2)\zeta}{\sigma_0^2\zeta_0}\big]$} &
\parbox[t]{6.6cm}{\raggedright
Strength of the logarithmic scale-breaking term (effective trace-anomaly contribution) in the scalar potential.
} \\

$m_\omega$ & vector &
\parbox[t]{6.6cm}{$\tfrac{1}{2}m_\omega^2\omega^2$} &
\parbox[t]{6.6cm}{\raggedright
Mass of isoscalar vector meson.
} \\

$m_\rho$ & vector &
\parbox[t]{6.6cm}{$\tfrac{1}{2}m_\rho^2\rho^2$} &
\parbox[t]{6.6cm}{\raggedright
Mass of isovector meson.
} \\

$m_\phi$ & vector &
\parbox[t]{6.6cm}{$\tfrac{1}{2}m_\phi^2\phi^2$} &
\parbox[t]{6.6cm}{\raggedright
Mass of hidden-strangeness meson.
} \\

$g_4$ & vector &
\parbox[t]{6.6cm}{$g_4\!\left(\omega^4+\tfrac{\phi^4}{4}+3\omega^2\phi^2+2\sqrt2\,\omega^3\phi+\sqrt2\,\omega\phi^3\right)$} &
\parbox[t]{6.6cm}{\raggedright
Common vector self-interaction strength.
} \\

$m_\pi,\ f_\pi$ & ESB/scalar &
\parbox[t]{6.6cm}{$-\,m_\pi^2 f_\pi\,\sigma + \tfrac{1}{\sqrt2}m_\pi^2 f_\pi\zeta$} &
\parbox[t]{6.6cm}{\raggedright
Pion mass and decay constant; set the explicit symmetry breaking scale and calibrate $\sigma_0$.
} \\

$m_K,\ f_K$ & ESB/scalar &
\parbox[t]{6.6cm}{$- \sqrt2\,m_K^2 f_K \zeta$} &
\parbox[t]{6.6cm}{\raggedright
Kaon mass and decay constant; set explicit SU(3) breaking and calibrate $\zeta_0$.
} \\

$g_1^X$ & $B$ int. &
\parbox[t]{6.6cm}{$-\bar\psi_N\,g_{N\sigma}\sigma\,\psi_N$} &
\parbox[t]{6.6cm}{\raggedright
SU(3) scalar singlet coupling controlling the overall scalar attraction (via baryon effective masses).
} \\

$g_8^X$ & $B$ int. &
\parbox[t]{6.6cm}{$-\bar\psi_N(g_{N\sigma}\sigma+g_{N\delta}\delta)\psi_N$} &
\parbox[t]{6.6cm}{\raggedright
SU(3) scalar octet coupling controlling flavor/isospin dependence of scalar attraction (mass splittings and isovector response).
} \\

$m_N$ & $B$ int. &
\parbox[t]{6.6cm}{$m_N=g_{N\sigma}\sigma_0+g_{N\zeta}\zeta_0+\Delta m_N$} &
\parbox[t]{6.6cm}{\raggedright
Nucleon vacuum mass; only enters \(\mathcal{L}\) via calibration of scalar couplings.
} \\

$g_{N\omega}$ & $B$ int. &
\parbox[t]{6.6cm}{$-\bar\psi_N\gamma_0\,g_{N\omega}\omega\,\psi_N$} &
\parbox[t]{6.6cm}{\raggedright
Primary isoscalar-vector coupling controlling repulsive mean-field contribution.
} \\

$g_{N\rho}$ & $B$ int. &
\parbox[t]{6.6cm}{$-\bar\psi_N\gamma_0\,g_{N\rho}\rho\,\psi_N$} &
\parbox[t]{6.6cm}{\raggedright
Isovector-vector coupling controlling symmetry energy and isospin-dependent pressure.
} \\

$g_{N\phi}$ & $B$ int. &
\parbox[t]{6.6cm}{$-\bar\psi_N\gamma_0\,g_{N\phi}\phi\,\psi_N$} &
\parbox[t]{6.6cm}{\raggedright
Hidden-strangeness vector coupling; controls additional repulsion in strange sector.
} \\

$m_0$ & $B$ int. &
\parbox[t]{6.6cm}{$-\sum_{i}\bar\psi_i\,m_0\,\psi_i$} &
\parbox[t]{6.6cm}{\raggedright
Chirally invariant (bare) mass parameter used in C4. Sets the non-vanishing baryon mass component as chiral condensates melt.
} \\

\bottomrule
\end{tabular}
\caption{21 parameters, sector, mean-field Lagrangian terms, and descriptions for the C4 \textit{npe} parameter set used in this work.}
\label{tab:cmf_params}
\end{table*}

The parameters listed in \Cref{tab:cmf_params} are input to our \texttt{CMF++} implementation. These quantities control both microscopic interactions and macroscopic observables of neutron-star matter, making them well-suited for the sensitivity analysis of \Cref{sec:Fisher_theory}.
All parameter variations explored in this study are performed around a fiducial reference point in the CMF parameter space, denoted by $\boldsymbol{\lambda^*}$. 
The explicit numerical values of all 21 parameters at this fiducial point are listed in \Cref{tab:fiducial_params}.
This parameter set corresponds to the current best-fit configuration of the model, providing an optimal description of nuclear saturation properties and astrophysical constraints from neutron-star observations~\cite{Dexheimer:2008ax}.

\begin{table}[t!]
\centering
\def\arraystretch{1.8}
\begin{tabular}{ccc}
\hline\hline
$\chi_0^* = 401.93~\text{MeV}$ &
$k_0^* = 2.373$ &
$k_1^* = 1.400$ \\
$k_2^* = -5.549$ &
$k_3^* = -2.652$ &
$\epsilon^* = 2/33$ \\
$m_\pi^* = 139~\text{MeV}$ &
$f_\pi^* = 93.30~\text{MeV}$ &
$m_K^* = 498~\text{MeV}$ \\
$f_K^* = 122~\text{MeV}$ &
$m_\omega^* = 780.56~\text{MeV}$ &
$m_\rho^* = 761.06~\text{MeV}$ \\
$m_\phi^* = 1019~\text{MeV}$ &
$g_4^* = 38.9$ &
$g_1^{X,*} = -7.324$ \\
$g_8^{X,*} = -2.339$ &
$g_{N\omega}^* = 11.9$ &
$g_{N\rho}^* = 4.03$ \\
$g_{N\phi}^* = 0$ &
$m_N^* = 937.24~\text{MeV}$ &
$m_0^* = 150~\text{MeV}$ \\
\hline\hline
\end{tabular}
\caption{Fiducial C4 \textit{npe} CMF parameter set, ordered by sector (scalar, ESB, vector, baryon-scalar, baryon-vector, and vacuum masses). This set of parameters defines $\boldsymbol{\lambda^*}$.}
\label{tab:fiducial_params}
\end{table}
For notational convenience, henceforth we collect all our model parameters into a vector, $\boldsymbol{\lambda}$, as presented in Eq.~\eqref{eq:param-vector}. This vector spans the 21-dimensional CMF parameter space, with $\boldsymbol{\lambda^*}$ representing the fiducial, C4 \textit{npe} point (see \Cref{tab:fiducial_params}), and $\lambda_k$ denoting the $k$th component of the $\boldsymbol{\lambda}$ vector. In later sections, we probe local parameter sensitivity by considering one-parameter variations about the fiducial point $\boldsymbol{\lambda}^*$.  For each index $k$, only the component $\lambda_k$ is varied, while all other components remain fixed at their fiducial values.

In this work, we perform a systematic and consistent exploration of a broad subset of the model’s parameter space for the first time within the CMF framework. Equation~\eqref{eq:param-vector} represents the subset relevant for \textit{npe} matter, compared to roughly $\sim100$ parameters in the full CMF parameter space. Specifically, we vary for the first time the following set
\begin{eqnarray}
\{\chi_0,\epsilon,m_\omega, m_\rho, m_\phi, m_\pi, f_\pi, \nonumber\\ m_K, f_K, g_1^X, g_8^X, m_N, m_0\} ,
\end{eqnarray}
spanning the scalar, explicit–symmetry–breaking, vector, and baryon–meson interaction sectors. To the best of our knowledge, no previous study has varied this many CMF parameters across wide, physically motivated ranges while consistently propagating their effects through the full mean-field solution and the resulting neutron-star observables. Other works, such as \cite{Singh:2025nri}, have explored variations of different parameter combinations not included in the set above; however, no local sensitivity analysis around a fiducial parameter point has been performed. Therefore, our work constitutes the first local sensitivity analysis of the CMF C4 model for \textit{npe} matter.

\subsubsection{Saturation properties}\label{sec:saturation}
The short-range aspect of the strong nuclear force gives rise to the phenomenon of saturation, a density at which the (negative) binding energy per nucleon has a maximum.
The value of $n_{sat}$ can be extracted from global analyses of experimentally-measured masses and radii of various nuclei (see eg.~\cite{MUSES:2023hyz} for details), where expected values are in the range $n_{sat}\sim (0.14$--$0.17)\,{\rm{fm}}^{-3}$. 
Then, at $n_{sat}$, the nuclear binding energy $E_B$ is the energy difference per nucleon that one expects from a free gas of nucleons versus the actual measured mass $m(Z,A)$, i.e.:
\begin{align}
   E_B&=\frac{m(Z,A)-\left[Zm_p+(A-Z)m_n\right]}{A} \sim \frac{m(Z,A)}{A}-m_{np}\,,
\end{align}
where $Z$ is the number of protons in the nucleus, $A$ is the total number of protons and neutrons in the nucleus, $m_p$ is the mass of the proton, $m_n$ is the mass of the neutron, and $m_{np}=\left(m_p+m_n\right)/2$ is the average nucleon mass. 
Since nuclei release energy when forming, the binding energy is negative.  Hence, stable nuclei have a larger binding energy (in absolute value) than unstable nuclei.
In the limit of an ideal, infinite system of symmetric nuclear matter, the binding energy at $n_{sat}$ is, extrapolated to be the measured volume term, $E_B\sim -16$ MeV. 

Within CMF, we can calculate both $(n_{sat},E_B)$, and we want to ensure that the results we obtain are reasonable for each parameter set value. This is especially a concern here, because we vary over a wide range of parameters, and thus, it is not a given that the values of $(n_{sat},E_B)$ we obtain are remotely reasonable for a given parameter set. To calculate these quantities, we must first calculate the CMF EoS evaluated along the symmetric-matter slice i.e. $\mu_Q=\mu_S=0$ (although see \cite{Noronha-Hostler:2026ykk} for a caveat regarding $\mu_S$ that we leave for future work). The saturation density is defined by the zero-pressure condition, $p(n_{\rm sat}) = 0$ at binding energy maximum, and the binding energy per baryon is then computed as
\begin{equation}
    E_B \equiv \frac{\varepsilon(n_{\rm sat})}{n_{\rm sat}} - m_{N},
\end{equation}
where $\varepsilon$ is the energy density, and $m_{N}$ denotes the nucleon vacuum mass for CMF. In practice, the pressure and energy density are interpolated from the tabulated CMF data using cubic splines. 
The function $n_B(p)$ is extrapolated to $p=0$ to locate $n_{\rm sat}$, and $\varepsilon(n_{\rm sat})$ is subsequently evaluated from the corresponding $\varepsilon(n_B)$ interpolation to compute $E_B$. 

\begin{figure*}[t]
    \centering
    \includegraphics[width=0.95\textwidth]{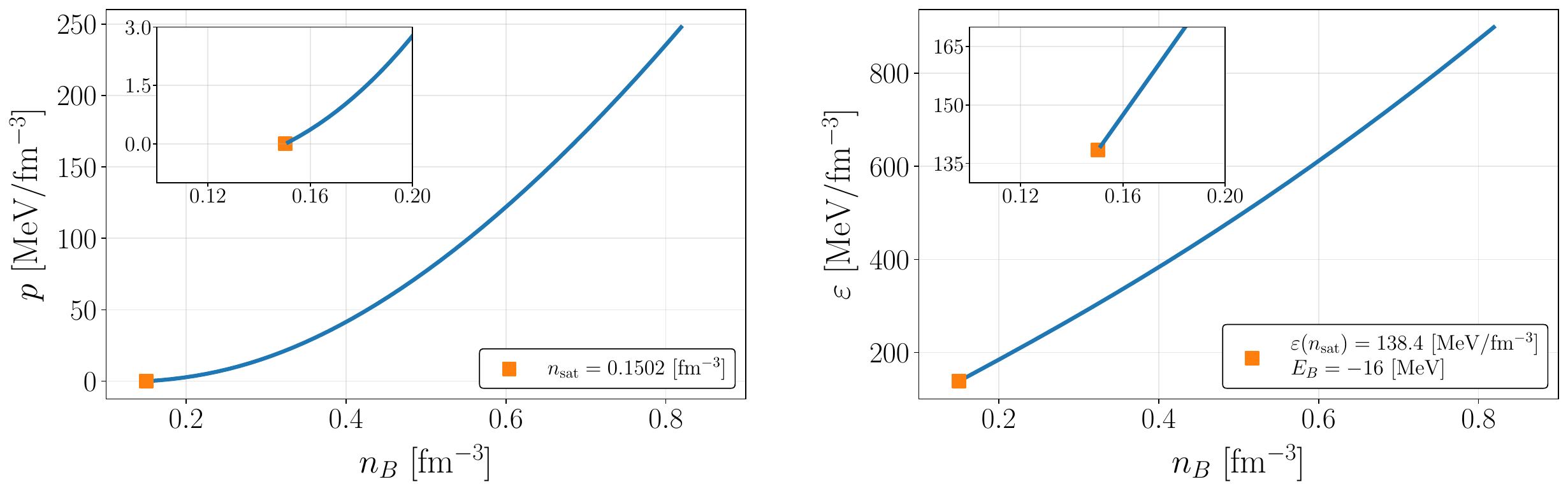}
    \caption{Diagnostics for the determination of $n_{\rm sat}$ (a) $p$ vs.\ $n_B$ and $E_B$ (b) $\varepsilon$ vs.\ $n_B$ for symmetric matter ($\mu_Q=\mu_S=0$). The saturation point is indicated by the square markers.}
    \label{fig:sat_diag}
\end{figure*}

\Cref{fig:sat_diag} illustrates the procedure. The left panel shows the zero-pressure crossing that defines $n_{\rm sat}$, while the right panel displays the corresponding point on $\varepsilon(n_B)$ used to compute the binding energy. The square markers indicate the extracted values of $(n_{\rm sat},\varepsilon_{\rm sat})$ for the reference CMF configuration.
These extracted values serve as the empirical reference for nuclear matter, defining the acceptable saturation window within which CMF parameters must lie. These quantities will be later used to establish global bounds on each parameter through an iterative scan that identifies the range consistent with saturation constraints.

\subsubsection{Range of parameters}
To ensure phenomenological viability of each CMF parameter choice, we impose broad empirical windows around the canonical values,
\begin{equation}
    n_{\rm sat}\in[0.10,\,0.20]~\mathrm{fm}^{-3}, 
    \qquad 
    E_B\in[-20,\,-10]~\mathrm{MeV},
\end{equation}
and define the admissible domain as the set of parameters for which both conditions hold. 
We purposefully include generous ranges in $n_B,E_B$ since the purpose of this study is not to find the most ideal parameter set for CMF, but rather to understand which sets of parameter lead to the largest changes in neutron-star observables. 

We chart this domain with a ``one-at-a-time'' scan. For the $i$th parameter $\lambda_i$ in the set $\lambda_k$ (other parameters fixed at their defaults, $\lambda_{k \neq i} = \lambda_{k \neq i}^*$), we locate the largest interval of $\lambda_i$ values that satisfies the saturation constraints. The search is conducted with a robust two-stage procedure:
\begin{enumerate}
    \item Outward bracketing. Starting from the default $\lambda_i^*$, we evaluate $(n_{\rm sat},E_B)$ and then expand geometrically in $\pm$ directions (step scaling factor $>1$) until the constraints are first violated, thereby bracketing each boundary.
    \item Bisection to the boundary. On each boundary of the interval, we bisect the last admissible/inadmissible bracket to machine tolerance, returning the extremal in-bounds values $\lambda_{i,\min}$ and $\lambda_{i,\max}$.
\end{enumerate}
We then repeat this procedure for the next parameter in the $\lambda_k$ set. Each $(\lambda_i, n_{\rm sat,i}, E_{B,i})$ evaluation is produced by an isolated CMF run and post-processed with the saturation extractor described above. 

\begin{figure*}[t]
    \centering
    \includegraphics[width=0.47\textwidth]{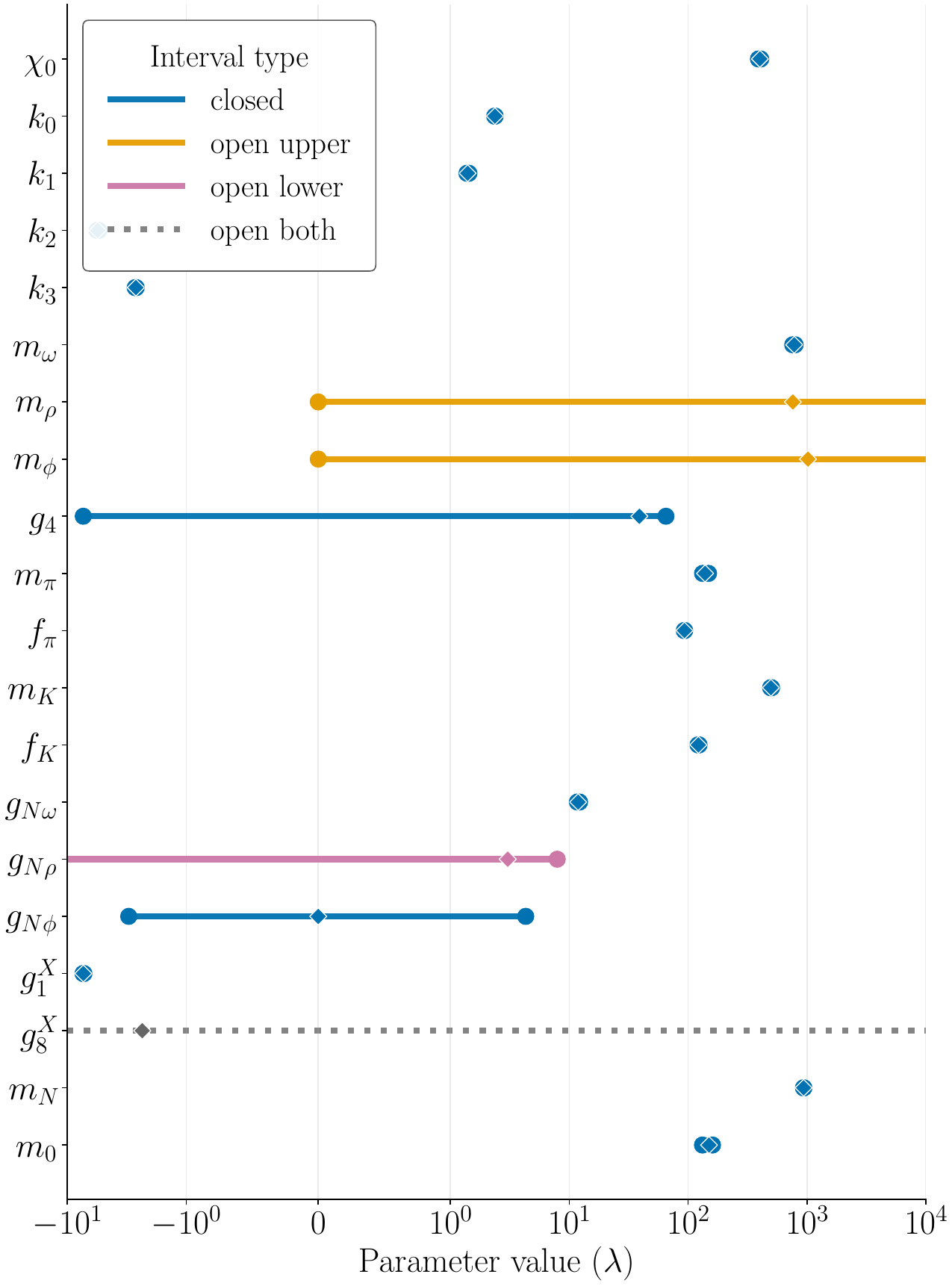} \quad
    \includegraphics[width=0.47\textwidth]{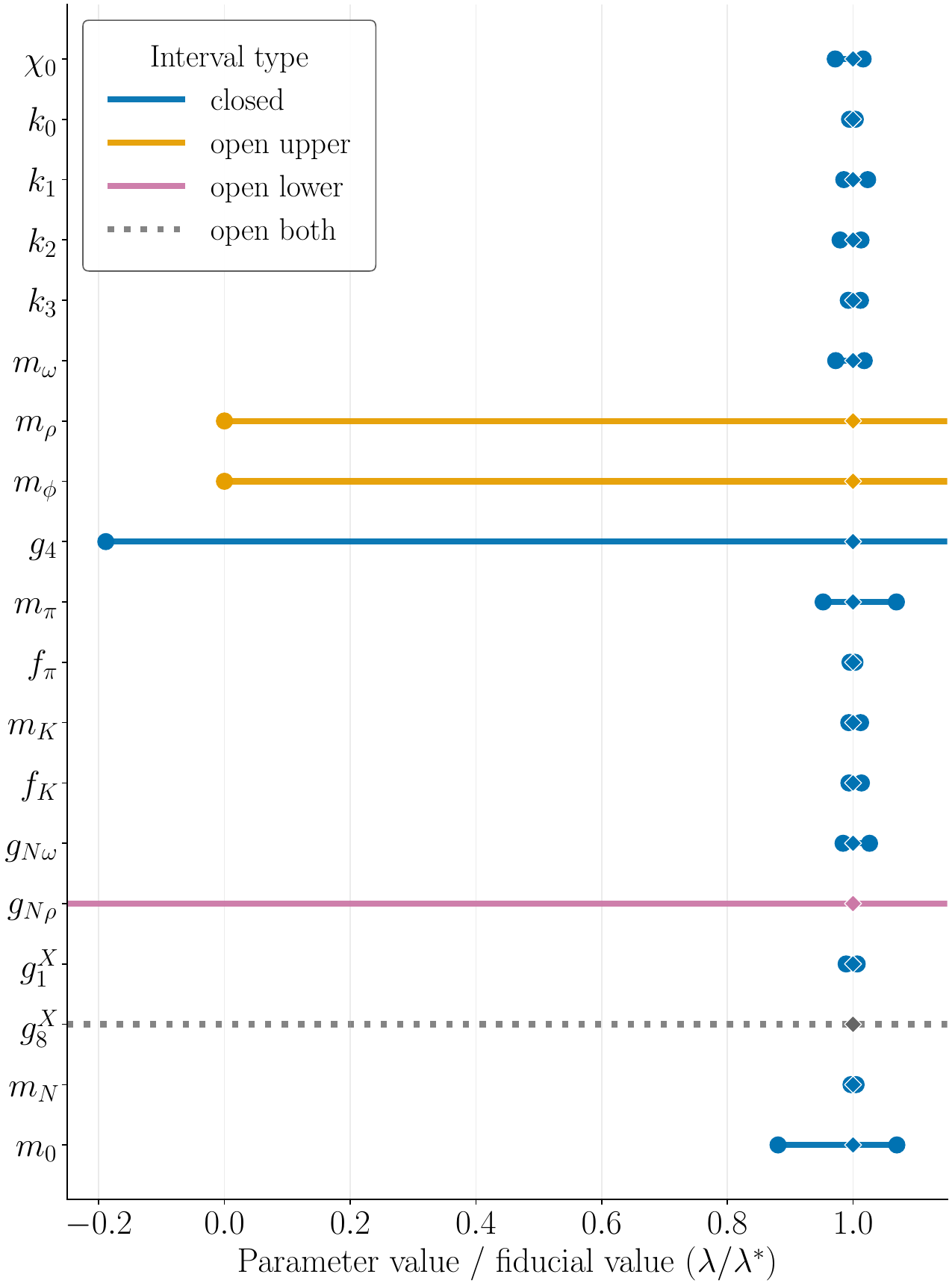}
    \caption{
    Raw parameter ranges (left) and parameter ranges rescaled by their fiducial values $(\lambda/\lambda^\ast)$ (right). These intervals are obtained from a bisection search that enforces the saturation conditions $n_{\mathrm{sat}}\!\in\![0.10,0.20]~\mathrm{fm}^{-3}$ and $E_B\!\in\![-20,-10]~\mathrm{MeV}$. The diamond symbols represent the fiducial values. Blue segments (“closed”) indicate parameters for which the saturation constraints determine both the lower and upper ends of the allowed interval. Orange (“open upper”) indicates that the constraints determine only the lower bound, with no upper violation found. Magenta (“open lower”) indicates that only the upper bound is set by the constraints. Gray, dotted segments (“open both”) correspond to parameters that satisfy the saturation constraints throughout the entire explored domain, with neither end constrained. The coupling $g_{N\phi}$ is omitted in the right panel because its fiducial value is zero.
    }
    \label{fig:param_bounds}
\end{figure*}

The resulting allowed intervals $[\lambda_{k,\min},\lambda_{k,\max}]$ for all $k$ parameters is stored and visualized as a horizontal segment (\Cref{fig:param_bounds}). This yields per-parameter global ranges consistent with saturation physics, which we will subsequently use to define priors and safe step sizes for the Fisher-like derivatives.
The left panel of \Cref{fig:param_bounds} shows that some of the parameters can vary widely and still yield reasonable saturation properties, although more stringent experimental constraints already exist for them. For instance, $m_\rho=775.11\pm0.34$ (for charged $\rho$ mesons) from the most recent Particle Data Group booklet \cite{ParticleDataGroup:2024cfk}; this range of allowed $m_\rho$ values is much tighter than what is shown in \Cref{fig:param_bounds}. That being said, in our work, it is illustrative to discover if varying $m_\rho$ widely strongly affects neutron star observables or not. 
The right panel of \Cref{fig:param_bounds} shows the same parameter ranges as the left panel, but rescaled by the fiducial values of the parameters $\lambda^*$. Such a rescaling is important to see how potentially sensitive a parameter may be to the range that we use in this work. For instance, some parameters might always stay close to the fiducial value, whereas others can vary widely.  This behavior is particularly visible for $g_8^X$, which appears effectively unbounded in the scan. 
This occurs because its contribution to the nucleon mass,
\begin{equation}
m_{N}=m_{0}
+\frac{g_{1}^{X}}{\sqrt{3}}(\sqrt{2}\sigma_{0}+\zeta_{0})
-\frac{g_{8}^{X}}{3}(4\alpha_{X}-1)(\sqrt{2}\zeta_{0}-\sigma_{0}) .
\end{equation}
can always be compensated by adjusting $\alpha_X$. As a consequence, the nucleon mass gap equation can remain satisfied while $g_8^X$ varies widely, leaving the saturation properties $n_{\mathrm{sat}}$ and $E_B$ largely unaffected.

\subsection{Crust EoS and matching to core EoS}
\label{sec:crust_core_match}
%

\subsubsection{Crust EoS}
At sub-saturation densities, we will describe the outer and inner crust of neutron stars with the SLy EoS, which we obtain from the \textsc{CompOSE} database~\cite{CompOSECoreTeam:2022ddl}. The SLy EoS is based on the Skyrme-type effective nuclear interaction SLy4, calibrated to reproduce empirical nuclear masses, charge radii, and microscopic neutron-matter calculations. The SLy EoS' unified treatment of the outer and inner crust ensures a thermodynamically consistent transition to homogeneous nuclear matter. 

\subsubsection{Matching strategy}
Let us now construct a smooth crust-to-core transition that yields a unified EoS. We do so here by ``merging'' the speeds of sound squared, $c_s^2$ computed with the core and the crust EoSs through a transition function, and then reconstructing $p(\varepsilon)$ by integration. As explained in~\cite{ReinkePelicer:2025vuh}, this merging method introduces the least number of undesired features into the $c_s^2$ of the merged EoS and avoids causality/stability issues. 

The merged speed-of-sound is defined to be
\begin{equation}
  c_{s,\mathrm{merged}}^2(x)
  \;=\; \left[1-w(x)\right]\,c_{s,\mathrm{SLy}}^2(x)
        \;+\; w(x)\,c_{s,\mathrm{CMF}}^2(x)\,,
\end{equation}
where $x \equiv {n_B}/{n_{\rm sat}}$, the individual speeds of sound are 
\begin{equation}
  c_{s,\mathrm{SLy,CMF}}^2(n_B) \equiv \left(\frac{dp}{d\varepsilon}\right)_{\mathrm{SLy,CMF}}\,,
\end{equation}
and $\omega(x)$ is a C$^{\infty}$ transition function. In this work, we choose the transition function 
\begin{equation}
  w(x) \;=\; \tfrac{1}{2}\Big[1+\tanh\!\Big(\frac{x-x_0}{\gamma}\Big)\Big]\,,
  \label{eq:tanh_weight}
\end{equation}
which is centered at some value $x_0$ and has a scale parameter $\gamma$ that defines the width of the transition region between the two EoS.

We wish to transition between the core and crust EoSs as smoothly as possible, and thus, we choose the transition function parameters as follows. The matching window is chosen according to the physical domains of validity of the two EoSs. The SLy crust is calibrated to nuclei and microscopic crust calculations, and thus, it is reliable up to densities past $n_{\rm sat}$, while the CMF model is intended for uniform nuclear matter and becomes accurate soon after $n_{\rm sat}$. By $n_B=2\,n_{\rm sat}$, more complex physics may appear (e.g. chiral symmetry, new degrees of freedom, phase transitions etc) that CMF can take into account. We therefore place the transition between $n_{\rm sat}$ and $2\,n_{\rm sat}$, corresponding to $x_{\min} = 1$ and $x_{\max} = 2$. To keep the transition symmetric about the midpoint, we then choose $x_0 = 1.5$. 
The scale function $\gamma$ is chosen to ensure that the merged speed of sound is close to the crust and the core EoSs at $x_{\rm min}$ and $x_{\rm max}$ respectively. We enforce this by defining
\begin{equation}
  w(x_{\min}) = \delta_{x_{\min}}, 
  \qquad 
  w(x_{\max}) = 1 - \delta_{x_{\max}},
  \label{eq:edge_values}
\end{equation}
for some small tolerances $\delta_{x_{\min}}$ and $\delta_{x_{\max}}$ (typically $\sim10^{-3}$), and then solving Eq.~\eqref{eq:edge_values} for $\gamma$, to obtain
\begin{equation}
  \gamma = 
  \min\!\left[
  \frac{x_{\min}-x_0}{\operatorname{arctanh}(2\delta_{x_{\min}}-1)},\,
  \frac{x_{\max}-x_0}{\operatorname{arctanh}(1-2\delta_{x_{\max}})}
  \right].
  \label{eq:gamma_anchor}
\end{equation}
The resulting $w(x_{\min})$ and $w(x_{\max})$ are exponentially small, ensuring $C^1$ smoothness in $c_s^2$ and consequently in $p(\varepsilon)$. See~\cite{ReinkePelicer:2025vuh} for more details on the EoS matching parameter influence on neutron-star observables.
Given \(c_{s,\mathrm{merged}}^2(x)\), we obtain the EoS by integrating the thermodynamic system (with independent variable \(n_B\)) 
\begin{eqnarray}
    \frac{dp}{dn_B}& =& c_{s,\mathrm{merged}}^2(n_B)\,\frac{\varepsilon+p}{n_B}=c_{s,\mathrm{merged}}^2(n_B)\,\mu_B ,\\
    \frac{d\varepsilon}{dn_B} &=& \frac{\varepsilon+p}{n_B}=\mu_B\,.
\end{eqnarray}
Recall that these equations only hold for a one-dimensional EoS (i.e. $T=0$, $\beta$-equilibrium), which is the case we study here. For the integration, we use a standard fourth-order Runge-Kutta method. The initial conditions are taken from the last SLy point \(\big(n_B^\mathrm{init}=n_{\rm SLy}\big)\): 
\(
p(n_B^\mathrm{init})=p_{\rm SLy},\;
\varepsilon(n_B^\mathrm{init})=\varepsilon_{\rm SLy}.
\)
We then evaluate \(\mu_B=(\varepsilon+p)/n_B\) along the merged branch to populate the output table.

\begin{figure*}[t]
    \centering
    \includegraphics[width=0.95\textwidth]{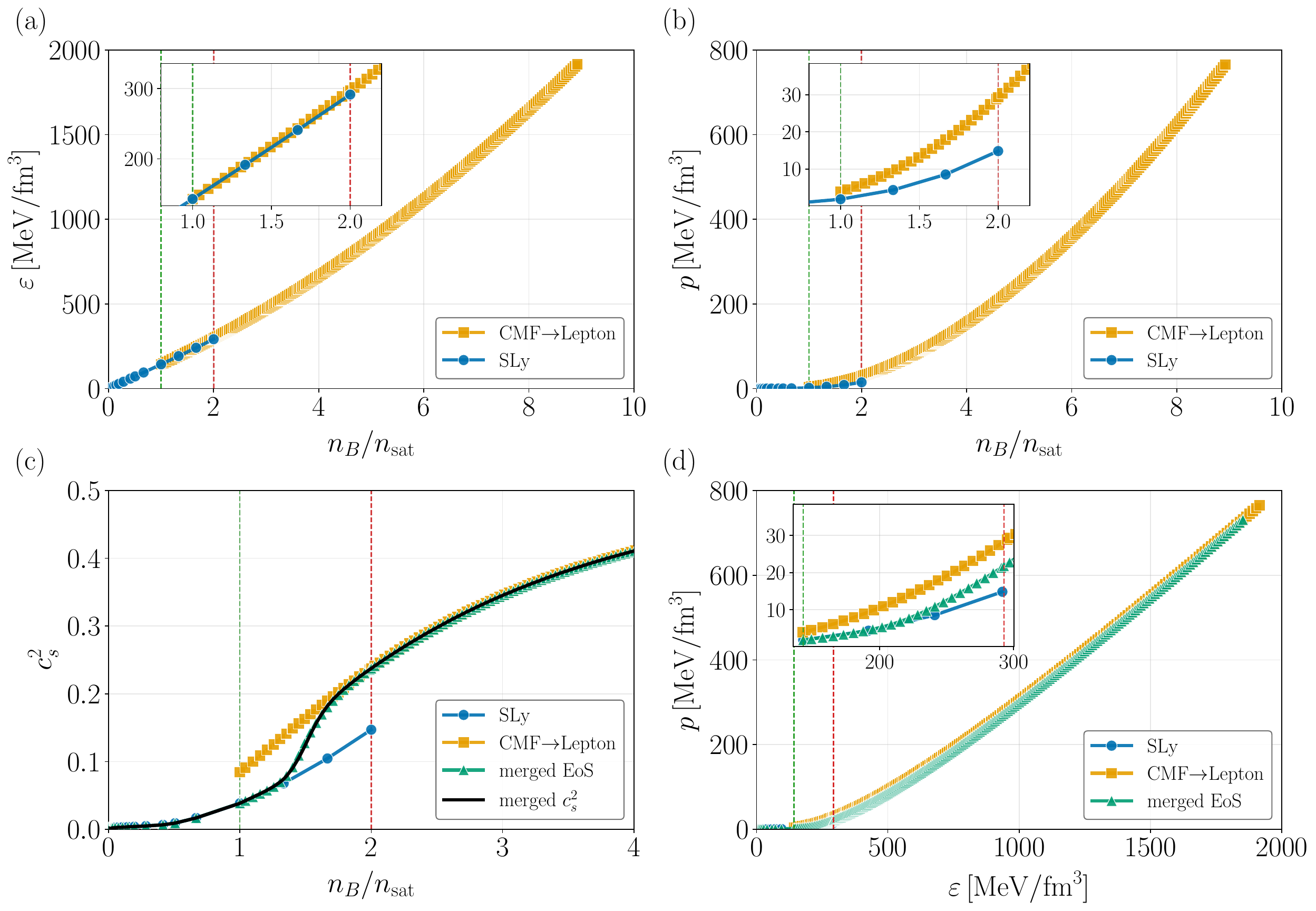}
    \caption{
    Diagnostic panels of the crust to core connection. 
    (a) Energy density versus normalized baryon density for the SLy (crust) and CMF$\to$Lepton (core) barotropes. 
    (b) Pressure versus normalized baryon density for the same EoSs. 
    (c) Squared speed of sound versus normalized baryon density, comparing the direct hyperbolic-tangent merge (black) 
    to the derivative of the merged EoS obtained from integration of the merged speed of sound (green triangles).
    (d) Pressure versus energy density for the merged EoS. 
    Insets in panels highlight the detailed behavior between $1$ and $2\,n_{\rm sat}$.
    }
    \label{fig:tanh_merge_result}
\end{figure*}

\Cref{fig:tanh_merge_result} summarizes the full construction of the crust to core interface. The upper panels display the input energy and pressure as functions of baryon density for the SLy crust and the CMF$\to$Lepton core, providing the boundary conditions for the matching window. Panel~(c) shows the speeds of sound of the SLy and CMF$\to$Lepton EoSs, as well as the speed of sound of the merged EoS. As a diagnostic, we also show here the speed of sound obtained by differentiating the merged EoS (after integrating the merged speed of sound). 
The near coincidence of the two merged curves demonstrates that the reconstruction procedure is thermodynamically consistent and that the merged branch preserves $C^1$ continuity in $p(\varepsilon)$. Panel~(d) presents the merged EoS (obtained after integrating the merged speed of sound), as well as the SLy and CMF EoSs. The insets in all panels highlight the $(1$--$2)\,n_{\rm sat}$ region, where the transition occurs, revealing smooth and monotonic behavior.
%

\subsubsection{Diagnostics and safeguards}
Before merging, both SLy and CMF tables are cubic-interpolated in $n_B$ to include (i) points at \(n_B=\{n_{\rm sat},\,2n_{\rm sat}\}\), and (ii) a set of \(n_B\) points that contain both SLy and CMF domains to compute $c_s^2$ on the same grid. After integration to recover pressure and energy density, we verify the following:
\begin{itemize}
  \item Causality and stability: \(0 \le c_s^2 \le 1\) across the merged EoS.
  \item Monotonicity: \(p(\varepsilon)\) increases monotonically across the merged EoS.
  \item Chiral Effective Field Theory consistency: for each CMF parameter set, we build the cold, $\beta$-equilibrated CMF$\to$Lepton barotrope, interpolate the pressure to \(n_B = 2\,n_{\rm sat}\), and require that it lies within the chiral-EFT high-plausibility band
  \(6 \le p(2n_{\rm sat}) \le 50~\mathrm{MeV/fm}^3\)~\cite{Huth:2021bsp}. Parameter sets that fail this check are discarded and not used for crust matching or TOV integration.
\end{itemize}
These steps ensure a physical and continuous, differentiable (\(C^1\) in \(p(\varepsilon)\)), and causal connection between the SLy crust and the CMF core, which is suitable for TOV integration and for the solution of the linearized Einstein Equations, deployed in QLIMR, as we discuss in the next section.
%

\section{Neutron Star Observables}
\label{sec:nsobs}
In this work, we aim to identify which parameters of the CMF model (while keeping the crust and the matching parameters fixed) have the strongest impact on neutron-star macroscopic observables when varied around a fiducial reference model. We focus on quantities that can be constrained by current astrophysical observations, namely the gravitational mass \(M\), the radius \(R\), and the static dimensionless tidal deformability \( \Lambda\)  of a neutron star. 
The first two quantities, \(M\) and \(R\), can be inferred from X-ray pulse-profile modeling performed by missions such as NICER~\cite{gendreau2016neutron} and the proposed eXTP~\cite{eXTP:2016rzs}. Tidal deformability \( \Lambda\) can be constrained from gravitational-wave observations of binary neutron-star mergers detected by advanced LIGO, Virgo and KAGRA~\cite{LIGOScientific:2007fwp,LIGOScientific:2014pky,VIRGO:2012dcp,Aso:2013eba}.  Tidal interactions during the inspiral contribute to the waveform phase and encode information about the stars' internal structure, including their compactness $C$~\cite{LIGOScientific:2017vwq}. We focus on these observables because they can be directly compared with astrophysical measurements. However, in this work we \textit{do not} use observational data explicitly; instead, we perform a sensitivity analysis relative to a chosen fiducial model, allowing us to quantify how variations in the CMF model parameters affect the predicted observables independently of observational measurements.

We consider a primary non-rotating neutron star in the presence of a distant secondary companion. Our focus is on extracting the gravitational mass \(M\), the radius \(R\), the compactness $C$, and the dimensionless tidal deformability $\Lambda$ of the primary star. We assume that the  gravitational field of the distant companion perturbs the originally-spherically-symmetric primary, producing predominantly a quadrupolar deformation. The \(+z\) axis is aligned along the line connecting the centers of the two stars, so that the system retains axial symmetry around this direction, simplifying the analysis of the perturbed geometry~\cite{Hinderer:2009ca}. 

The spacetime of the primary star is described using a small-deformation expansion \cite{Hinderer:2007mb}. In Schwarzschild-type coordinates \((t, r, \theta, \phi)\), the metric can be written as
\begin{align}
\label{eq:pmetric}
ds^{2} &= -e^{\nu(r)}\left[ 1 + \epsilon\, \tilde{H}(r,\theta) \right] dt^{2} 
+ e^{ \beta(r) }\left[ 1 + \epsilon\, H(r,\theta) \right] dr^{2} \nonumber \\
&\quad + r^{2}\left[ 1 + \epsilon\, K(r,\theta) \right] 
\left( d\theta^{2} + \sin^{2}\theta\, d\phi^{2} \right) \, ,
\end{align}
where \(\epsilon\) is a bookkeeping parameter that tracks the linear correction away from spherical symmetry due to the tidal deformation induced by the companion. The perturbed metric functions \(\tilde{H}\), \(H\), and \(K\) depend only on radial \(r\) and angular \(\theta\), since the configuration preserves axial symmetry about the \(+z\)-axis aligned with the line connecting the centers of the two stars. 

The neutron-star stellar matter is modeled as a perfect fluid, neglecting viscous and heat-conduction effects. The EoS is assumed to be barotropic, \(p = p(\varepsilon)\), since the ``Fermi temperature'' of the star is much higher than its thermal temperature, making thermal excitations negligible. Magnetic fields are also neglected, as their contribution to the total energy density is generally small, except in magnetars or proto-neutron stars, which are not considered here. With this in mind, the stress-energy tensor is 
\begin{align}
    T_{\mu\nu} = (\mathcal{E} + \mathcal{P})\, U_{\mu}U_{\nu} + \mathcal{P}\, g_{\mu\nu} \, ,
\end{align}
where $\mathcal{E}$ is the energy density, $\mathcal{P}$ is the pressure, and $u^{\mu}$ is the (timelike) four velocity. These fields acquire small corrections due to the external gravitational field, and so, they can be expanded as
\begin{align}
\label{eq:ppert}
\mathcal{E}(r,\theta) &= \varepsilon(r) + \epsilon\, \delta \varepsilon(r,\theta) \, , \\
\label{eq:epert}
\mathcal{P}(r,\theta) &= p(r) + \epsilon\, \delta p(r,\theta) \,, \\
U_{\mu}(r,\theta) &= u_{\mu}(r) + \epsilon\, \delta u_{\mu}(r,\theta) \,,
\label{eq:upert}
\end{align}
where \(\varepsilon\), \(p\), and $u^\mu$ are the unperturbed fields and depend only on the radial coordinate \(r\) due to spherical symmetry. Since the (perturbed or unperturbed) four-velocity of the fluid is purely timelike, we can write
\begin{equation}
    U^{\mu} = (U^{t}, 0, 0, 0) \, ,
\end{equation}
where \(U^{t}\) is determined from the normalization condition \(U^{\alpha}U_{\alpha} = -1\). 
The fields \(\delta \varepsilon\), \(\delta p \), and \(\delta u_{\mu}\) represent the linear deviations induced by the external tidal field. In the following two subsections, we outline the equations required to obtain the observables \(M\), \(R\), $C$, and \(\Lambda\) of the primary star.

\subsection{Order $\mathcal{O}(\epsilon^{0})$: Mass $M$, radius $R$ and compactness $C$}
The unperturbed configuration is fully described by the well-known Tolman-Oppenheimer-Volkoff (TOV) equations, which are recovered by setting \(\epsilon = 0\).
From the \((t,t)\) and \((r,r)\) components of Einstein’s field equations, we obtain
\begin{align}
\label{eq:m}
\dfrac{dm}{dr} &= 4 \pi r^{2} \varepsilon \, , \\ 
\label{eq:nu}
\dfrac{d\nu}{dr} &= 2\, \dfrac{m + 4\pi r^{3} p}{r(r - 2m)} \, ,
\end{align}
where \(m = m(r)\) is the enclosed gravitational mass-energy function of the unperturbed star within radius $r$, related to the metric function \(\beta(r)\) through
\begin{equation}
e^{\beta(r)} \equiv \left[ 1 - \dfrac{2m(r)}{r} \right]^{-1} \, .
\end{equation}
Additionally, the \(r\)-component of the energy-momentum conservation equation, \(\nabla^{\mu} T_{r\mu} = 0\), together with Eq.~\eqref{eq:nu}, yields the relativistic equation of hydrostatic equilibrium,
\begin{equation}
\dfrac{dp}{dr} = -(\varepsilon + p)\, \dfrac{m + 4\pi r^{3} p}{r(r - 2m)} \, .
\label{eq:p}
\end{equation}
The set of Eqs.~\eqref{eq:m}, \eqref{eq:nu}, and \eqref{eq:p}, together with the EoS \(p = p(\varepsilon)\), form a closed system that determines the metric functions \(m(r)\), \(\nu(r)\), \(p(r)\), and \(\varepsilon(r)\). In practice, to compute the radius $R$ and the gravitational mass $M$, only Eqs.~\eqref{eq:m} and \eqref{eq:p} are required, since they decouple from Eq.~\eqref{eq:nu}. The stellar radius \(R\) is defined as the point where the pressure vanishes at the surface, \(p(R) = 0\), while the total gravitational mass \(M\) is obtained from the enclosed mass function evaluated at that radius, \(M = m(R)\).

For convenience, we introduce a change of variable that improves the numerical accuracy of the computation of the stellar radius \(R\) and gravitational mass \(M\) when integrating the TOV equations, as done in~\cite{lindblom1992determining}. We define a new dimensionless variable \(h \) such that
\begin{equation}
\label{eq:h}
dh \equiv \dfrac{dp}{\varepsilon + p} = \dfrac{d\mu_{B}}{\mu_{B}},
\end{equation}
where \(\mu_{B} = (\varepsilon + p)/n_{B}\) is the baryon chemical potential and \(n_{B}\) the baryon number density at $T=0$~\cite{Lattimer:2010uk}. Accordingly, Eqs.~\eqref{eq:p} and~\eqref{eq:m} can be rewritten, respectively, as
\begin{align}
\label{eq:rh}
\dfrac{dr}{dh} &= - \dfrac{r(r - 2m)}{m + 4\pi r^{3} p} \, , \\
\dfrac{dm}{dh} &= - 4 \pi r^{2} \varepsilon \, \dfrac{r(r - 2m)}{m + 4\pi r^{3} p} \, .
\label{eq:mh}
\end{align}
The set of Eqs.~\eqref{eq:h}, \eqref{eq:rh}, and~\eqref{eq:mh}, together with the EoS $p=p(\varepsilon)$, form a closed system for the variables $p(h)$, $\varepsilon(h)$, $r(h)$, and $m(h)$. 
The relations $\varepsilon = \varepsilon(h)$ and $p = p(h)$ can be obtained by integrating Eq.~\eqref{eq:h}, yielding
\begin{align}
\label{eq:hofp}
h(p) &= \int_{0}^{p} \dfrac{dp'}{\varepsilon(p') + p'} \, , \\
h(\varepsilon) &= \int_{0}^{\varepsilon} \dfrac{c_{s}^{2}(\varepsilon')}{\varepsilon' + p(\varepsilon')} \, d\varepsilon' \, ,
\label{eq:hofe}
\end{align}
where $c_{s}^2$ denotes the speed of sound squared in the stellar matter, calculated at $\beta$-equilibrium. In practice, we first evaluate the integral in Eq.~\eqref{eq:hofe} to obtain $h(\varepsilon)$ and then invert the result to express $\varepsilon$ as a function of $h$, i.e., $\varepsilon = \varepsilon(h)$. The pressure $p(h)$ is then computed from the EoS as $p(h) = p[\varepsilon(h)]$.

We solve Eqs.~\eqref{eq:rh} and~\eqref{eq:mh} numerically from the center of the star out to its surface. Since these equations become singular at the stellar center, we perform an asymptotic expansion around the value of $h$ at the center, denoted by \(h_{c}\), in order to obtain regular boundary conditions. Because Eqs.~\eqref{eq:rh} and~\eqref{eq:mh} diverge exactly at \(h = h_{c}\), we start the integration from a nearby value, \(h = h_{\epsilon}\gtrsim h_c\), slightly away from the center. The boundary conditions are then specified as
\begin{align}
\label{eq:rofh}
r(h_{\epsilon}) &\simeq  \left[ \dfrac{3(h_{c} - h_{\epsilon})}{2\pi (\varepsilon_{c} + 3p_{c})} \right]^{1/2} \equiv r_{\epsilon}, \\[0.2cm]
m(h_{\epsilon}) &\simeq \frac{4\pi}{3} \varepsilon_{c} r_{\epsilon}^{3},
\end{align}
where \(\varepsilon_{c}\) is the energy density at the stellar center, and \(r_{\epsilon}\) is a small radial distance from the center, typically of order \(\sim 40~\mathrm{cm}\), used to initialize the integration. 

Since Eqs.~\eqref{eq:rh} and~\eqref{eq:mh} must be integrated from the center up to the stellar surface, the boundary of the star is defined by the condition \(p(h=0) = 0\), as can be seen from Eq.~\eqref{eq:hofp}. Therefore, the integration domain in the new variable \(h\) extends from \(h=h_{\epsilon}\) to \(h=0\), where \(h_{\epsilon}\) can be obtained from Eq.~\eqref{eq:rofh} as
\begin{equation}
h_{\epsilon} = h_{c} - \dfrac{2\pi}{3} r_{\epsilon}^{2} \left( \varepsilon_{c} + 3 p_{c} \right),
\end{equation}
with \(h_{c} \equiv h(\varepsilon = \varepsilon_{c})\) as given by Eq.~\eqref{eq:hofe}. Once Eqs.~\eqref{eq:rh} and~\eqref{eq:mh} are solved, the gravitational mass $M$, the radius $R$ and the compactness $C$ of the star are found as
\begin{align}
M &= m(h=0) \, , \\
R &= r(h=0) \, , \\
C &= M/R \, .
\end{align}

\subsection{Order $\mathcal{O}(\epsilon)$: Dimensionless tidal deformability $\Lambda$}
The perturbed configuration is obtained by solving Einstein's equations for the metric perturbation given in Eq.~\eqref{eq:pmetric}. To decouple the radial and angular sectors, we decompose the quadrupolar ($\ell=2$) perturbations as
\begin{align}
\tilde{H}(r,\theta) &= \tilde{H}_{2}(r) P_{2}(\cos\theta), \\
H(r,\theta) &= H_{2}(r) P_{2}(\cos\theta), \\
K(r,\theta) &= K_{2}(r) P_{2}(\cos\theta),
\end{align}
while the matter perturbations are expressed as
\(\delta \varepsilon(r,\theta) = \delta \varepsilon(r) P_{2}(\cos\theta)\)
and
\(\delta p(r,\theta) = \delta p(r) P_{2}(\cos\theta)\),
where, for simplicity, we have omitted the \(\ell = 2\) subscript in
\(\delta \varepsilon\) and \(\delta p\). 
Defining the perturbed Einstein equations as
\(\delta E^{\alpha}_{\ \beta} \equiv \delta G^{\alpha}_{\ \beta} - \delta T^{\alpha}_{\ \beta}\),
where \(\delta G^{\alpha}_{\ \beta}\) and \(\delta T^{\alpha}_{\ \beta}\) are the linear perturbations to the Einstein tensor and the matter stress-energy tensor, respectively, we obtain relations between the perturbed metric functions for the quadrupolar mode:
\begin{align}
\label{eq:h2rule}
\delta E^{\theta}_{\ \theta} - \delta E^{\phi}_{\ \phi} = 0 &:&  \tilde{H}_{2} &= -H_{2}, \\
\delta E^{r}_{\ \theta} = 0 &:& K'_{2} &= H'_{2} + H_{2} \nu', \\
\delta E^{\theta}_{\ \theta} + \delta E^{\phi}_{\ \phi} = 0 &:& \delta p &= \frac{1}{2} H_{2} \left( \varepsilon + p \right),
\label{eq:dprule}
\end{align}
where the prime $(\, ' \,)$ denotes differentiation with respect to \(r\).
The perturbation of the energy density follows from the EoS,
\(c_{s}^{2} = dp/d\varepsilon\), giving \(\delta \varepsilon = \delta p / c_{s}^{2}\).
Finally, combining \(\delta E^{t}_{\ t} - \delta E^{r}_{\ r} = 0\) with
Eqs.~\eqref{eq:h2rule}-\eqref{eq:dprule}, we obtain
\begin{align}
\nonumber
&\dfrac{d^{2}H_{2}}{dr^{2}}
+ \left\{ \dfrac{2}{r}
+ \left[ \dfrac{2m}{r^{2}} + 4 \pi r (p - \varepsilon) \right] e^{\beta} \right\}
\dfrac{dH_{2}}{dr} \\
&- \left\{ \dfrac{6 e^{\beta}}{r^{2}}
- 4 \pi \left[ 5 \varepsilon + 9p
+ \dfrac{\left( \varepsilon + p \right)}{c_{s}^{2}} \right] e^{\beta}
+ \left( \dfrac{d\nu}{dr} \right)^{2} \right\} H_{2} = 0.
\label{eq:tidaleq}
\end{align}
Solving Eq.~\eqref{eq:tidaleq} throughout the entire spacetime allows us to extract the static dimensionless tidal deformability $\Lambda$. This quantity is formally defined in the \textit{buffer zone}, a region of spacetime sufficiently far from the surface of the star \textit{and} sufficiently from the source of the metric perturbation. Mathematicaly, the buffer zone is then defined through $\mathcal{R} \gg r \gg R$, where $\mathcal{R}$ is the radius of curvature associated with the external perturbation. We will now explain how to extract $\Lambda$ from the solution to Eq.~\eqref{eq:tidaleq}.

\subsubsection{Interior problem}
To obtain a solution for \( H_{2}(r) \) inside the star, we integrate Eq.~\eqref{eq:tidaleq} 
numerically from the stellar center to the surface. For numerical convenience, we simplify the calculation of $\Lambda$ by introducing the auxiliary variable \( y(r) = r\, H'_{2}(r) / H_{2}(r) \). In terms of \( y(r) \), Eq.~\eqref{eq:tidaleq} can be rewritten as \cite{Postnikov:2010yn}  
\begin{equation}
\label{eq:tidaleq-new}
r \dfrac{dy}{dr} + y^{2} + y\, e^{\beta} \left[ 1 + 4 \pi r^{2} ( p - \varepsilon ) \right] + r^{2} Q = 0 \, ,
\end{equation}
where  
\begin{equation}
Q = 4 \pi e^{\beta} \left( 5 \varepsilon + 9p + \dfrac{\varepsilon + p}{c^{2}_{s}} \right) - 6 \dfrac{e^{\beta}}{r^{2}} - \left( \dfrac{d\nu}{dr} \right)^{2} \, .
\end{equation}
Near the center, an asymptotic analysis yields the boundary condition \( y(r_{\epsilon}) = 2 \). The numerical integration is then carried out in the domain \( r \in [r_{\epsilon}, R] \), since Eq.~\eqref{eq:tidaleq-new} diverges exactly at \( r = 0 \).

\subsubsection{Exterior problem}
In the exterior region, the pressure and energy density vanish, i.e., \(p = \varepsilon = 0\), and Eq.~\eqref{eq:tidaleq-new} admits the analytical solution 
\begin{align} 
\label{eq:h2ext}
\nonumber h^{\textsf{ext}}_{2}(r) &= c_{1} \left( \dfrac{r}{M}\right)^{2}\left( 1 - \dfrac{2M}{r} \right) \bigg[ \dfrac{2M(r-M)(6M -3r^{2})}{r^{2}(r-2M)^{2}} \\ &+ 3 \ln \left( \dfrac{r}{r-2M} \right) \bigg] + c_{2} \left(\dfrac{r}{M} \right)^{2}\left( 1-\dfrac{2M}{r} \right) , \end{align} 
where \(c_{1}\) and \(c_{2}\) are integration constants. Using Eq.~\eqref{eq:h2ext} in the metric given by Eq.~\eqref{eq:pmetric}, the $g^{\textsf{ext}}_{tt}$ component in the \textit{buffer zone} reads
\begin{align}
\label{eq:gttext}
\nonumber
g^{\textsf{ext}}_{tt}(r,\theta) = &- 1 + \dfrac{2M}{r} + \dfrac{16}{5} c_{1} \dfrac{M^{3}}{r^{3}} P_{2}(\cos\theta)
+ \mathcal{O}\!\left( \dfrac{R^{4}}{r^{4}} \right) \\
&+\, \dfrac{c_{2}}{M^{2}} r^{2} P_{2}(\cos\theta)
+ \mathcal{O}\!\left( \dfrac{r^{3}}{\mathcal{R}} \right) \, .
\end{align}
In terms of the quadrupole moment $Q$ and the quadrupolar tidal field $\mathcal{E}$, the $g_{tt}$ component can be expanded in the \textit{buffer zone} as \cite{Yagi:2013awa, Thorne:1980ru}
\begin{align}
\label{eq:gtt}
\nonumber
g_{tt}(r,\theta) = &- 1 + \dfrac{2M}{r} + \dfrac{2Q}{r^{3}} P_{2}(\cos\theta)
+ \mathcal{O}\!\left( \dfrac{R^{4}}{r^{4}} \right) \\
&-\, \dfrac{2}{3} \mathcal{E}\, r^{2} P_{2}(\cos\theta)
+ \mathcal{O}\!\left( \dfrac{r^{3}}{\mathcal{R}} \right) \, .
\end{align}
Comparing Eq.~\eqref{eq:gttext} with Eq.~\eqref{eq:gtt} yields
\(
Q = (8/5) c_{1} M^{3}
\)
and
\(
\mathcal{E} = -(3/2)(c_{2}/M^{2}).
\)
The tidal deformability is defined as
\(
\lambda^{\textrm{tid}} \equiv - Q/\mathcal{E},
\)
which gives
\(
\lambda^{\textrm{tid}} = (16/15)(c_{1}/c_{2}) M^{5}.
\)
This quantity is related to the quadrupolar dimensionless Love number $k_{2}$ as
\begin{align}
k_{2} = \dfrac{3}{2}\, \dfrac{\lambda^{\textrm{tid}}}{R^{5}} = \dfrac{8}{5}\, C^{5}\, \dfrac{c_{1}}{c_{2}} \, .
\label{eq:k2}
\end{align}

\subsubsection{Tidal deformability extraction}
From Eq.~\eqref{eq:k2}, the Love number $k_{2}$ depends on the ratio $c_{1}/c_{2}$, which can be obtained by matching the interior and exterior solutions at the stellar surface $r = R$. Imposing continuity at the boundary gives
\begin{align}
H^{\textsf{int}}_{2}(R) = H^{\textsf{ext}}_{2}(R) \hspace{0.25cm} ; \hspace{0.25cm}
H'^{\textsf{ int}}_{2}(R) = H'^{\textsf{ ext}}_{2}(R) \, .
\end{align}
Using these conditions, one can express the ratio $c_{1}/c_{2}$ in terms of the auxiliary variable 
$y(r=R) \equiv y_{R} = R H'_{2}(R)/H_{2}(R)$
and the compactness $C$. 
Therefore, from Eq.~\eqref{eq:k2}, $k_{2}$ can be written as
\begin{align}
\nonumber
k_{2} &= \dfrac{8}{5} C^{5} (1 - 2C)^{2}
\left[ 2 + 2C (y_{R} - 1) - y_{R} \right] \\
\nonumber
&\times
\Big\{
2C \left[ 6 - 3y_{R} + 3C (5y_{R} - 8) \right] \\
\nonumber
&\quad +\, 4C^{3} \left[ 13 - 11y_{R} + C (3y_{R} - 2) + 2C^{2} (1 + y_{R}) \right] \\
&\quad +\, 3(1 - 2C)^{2}
\left[ 2 - y_{R} + 2C (y_{R} - 1) \right]
\ln (1 - 2C)
\Big\}^{-1} \, .
\end{align}
We are interested in the dimensionless tidal deformability $\Lambda$ which can be found once $k_{2}$ is found as,
\begin{equation}
\Lambda \equiv \dfrac{\lambda^{\textrm{tid}}}{M^{5}} = \dfrac{2}{3}k_{2}C^{-5} \, .
\label{eq:tidal_and_C}
\end{equation}

\subsection{QLIMR: Numerical integrator solver}
To compute the macroscopic observables of neutron stars, namely the mass $M$, radius $R$, compactness $C$, and dimensionless tidal deformability $\Lambda$, we employ the \textsc{QLIMR} module (Quadrupole moment, tidal Love number, moment of Inertia, Mass, and Radius). 
\textsc{QLIMR} is an optimized, open-source solver developed within the \textsc{MUSES} cyberinfrastructure project~\cite{ReinkePelicer:2025vuh}, designed to compute neutron-star observables from arbitrary EoSs with high numerical precision~\cite{zenodo_qlimr}.
The module is written in \texttt{C++} and employs the adaptive Runge-Kutta-Fehlberg (RK45) integrator from the GNU Scientific Library (GSL). It solves the coupled structure and perturbation equations governing non-rotating and slowly-rotating neutron stars. Specifically, \Cref{eq:rh}--\Cref{eq:mh} are integrated to obtain $M$, $R$, and $C$, while \Cref{eq:tidaleq-new} yields the tidal Love number and the corresponding tidal deformability $\Lambda$. The integration begins at a small radial coordinate $r_{\epsilon}$ using boundary conditions parametrized by the central energy density, $\varepsilon_c$, which uniquely defines each stellar model. The system is evolved outward until the stellar surface is reached, defined by the condition $p(h=0)=0$. In practice, for the equations of state considered here the tabulated pressure does not reach an exact zero. To ensure a well-defined crust surface, \textsc{QLIMR} appends an additional point to the EoS table with $\varepsilon = 0$ and $p = 0$. This construction enforces a linear decrease of the pressure with energy density between the lowest tabulated point and the origin, allowing the integrator to terminate precisely at the surface. Alternative approaches introduce a low-density polytropic extension in this region; however, because it represents a negligibly small fraction of the EoS, it is not expected to produce appreciable differences in the resulting macroscopic observables.

An important feature of \textsc{QLIMR} is its control over the \textit{resolution} of the mass--radius curve, as explained in~\cite{ReinkePelicer:2025vuh} and in the module's documentation website~\cite{muses_qlimr}. The user may specify step sizes in mass and radius, $\Delta M$ and $\Delta R$, which determine the density of points along the mass--radius curve. In this work we adopt $r_{\epsilon} = 40~\mathrm{cm}$, $\Delta M = 0.001\,M_{\odot}$, and $\Delta R = 0.001~\mathrm{km}$, resulting in sequences with $N_{\mathrm{points}} \sim 3.5\times10^{3}$. This resolution is sufficient to accurately locate the maximum-mass configuration, which is required for the sensitivity analyses discussed in the following sections.

A schematic overview of the full data-generation pipeline, from the input CMF parameter vector $\boldsymbol{\lambda}$ to the extracted stellar observables, is provided in Appendix~\ref{app:workflow}.
%

\section{A Fisher-inspired sensitivity analysis of neutron star observables}
\label{sec:Fisher_theory}
In this section, we introduce a Fisher-information--inspired framework to quantify how neutron-star observables respond to variations in the CMF model parameters. We first review the basic Fisher and PCA concepts, and then define the logarithmic sensitivity matrices that we actually construct and analyze in this work.

\subsection{Basics of Fisher Information and Principal Component Analysis}
The Fisher Information Matrix (FIM) provides a quantitative measure of how well a set of model parameters $\boldsymbol{\theta}=\{\theta_1,\dots,\theta_N\}$ can be infered, given a set of observations $\mathbf{O}=\{O_1,\dots,O_{N_{\mathrm{obs}}}\}$ (assuming high signal-to-noise ratio~\cite{Vallisneri:2007ev}). In the present context, we do not intend to compare the EoS model of \Cref{sec:workflow} to observational data. Instead, we wish to ask how sensitive the observables are to variations in the model parameters. In order to employ a Fisher-like formalism to answer this question, let us first review the standard Fisher framework in general terms.

Under standard regularity conditions (differentiability and integrability), certain assumptions on the signal-to-noise ratio and on the noise properties~\cite{Vallisneri:2007ev}, the Fisher matrix is the expectation value of the curvature of the log-likelihood $\mathcal{L}(\boldsymbol{\theta})$ (of the data given the model) around its maximum $\boldsymbol{\theta}=\boldsymbol{\theta}_{\max}$:
\begin{equation}
    F_{ij}\equiv-\left.\Big\langle\frac{\partial^{2}\ln\mathcal{L}}{\partial\theta_{i}\,\partial\theta_{j}}\Big\rangle\right|_{\boldsymbol{\theta}_{\max}}.
    \label{eq:fisher_general}
\end{equation}
For a single observable $O$, the stationary and Gaussian likelihood of obtaining a measurement 
$O_{\mathrm{obs}}$ given the model parameters $\boldsymbol{\theta}$ is
\begin{equation}
    \mathcal{L}(\boldsymbol{\theta})
    \propto
    \exp\!\left[
        -\frac{(O_{\mathrm{obs}} - O(\boldsymbol{\theta}))^2}{2\sigma^2}
    \right],
    \label{eq:gaussian_like}
\end{equation}
where $\sigma$ represents the associated uncertainty on the measurement of $O_{\rm obs}$ (which depends on the noise properties of the associated observation).
Taking the logarithm gives the log-likelihood (up to an additive constant), and twice-differentiating this with respect to the model parameters yields
\begin{equation}
    F_{ij}
    = -\Big\langle
        \frac{\partial^2 \ln \mathcal{L}}{\partial \theta_i \, \partial \theta_j}
      \Big\rangle\bigg|_{\boldsymbol{\theta}_{\max}}
    = \frac{1}{\sigma^2}
      \frac{\partial O}{\partial \theta_i}
      \frac{\partial O}{\partial \theta_j}\bigg|_{\boldsymbol{\theta}_{\max}}\,.
    \label{eq:fisher_single}
\end{equation}
In deriving this expression, we have used that Gaussian residuals have zero mean. If the likelihood has a single peak, one can relate the (inverse of the) Fisher matrix to (an upper bound on) the accuracy to which the model parameters could be measured, given an observation (a fact established through the so-called Cram\'er-Rao bound~\cite{rao1945information,cramer1946mathematical}).

The Fisher matrix allows one to determine the combination of parameters that contain the most information in the model through a principal component analysis (PCA). Let $v_i^{(\mu)}$ denote the $\mu$th normalized eigenvector of $F_{ij}$, with eigenvalue
$\tau_\mu$ ($\mu=1,\dots,N_\theta$). In component notation, the eigen-decomposition is
\begin{equation}
    F_{ij} = \sum_{\mu=1}^{N_\theta} 
    v_i^{(\mu)}\, \tau_\mu\, v_j^{(\mu)} .
    \label{eq:F_PCA}
\end{equation}
The eigenvectors $v_i^{(\mu)}$ reveal the independent parameter combinations that the observable is most sensitive to. Large eigenvalues $\tau_{\mu}$ correspond to combination of parameters that could, in principle, be measured well. 
Ranking the eigenvalues, thus, identifies the most influential parameter combinations for the given model considered. 

\subsection{Application to neutron star EoS models and astrophysical observables}
In this work, we will draw \textit{inspiration} on the Fisher information to develop a \emph{local sensitivity diagnostic}, and, then, we will perform a PCA on it. Our diagnostic will aim to quantify how sensitive a given astrophysical observable (like the mass, radius, or tidal deformability of a neutron star) is to variations in the EoS model parameters. Given this diagnostic, we will then diagonalize it to find the combination of EoS parameters that yields the largest changes in astrophysical observables.

Let us begin by defining the model, its parameters, and the observables. The model itself is the EoS described in \Cref{sec:workflow}, so the model parameters are $\boldsymbol{\theta} = \boldsymbol{\lambda} \, \cup \, \{\varepsilon_c\}$. Here, recall that $\boldsymbol{\lambda}$ are the CMF parameters, while $\varepsilon_c$ is the central energy density of the neutron star. In this work, we will imagine that some astrophysical data set has yielded the following observations: 
\[
\mathbf{O} = \{ M,\, R,\, \Lambda,\, C,\, M_{\mathrm{max}},\, R_{\mathrm{max}} \},
\]
representing the stellar mass, radius, tidal deformability, compactness, maximum mass, and the radius at the maximum mass of a neutron star, respectively. Note that one astrophysical observation of a given neutron star (with a given central denstiy) may yield values for $\mathbf{O}' = \{ M,\, R,\, \Lambda,\, C\}$ for that star, while $\mathbf{O}'' = \{M_{\mathrm{max}},\, R_{\mathrm{max}}\}$ are properties typically derived from an ensemble of observations.  Let us then denote the $\alpha$-th element of $\mathbf{O}'$ via $O^{\alpha}$.

Imagine now that $N$ (independent) astrophysical observations have yielded measurements $O^{\alpha}$, but for $N$ different neutron stars (and thus, for $N$ different central densities).  Given all of this data, we can define the \textit{global sensitivity measure} for the $\alpha$-th observable via
\begin{equation}
    F_{ij}^{\alpha} =
\sum_{\epsilon_{c,k}=\epsilon_{c,1}}^{\epsilon_{c,N}}
    \frac{1}{\sigma_\alpha^2}
    \frac{\partial O_\alpha}{\partial \theta_i}
    \frac{\partial O_\alpha}{\partial \theta_j}\bigg|_{ 
    \boldsymbol{\lambda}^* \cup \varepsilon_{c,k}}\,,
    \label{eq:fisher_sum_eps}
\end{equation}
where the derivatives are evaluated at the $\epsilon_c$ associated with the $k$-th neutron star observation, and we have assumed the ``maximum of the likelihood'' occurs at the fiducial value of the EoS parameters $\boldsymbol{\theta}_{\max} = \boldsymbol{\lambda}^*$. The expression above forms the basis of our Fisher-inspired sensitivity analysis: it measures how infinitesimal perturbations in each element of $\boldsymbol{\theta} = \boldsymbol{\lambda} \cup \{\varepsilon_{c}\}$ affect the collective set of neutron-star observables, weighted by their respective uncertainties. Henceforth, we set these uncertainty weights to unity, as we do not wish to privilege a certain observation (at a given central density) over another. 

Because the EoS parameters span multiple orders of magnitude and carry different physical units, we reformulate the sensitivity measure in terms of \textit{relative or logarithmic, global sensitivity matrices}, namely 
\begin{equation}
   S_{ij}^{\alpha} =
    \sum_{\epsilon_{c,k}=\epsilon_{c,1}}^{\epsilon_{c,N}}
    \frac{\partial \ln O_\alpha}{\partial \ln \theta_i}
    \frac{\partial \ln O_\alpha}{\partial \ln \theta_j}\bigg|_{ 
    \boldsymbol{\lambda}^* \cup \varepsilon_{c,k}}\,.
    \label{eq:fisher_relative}
\end{equation}
Doing so achieves scale invariance and eliminates unit dependence across the parameter space, allowing us to compare different parameters on equal footing, regardless of their absolute magnitudes.

Diagonalization of $S_{ij}^{\alpha}$ proceeds analogously to Eq.~\eqref{eq:F_PCA}. Let $w_i^{(\mu,\alpha)}$ denote the $\mu$th normalized eigenvector of $S_{ij}^{\alpha}$, and let $\eta_\mu^{(\alpha)}$ be the corresponding eigenvalue ($\mu=1,\dots,N_\theta$). In component notation, the decomposition reads
\begin{equation}
    S_{ij}^{\alpha}
    = \sum_{\mu=1}^{N_\theta}
      w_i^{(\mu,\alpha)}\,
      \eta_\mu^{(\alpha)}\,
      w_j^{(\mu,\alpha)} .
    \label{eq:S_PCA}
\end{equation}
The eigenvalues $\eta_\mu^{(\alpha)}$ quantify the strength of independent sensitivity modes for the observable $O_\alpha$, while the eigenvectors $w_i^{(\mu,\alpha)}$ identify the effective linear combinations of EoS parameters that govern the astrophysical observable. Large eigenvalues correspond to dominant directions in parameter space.

\subsection{Numerical implementation}
The logarithmic sensitivity matrices used in this work are constructed from numerical derivatives of neutron-star observables with respect to the extended parameter set $\boldsymbol{\theta} = \boldsymbol{\lambda} \cup \{\varepsilon_c\}$. All derivatives are evaluated around the fiducial EoS parameters $\boldsymbol{\lambda}^*$ (see \Cref{sec:parameter_set}) and eleven sampled central densities $\{\epsilon_{c,1}^*, \ldots, \epsilon_{c,11}^*\}$ that correspond to neutron stars from 1 to 2 $M_\odot$ in steps of 0.1 $M_\odot$ for the fiducial case. For each EoS parameter $\lambda_j$ and sampled central energy density $\varepsilon_{c,j}$ needed to compute the numerical derivatives, the observables $\mathbf{O}'$ are computed through the coupled CMF-QLIMR workflow.

To estimate the derivatives $\partial O_\alpha / \partial \lambda_j$, a diverse collection of numerical differentiation schemes was studied first to ensure robustness and cross-validation. These included forward and backward finite differences, two- and four-point central differences, linear fits over two to five sampling points, and cubic polynomial fits evaluated analytically at the fiducial point, all using multiple step sizes to verify numerical stability. All methods were systematically compared across a wide range of logarithmically-spaced step sizes to assess convergence and numerical consistency. After experimentation across observables, the four-point (fourth-order) central-difference scheme was selected for CMF parameter derivatives. Other methods give consistent results (albeit slightly noisier). In contrast, the derivatives with respect to the central energy density, $\partial O_\alpha / \partial \varepsilon_c$, are obtained using a standard second-order central-difference scheme, as the corresponding observables vary smoothly along the fiducial stellar sequence. A detailed description of the derivative algorithms and stencil formulas is provided in \Cref{app:numerical_derivatives}. 

\section{Results}
\label{sec:results}
In this section, we present our EoS sensitivity results. We first define the different, astrophysical-motivated cases we will consider. We then present a first look at the parameter sensitivity of the EoS itself, the observables themselves and, the logarithmic sensitivity matrices. We conclude with the calculation of the principal directions in the EoS parameter space. 

\subsection{Cases studied}
In order to estimate EoS sensitivity and principal directions in the EoS parameter space, we must choose a set of astrophysical observables to study. In this work, we construct the following logarithmic sensitivity matrices [using Eq.~\eqref{eq:fisher_relative}] :
\begin{align}
  S^{M}_{ij}           &\quad \text{(stellar mass)}, \nonumber\\
  S^{R}_{ij}           &\quad \text{(radius)}, \nonumber\\
  S^{\Lambda}_{ij}     &\quad \text{(tidal deformability)}, \nonumber\\
  S^{C}_{ij}           &\quad \text{(compactness)}.
\end{align}
For each of these, we consider 11 observed astrophysical data points, corresponding to 11 different neutron stars, with 11 different central densities $\epsilon_{c,k}$. Thus, in Eq.~\eqref{eq:fisher_relative}, the sum runs over $N=11$ central energy densities, $\epsilon_{c,k}$ with $k=1,\dots,11$.

In addition to these logarithmic sensitivity matrices, we here also consider the following
\begin{align}
  S^{M_{\max}}_{ij}    &\quad \text{(maximum mass)}, \nonumber\\
  S^{R_{\max}}_{ij}    &\quad \text{(radius at maximum mass)},
\end{align}
which we assume has been inferred. In this case, the logarithmic sensitivity matrices of Eq.~\eqref{eq:fisher_relative} do not contain a sum over central densities, but rather, the derivatives are evaluated at the maximum central density of the sequence. We here do not consider the possibility of mass twins due to first-order phase transitions. 

Finally, we consider the combined logarithmic sensitivity matrix
\begin{align}
  S^{M,\Lambda}_{ij}   &\quad \text{(mass-tidal deformability)} \nonumber,
\end{align}
which is defined through Eq.~\eqref{eq:fisher_relative_ML}, but with $O_\alpha$ summed over both mass and tidal deformability, namely
\begin{align}
   S^{M,\Lambda}_{ij} =
    \sum_{\epsilon_{c,k}=\epsilon_{c,1}}^{\epsilon_{c,N}}
    & \left(  \frac{\partial \ln M}{\partial \ln \theta_i}
    \frac{\partial \ln \Lambda}{\partial \ln \theta_j}
    \right. \nonumber \\
    &+ \left.  \frac{\partial \ln \Lambda}{\partial \ln \theta_i}
    \frac{\partial \ln M}{\partial \ln \theta_j} \right)\bigg|_{ 
    \boldsymbol{\lambda}^* \cup \varepsilon_{c,k}}
    \,.
    \label{eq:fisher_relative_ML}
\end{align}
This combined logarithmic sensitivity matrix is a way to quantitatively assess the simultaneous sensitivity $M$ and $\Lambda$ to variations in the EoS parameters.

To explore combined sensitivities of different observables, we define several aggregate logarithmic sensitivity matrices corresponding to distinct possible/future observational channels, namely
\begin{description}
  \item[\textbf{LIGO case}] 
  combines information from the stellar mass, tidal deformability, and their cross term,
  \[
    S^{\mathrm{LIGO}} = S^{M} + S^{\Lambda} + S^{M,\Lambda}.
  \]
  This case attempts to represent the EoS parameter sensitivity accessible to current gravitational-wave observations of binary neutron-star coalescence.
  \item[\textbf{N+LIGO case}]
  extends the LIGO case by including compactness,
  \[
    S^{\mathrm{N+LIGO}} = S^{\mathrm{LIGO}} + S^{C}.
  \]
  The name “N+LIGO’’ attemps to reflect the combined information from NICER and LIGO (i.e.~gravitational-wave and X-ray pulse-profile) observations, providing complementary constraints on both tidal deformability and radius (or compactness).  
  This setup attempts to approximate a near-future multi-messenger regime, where these observables are jointly measured with high precision.
  \item[\textbf{Futuristic case}]
  aggregates all macroscopic observables, including maximum-mass and radius-at-maximum configurations,
  \[
    S^{\mathrm{Futuristic}} = S^{\mathrm{N+LIGO}} + S^{M_{\max}} + S^{R_{\max}}.
  \]
  This comprehensive case encapsulates the full phenomenological sensitivity of the EoS model across the neutron-star sequence, representing an idealized “complete-information’’ limit.
\end{description}
Each case provides a different level of physical completeness: from gravitational-wave-based observations (LIGO), through multi-messenger extensions that combine LIGO and NICER (N+LIGO), to a full theoretical bound on the model’s macroscopic variability (Futuristic). 

\subsection{A first look at parameter sensitivity of the EoS}
\begin{figure*}[t]
  \centering
  \includegraphics[width=0.95\textwidth]{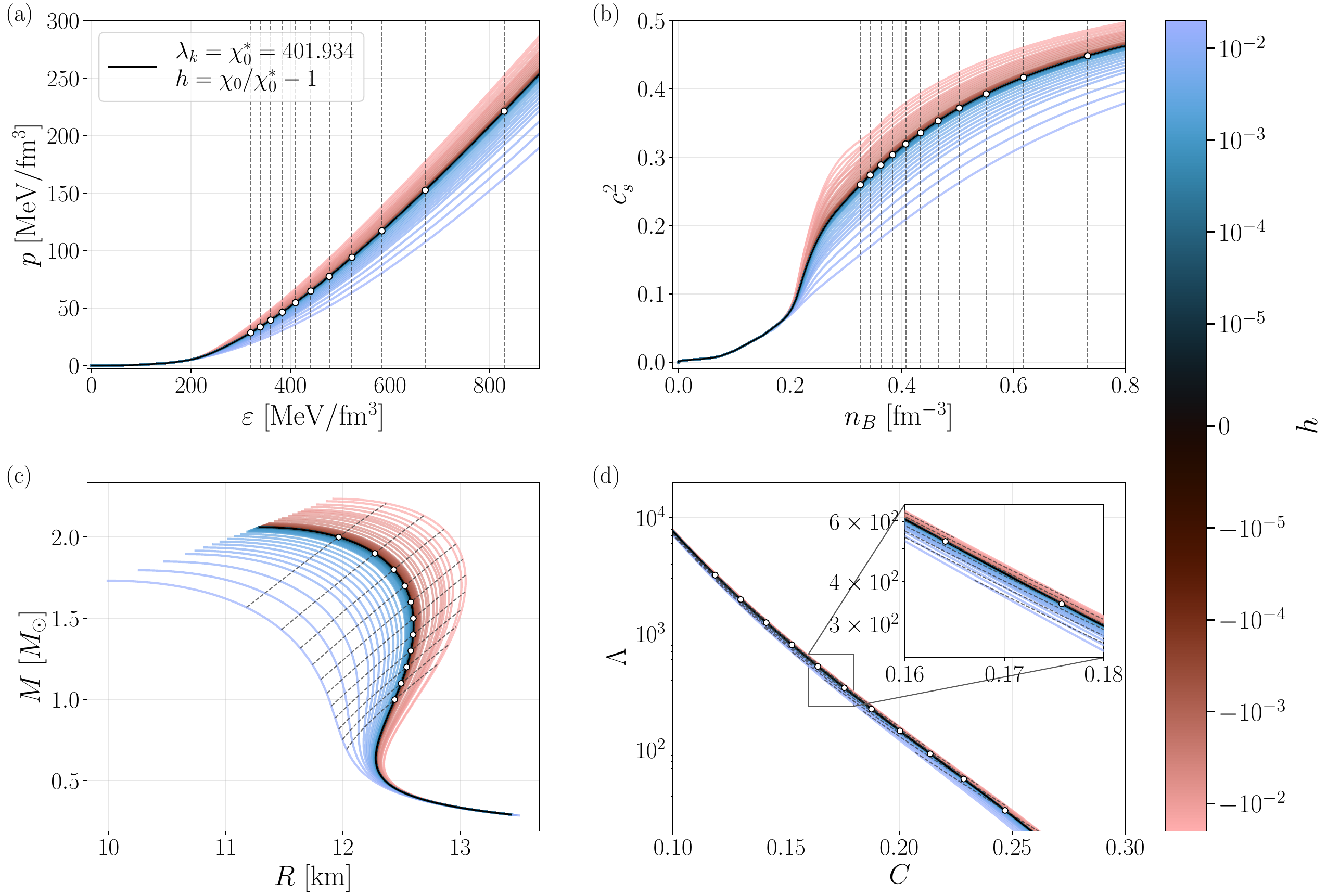}
  \caption{Variations of the CMF parameter $\chi_0$ on (a) the EoS, pressure as a function of energy density, (b) the squared speed of sound as a function of baryon density, (c) the mass--radius curves, and (d) the tidal deformability as a function of compactness. The fiducial model, corresponding to $\chi_0^* = 401.934$, is shown in black, while colored curves represent variations of $\chi_0$. Colors encode the signed fractional stepsize $h$, with blue (red) tones indicating increasing (decreasing) $\chi_0$. Vertical dashed lines in panels (a) and (b) denote the eleven central energy densities $\varepsilon_{c,k}$ associated with stellar configurations uniformly sampled between $1$ and $2\,M_\odot$ (white markers). In panel (c), dashed trajectories trace the evolution of these fixed $\varepsilon_{c,k}$ configurations under parameter variations. Panel (d) shows the corresponding universal tidal deformability compactness relation; the inset magnifies the local ordering near the fiducial sequence.
}
  \label{fig:chi_panel}
\end{figure*}
As a first step toward quantifying parameter sensitivities, we examine how the EoS responds to variations of a single CMF parameter about its fiducial value. \Cref{fig:chi_panel} illustrates this for the vacuum dilaton value $\chi_0$, while holding all other parameters fixed at $\boldsymbol{\lambda}^*$.

Panel~(a) shows the EoS, and panel~(b) displays the squared speed of sound $c_s^2$ as a function of baryon number density $n_B$. The fiducial EoS corresponds to $\chi_0^* = 401.934$ and is shown in black. Colored curves represent fractional variations $h = \chi_0/\chi_0^* - 1$ on a symmetric logarithmic scale; blue (red) tones indicate increasing (decreasing) values of $\chi_0$.
The vertical dashed lines mark the eleven central energy densities $\varepsilon_{c,k}$ used throughout this work to evaluate sensitivities. These correspond to stellar configurations uniformly sampled between $1$ and $2\,M_\odot$ along the fiducial equilibrium sequence. The white markers identify these sampling points and define the discrete configurations entering the logarithmic sensitivity matrices of Eq.~\eqref{eq:fisher_relative}.

A systematic trend is evident. Increasing $\chi_0$ softens the EoS, reducing the pressure at fixed energy density, while decreasing $\chi_0$ stiffens it. This behavior is reflected in $c_s^2$: larger $\chi_0$ leads to a smaller $\partial p / \partial \varepsilon$ over intermediate densities, signaling reduced stiffness. The spread of curves near $n_B \simeq (1$--$2)\,n_{\rm sat}$ corresponds to the density region where the core EoS is matched to the SLy crust.
The physical origin of this behavior lies in the role of $\chi_0$ as the vacuum expectation value of the dilaton field, which sets the scale of explicit scale symmetry breaking in the model. Variations of $\chi_0$ modify the scalar mean fields and therefore the effective baryon masses. Larger $\chi_0$ enhances the scalar contribution to the energy density relative to the pressure, shifting the balance between scalar attraction and vector repulsion toward a softer EoS. Conversely, smaller $\chi_0$ reduces this softening effect, resulting in a stiffer pressure response at a given energy density.

The macroscopic consequences of this microscopic modification are shown in panels~(c) and~(d). Panel~(c) displays the corresponding mass--radius (MR) sequences. The white markers again denote the eleven fiducial sampling configurations. The dashed trajectories trace how configurations at fixed central energy density $\varepsilon_{c,k}$ shift in the MR plane under parameter variations, directly mapping the vertical lines in panel~(a) onto curves in the MR diagram.
Increasing $\chi_0$ systematically reduces the maximum mass and shifts the MR curve toward smaller radii, consistent with a softer EoS providing less pressure support against gravity. Decreasing $\chi_0$ has the opposite effect, producing larger radii and higher maximum masses.
This perfectly ilustrates immediate benefits of our analysis, as it becomes apparent that modifying only one parameter (in this case $\chi_0$) is not sufficient if one wants to produce e.g., massive and small stars in agreement with observational data~\cite{MUSES:2023hyz}.
Panel~(d) shows the tidal deformability $\Lambda$ as a function of compactness $C$. The curves display a nearly monotonic ordering with $\chi_0$, indicating that variations of this parameter primarily modify the global stiffness of the EoS without introducing qualitative structural changes. Since, to leading order, $\Lambda \propto C^{-5}$, the tidal deformability is extremely sensitive to even modest variations in compactness.

The inset highlights the local ordering of the curves at fixed $C$. When compactness is held constant, larger values of $\chi_0$ (blue curves) correspond to slightly smaller values of $\Lambda$. This behavior is consistent with the expectation that a softer EoS reduces the tidal deformability at fixed compactness through modifications of the internal density profile and Love number.

However, the physically relevant comparison in this work follows trajectories of fixed central energy density $\varepsilon_{c,k}$, represented by the white markers and their corresponding paths (dashed lines). Along these constant-$\varepsilon_c$ slices, increasing $\chi_0$ shifts the stellar configuration toward lower $M$ and lower $R$, with the fractional reduction in $M$ exceeding that in $R$. Consequently, the compactness decreases. Because $\Lambda \propto C^{-5}$, this reduction in $C$ induces an increase in $\Lambda$ along the fixed-$\varepsilon_c$ trajectory. Thus, although $\Lambda$ decreases when compactness is held fixed, it increases along constant-central-energy slices as $\chi_0$ grows.

\subsection{Generalized sensitivity of observables}
\begin{figure*}[t]
  \centering
  \includegraphics[width=0.95\textwidth]{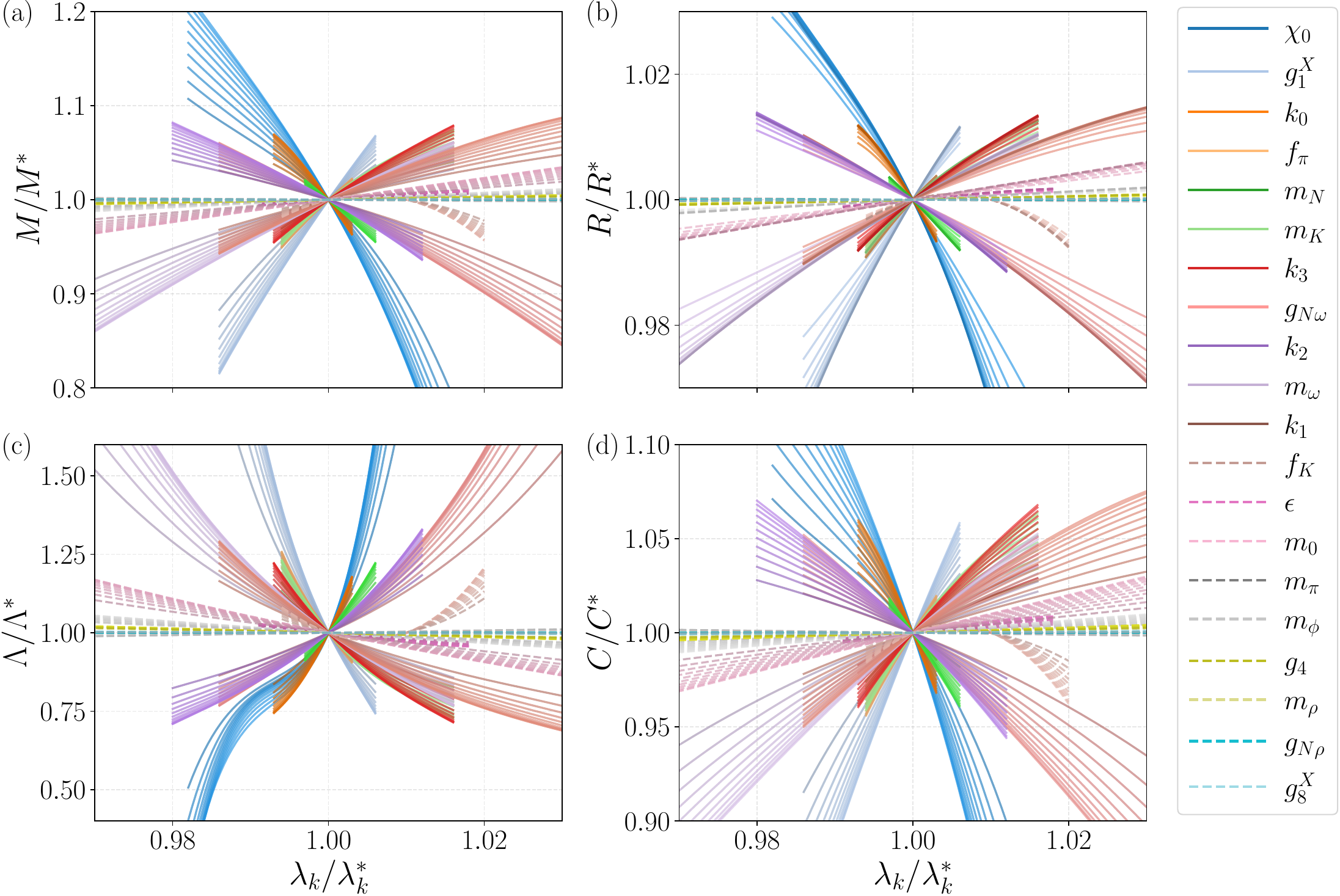}
  \caption{Relative responses for stellar mass, radius, tidal deformability, and compactness as functions of the relative parameter variation $\lambda_k/\lambda_k^*$ (wide view). Each color bundle corresponds to variations of a single CMF parameter $\lambda_k$, while individual curves within each bundle represent different central energy densities $\varepsilon_c$, sampled consistently between $1$ and $2\,M_\odot$. Lighter (darker) shades denote lower (higher) $\varepsilon_c$. The domain of each color might be different as each parameter has its own global bounds (see \Cref{fig:param_bounds}). The legend is ordered by the average absolute slope of the mass response. The coupling $g_{N\phi}$ is excluded because its fiducial value is zero. }
  \label{fig:relative_response_Z1}
\end{figure*}

\begin{figure*}[t]
  \centering
  \includegraphics[width=0.95\textwidth]{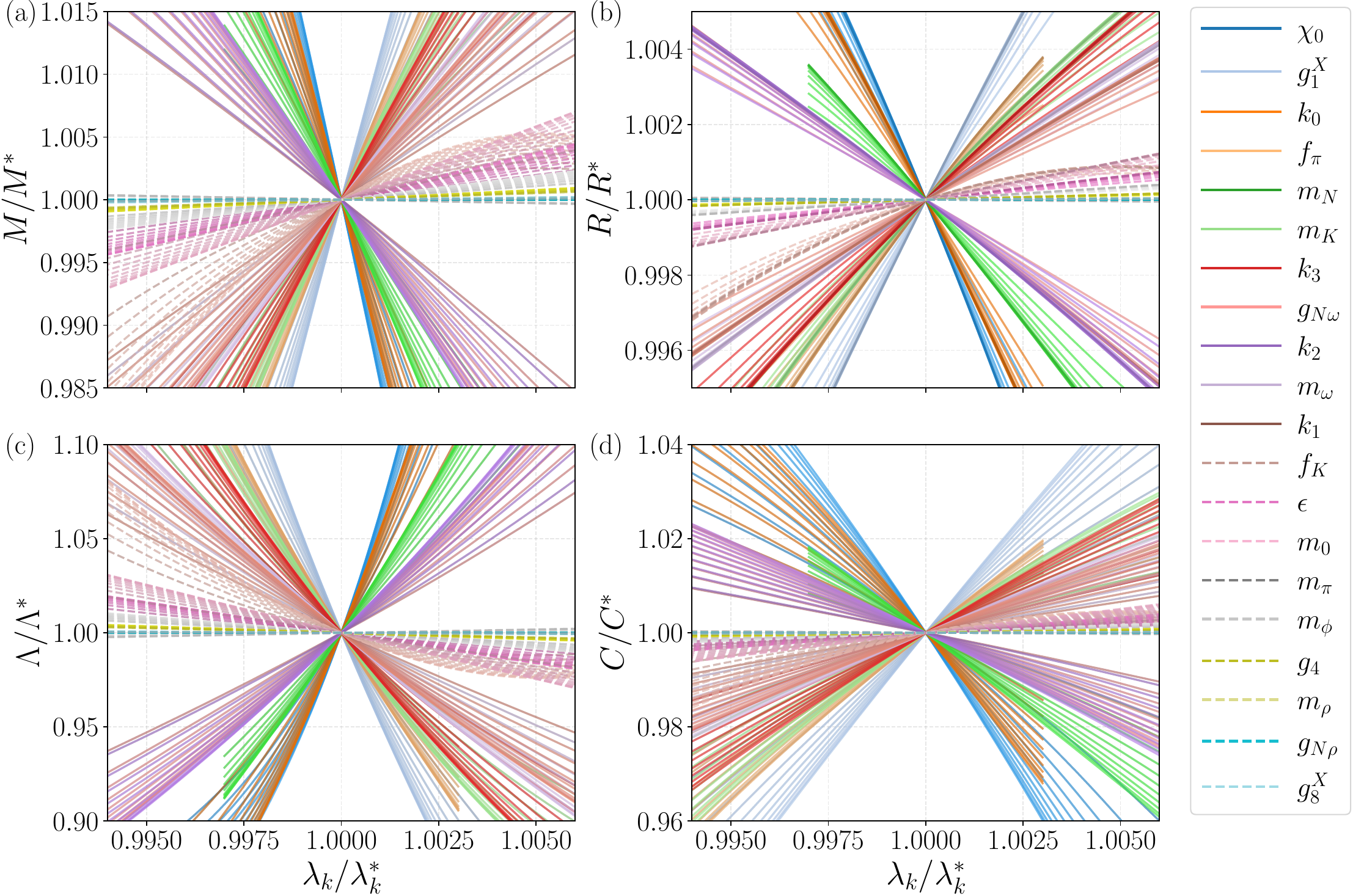}
  \caption{Zoom into the region $\lambda_k/\lambda_k^*\in[0.994,1.006]$ of \Cref{fig:relative_response_Z1} to highlight the local sensitivities used in the Fisher construction. In this regime, the response of each observable is well approximated by a low-order expansion around the fiducial model. The domain of each color might be different as each parameter has its own global bounds (see \Cref{fig:param_bounds}). The legend is ordered by the average absolute slope of the mass response. The coupling $g_{N\phi}$ is omitted again because its fiducial value is zero.}
  \label{fig:relative_response_Z2}
\end{figure*}

Having established how a single parameter modifies the EoS and MR relation, we now generalize the analysis to examine how all CMF parameters affect multiple neutron-star observables. \Cref{fig:relative_response_Z1} and \Cref{fig:relative_response_Z2} show the relative responses of the gravitational mass $M$, radius $R$, dimensionless tidal deformability $\Lambda$, and compactness $C$ to independent realizations of the 21 parameter set, varying each parameter $\lambda_k$ at a time (while holding all others fixed at their fiducial values $\lambda_k^*$). All quantities are normalized to their fiducial values, such that unity on both axes corresponds to the fiducial stellar sequence. The same results are shown in both \Cref{fig:relative_response_Z1} and \Cref{fig:relative_response_Z2}, but \Cref{fig:relative_response_Z1} displays the full range of parameter variation, whereas \Cref{fig:relative_response_Z2} zooms into the narrow interval $\lambda_k/\lambda_k^* \in [0.994,1.006]$ relevant for the local sensitivity analysis. Due to the intrinsic non-linearity of the CMF model, parameter responses may differ significantly far from the fiducial point compared to the immediate neighborhood of $\boldsymbol{\lambda}^*$.


In the wide view shown in \Cref{fig:relative_response_Z1}, the responses of the macroscopic observables to variations of each CMF parameter are organized into distinct color bundles. Each bundle corresponds to variations of a single parameter $\lambda_k$, while the individual curves within a bundle represent different central energy densities $\varepsilon_{c,k}$, sampled between that which leads to a $1 M_\odot$ star (lightest shade) and a $2\,M_\odot$ star (darkest shade) in the fiducial case. The relatively tight clustering of curves within each bundle indicates that, for a given parameter, the dependence of the response on $\varepsilon_c$ is modest compared to the overall sensitivity to the parameter itself.

The slopes of the curves quantify how strongly a given observable depends on $\lambda_k$. Flat slopes indicate parameters that have little influence on that observable, while steep slopes correspond to strong sensitivity. Positive slopes signify direct correlations, whereas negative slopes indicate anti-correlations. A clear hierarchy emerges near $\lambda_k/\lambda_k^* = 1$: scalar-sector parameters (most notably the vacuum dilaton field $\chi_0$, the scalar singlet coupling $g_1^X$, and the quadratic scalar self-interaction coefficient $k_0$) produce the largest and most coherent variations across all observables. In contrast, parameters such as the vector coupling $g_{N\rho}$, the vector meson mass $m_\rho$, and the scalar coupling $g_8^X$ generate comparatively weak responses, with curves remaining close to unity.

Away from the fiducial point, non-linear effects become apparent. For example, the dependence of $\Lambda$ on $k_0$ weakens for values below $\lambda_k^*$, while the response to variations in $f_k$ becomes more pronounced at large deviations from $\lambda_k^*$, with slopes steepening significantly relative to their local behavior near the fiducial point. These features reflect genuine curvature in the parameter space and highlight the limitations of strictly linear sensitivity estimates for large excursions from $\boldsymbol{\lambda}^*$.

For our sensitivity analysis, however, we focus on the local regime near the fiducial model. In the zoomed view of \Cref{fig:relative_response_Z2}, each curve is well approximated by a low-order expansion about $(\lambda_k/\lambda_k^*, O/O^*) = (1,1)$. The local slope in this region corresponds directly to the logarithmic derivative $\partial \ln O / \partial \ln \lambda_k$, which defines the entries of the sensitivity matrices constructed in subsequent sections. The visual ordering of slopes (same as in the previous figure) already anticipates the quantitative ranking obtained from the Fisher matrix and principal component analysis, with $\chi_0$ exhibiting the strongest overall impact.

Taken together, \Cref{fig:relative_response_Z1} and \Cref{fig:relative_response_Z2} show that the dominant variations of neutron-star observables are controlled by a limited subset of CMF parameters and that this ordering remains stable across the stellar mass range considered. This motivates the systematic sensitivity analysis that follows, where these qualitative trends are quantified and the effective dimensionality of the CMF parameter space is assessed.

A notable feature of the results is the strong structural similarity among the observables. Near the fiducial point, the relative variations of $M$, $R$, and $C$ exhibit comparable slope magnitudes and consistent sign patterns across most parameter directions. Parameters that stiffen (soften) the EoS generally increase (decrease) both $M$ and $R$, while the compactness responds according to the relative fractional changes of these two quantities. We emphasize that these trends are evaluated at fixed central energy density $\varepsilon_c$, so the correlations reflect how each parameter modifies the global TOV balance at a given $\varepsilon_c$, rather than comparisons at fixed mass or fixed radius.

The tidal deformability $\Lambda$ follows a closely related pattern but often appears with the opposite sign relative to $C$. This behavior is consistent with the approximate scaling $\Lambda \propto C^{-5}$ (see \Cref{eq:tidal_and_C}), which amplifies even modest changes in compactness. Consequently, $\Lambda$ may display visually steeper slopes than $M$ or $R$, even when driven by the same underlying structural modification of the star.

Quantitative differences among the observables are nevertheless present. For example, the $\chi_0$ dependence of $\Lambda$ exhibits noticeable curvature away from the fiducial point for smaller $\chi_0$ (\ref{fig:relative_response_Z1}), indicating non-linear structure in that region of parameter space. In addition, the spread of compactness curves across different $\varepsilon_c$ values is somewhat broader than for $M$ or $R$, reflecting the fact that $C$ depends on the ratio $M/R$ and, therefore, inherits sensitivity to variations in both quantities along the stellar sequence.

\subsection{Parameter Sensitivity from Fisher-like Matrices}
\label{sec:result_matrices}

The global sensitivity matrices $S_{ij}^{\alpha}$, as defined in \Cref{eq:fisher_relative}, quantify how information about each observable is distributed across the parameter space. Each entry
measures the accumulated covariance of logarithmic responses over the sampled central energy densities. The diagonal elements encode individual (self-) sensitivity, while off-diagonal elements trace correlated parameter directions that jointly influence the EoS and the resulting stellar observables. 

In all figures shown below, the matrices are sorted by decreasing diagonal magnitude. Parameters therefore appear ordered by decreasing individual relevance. Light colors indicate large sensitivities, dark colors small ones, and hatched cells denote negative correlations. The top eleven parameters are highlighted by a black box. In addition to the CMF model parameters, the stellar central energy density $\varepsilon_c$ is treated as an independent parameter in the analysis. Its variation is therefore included in the sensitivity matrices, increasing the dimensionality of the parameter space from 21 to 22.

\subsubsection*{Mass Sensitivity Matrix}
\begin{figure*}[t]
  \centering
  \includegraphics[width=0.95\textwidth]{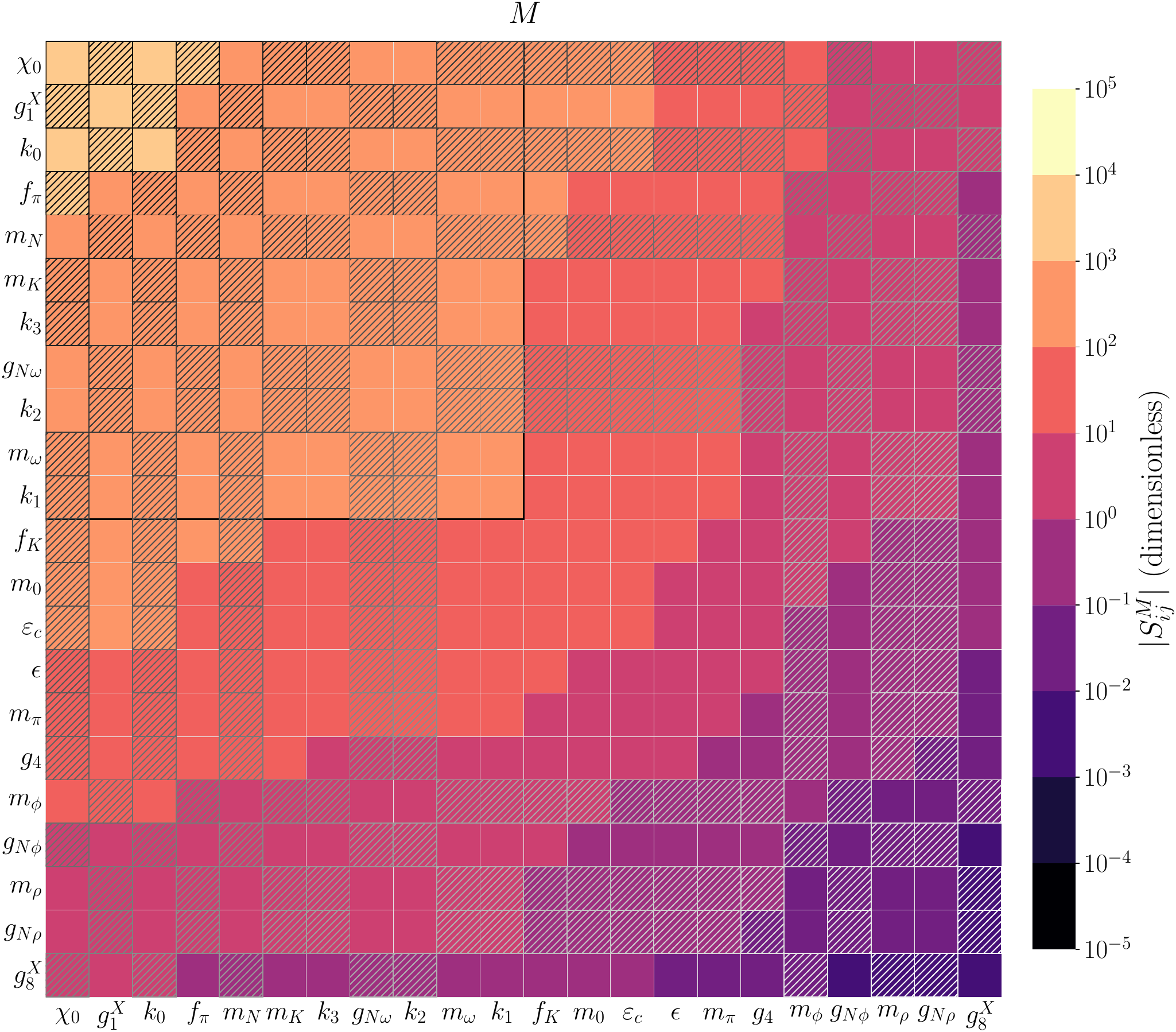}
  \caption{Sensitivity matrix $S^{M}$ for the stellar mass, summed over central energy densities and sorted by decreasing diagonal magnitude. Light (dark) colors indicate large (small) $|S_{ij}|$ on a logarithmic scale; hatched cells denote negative correlations. The black box highlights the eleven most self-sensitive parameters. The dominant block corresponds to the scalar sector  $\{\chi_0, g_1^X, k_0, k_1, k_2, k_3\}$, demonstrating that the stellar mass is primarily controlled by nonlinear scalar self-interactions and their associated baryon couplings. Vector-sector parameters appear only at the lower edge of the dominant block, while isovector and hidden-strangeness directions contribute negligibly in \textit{npe} matter. The nearly block-diagonal structure indicates that mass variations are governed by a coherent scalar-dominated stiffness direction in parameter space.}
  \label{fig:S_mass}
\end{figure*}
We begin with the simplest case, the mass sensitivity matrix $S_{ij}^{M}$ shown in \Cref{fig:S_mass}. Its structure is nearly block-diagonal and reveals a clear hierarchy in parameter relevance.
In that figure, we separate the relevance of parameters by the order of magnitude contribution to the sensitivity matrix. Outlined in black are the top 11 parameters that contribute to the matrix the most. 

The three largest diagonal entries correspond to $
\{\chi_0,\; g_1^X,\; k_0\}$ and this ordering agrees with the expectation from the previous subsection. The vacuum dilaton value $\chi_0$ sets the overall scale of the scalar potential and appears multiplicatively in multiple attractive contributions, including the quadratic scalar curvature term $k_0 \chi_0^2$, the cubic mixing term $k_3 \chi_0$, and the logarithmic trace-anomaly contribution proportional to $\varepsilon \chi_0^4$. A variation of $\chi_0$ therefore rescales several scalar attraction channels simultaneously. It controls the global strength and density evolution of the scalar condensates and hence the overall stiffness scale of the EoS.

The scalar singlet coupling $g_1^X$ governs the baryon--$\sigma$ interaction and directly determines how the effective nucleon mass evolves with density. Since the effective mass controls the attractive component of dense matter, its large sensitivity reflects the central role of scalar attraction in determining stellar mass.

The parameter $k_0$ sets the curvature of the scalar potential near the origin. Its prominence indicates that the detailed shape of the scalar potential is a primary driver of mass variations.
More generally, all scalar self-interaction parameters $\{k_0, k_1, k_2, k_3\}$ appear within the top eleven. This demonstrates that stellar mass is primarily governed by nonlinear scalar self-interactions. This interpretation is consistent with \Cref{fig:relative_response_Z1} and \Cref{fig:relative_response_Z2}, where variations of $\chi_0$ and the $k_i$ parameters produce the largest coherent shifts in the mass--radius bundles. Those variations visibly control the principal stiffness direction in parameter space.

The explicit symmetry-breaking parameters $\{f_\pi, m_K\}$ also lie within the dominant block. These parameters calibrate the vacuum values of the scalar condensates and therefore shift the baseline around which dense matter evolves. Their sensitivity is indirect but global, affecting the normalization of the effective masses and scalar fields.

Among baryonic parameters, only $m_N$ appears within the top eleven. Unlike the scalar self-interactions, it does not directly shape the density evolution of the condensates, but rather calibrates the mapping between condensate values and the physical nucleon mass. Its moderate but coherent sensitivity reflects this structural role.

From the vector sector, only $\{m_\omega, g_{N\omega}\}$ enter the top block of 11 most relevant parameters, and at its lower edge. This indicates that isoscalar vector repulsion contributes to high-density stiffness but remains subdominant relative to scalar attraction in determining stellar mass. In contrast, $\{g_4, m_\phi, g_{N\phi}, m_\rho, g_{N\rho}\}$ lie near the bottom of the matrix, demonstrating that hidden-strangeness and isovector directions play a negligible role in \textit{npe} matter for the mass observable.

It is also noteworthy that the extended parameter $\varepsilon_c$ does not appear among the dominant entries. This indicates that the sensitivity is not driven by a narrow density slice. Instead, variations of the CMF parameters induce coherent changes across the entire density range. The response of the mass is therefore controlled by structural features of the Lagrangian rather than localized density effects.

Finally, we observe no clear example of a parameter with small diagonal magnitude but large off-diagonal correlations. The dominant off-diagonal structure occurs within the scalar block itself, reflecting the fact that these parameters jointly define the scalar potential. The matrix is therefore structured but not dominated by hidden correlated directions among individually weak parameters.

The remaining sensitivity matrices for $R$, $\Lambda$, and compactness exhibit qualitatively similar block structures and are presented in \Cref{app:extra_matrices}. No qualitatively new dominant parameter directions appear beyond those identified in the mass and composite matrices.

\subsubsection*{Composite LIGO Sensitivity}
\begin{figure*}[t]
  \centering
  \includegraphics[width=0.95\textwidth]{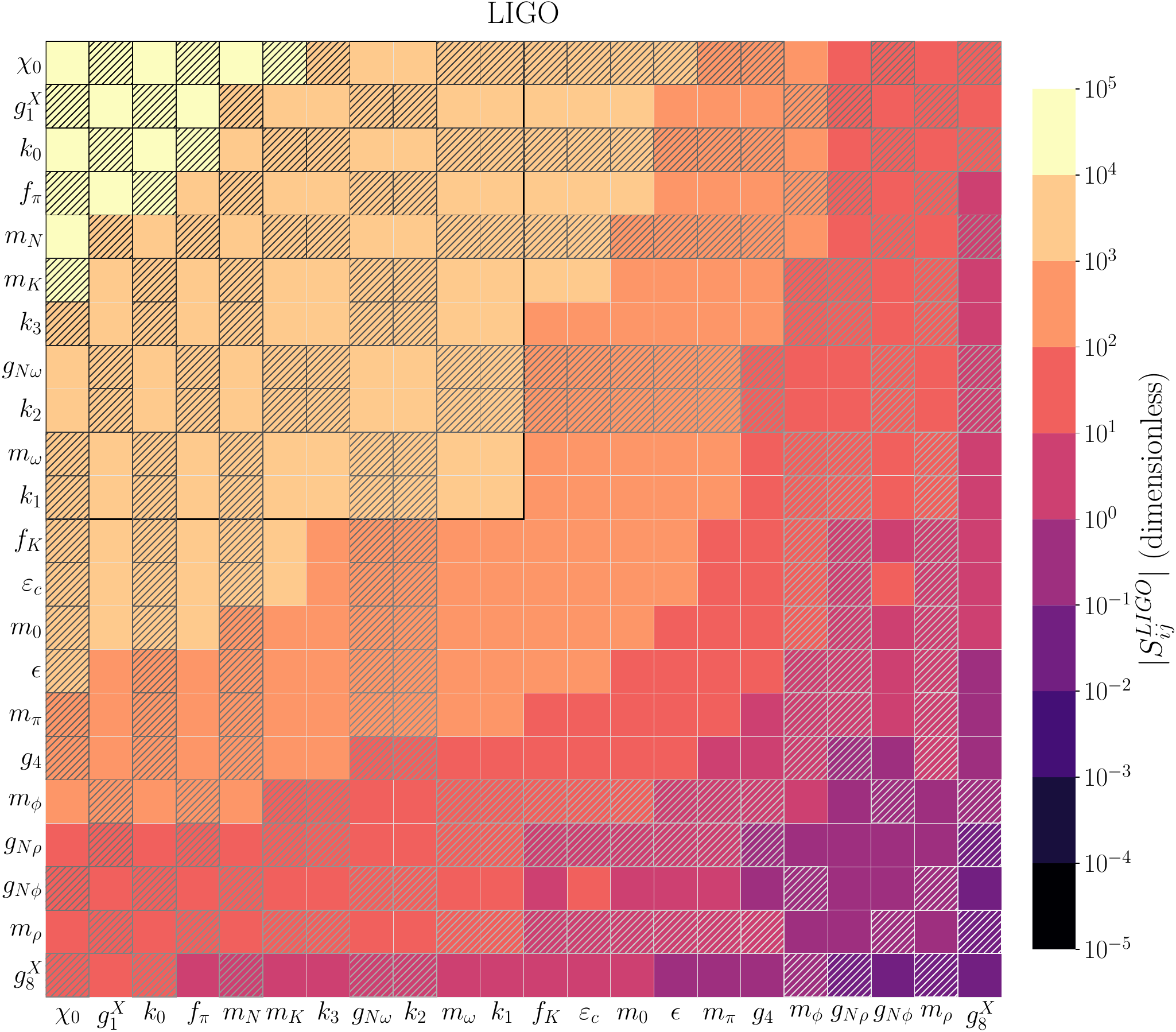}
  \caption{Composite sensitivity matrix $S^{\mathrm{LIGO}} = S^{M} + S^{\Lambda} + S^{M,\Lambda}$, sorted by decreasing diagonal magnitude. The visual encoding is identical to \Cref{fig:S_mass}. The dominant eleven parameters coincide with those of the mass matrix and appear in the same order, with only minor permutations among subdominant entries. No new leading parameter directions emerge. This demonstrates that current gravitational-wave observables primarily probe the same scalar-dominated stiffness directions that control the mass--radius relation, while orthogonal directions (e.g.\ isovector or hidden-strangeness couplings) remain weakly constrained.}
  \label{fig:S_LIGO}
\end{figure*}
\Cref{fig:S_LIGO} shows the composite matrix
$S^{\mathrm{LIGO}} = S^{M} + S^{\Lambda} + S^{M,\Lambda}$.
Remarkably, it exhibits essentially the same structure as $S^{M}$. The top eleven parameters are identical and appear in the same order. The only differences are minor permutations among subdominant entries (e.g.\ the ordering of $m_0$ and $\varepsilon_c$, and of $g_{N\phi}, g_{N\rho}, m_\rho$ near the bottom). No new dominant directions emerge.

This demonstrates that current gravitational-wave observables primarily probe the same stiffness-controlling directions already identified in the mass matrix. In other words, both mass and tidal deformability constrain the scalar attraction--vector repulsion balance that sets the global stiffness scale of dense matter. Orthogonal directions, such as isovector couplings or hidden-strangeness contributions, remain weakly constrained.
%

\subsection{Principal directions}
\label{sec:results_pca}
The matrix $S^{\alpha}_{ij}$ quantifies the sensitivity of neutron-star observables to variations of the model parameters. However, it does not by itself reveal which combinations of parameters correspond to directions in parameter space along which the observables exhibit the strongest response. In general, the impact of different parameters on the observables is correlated, implying that only specific linear combinations can be independently constrained. To identify these directions, referred to as principal directions, we perform a Principal Component Analysis (PCA) on $S^{\alpha}_{ij}$ for each case considered in this work.

Diagonalizing $S^{\alpha}_{ij}$ rotates the parameter space into an orthogonal basis whose eigenvectors define the principal directions. Each eigenvector can be expressed as a linear combination of the original parameters, and in this basis, variations along distinct directions are statistically uncorrelated. The associated eigenvalues quantify the importance of each parameter combination:
Larger eigenvalues correspond to parameter combinations for which small variations induce larger fractional changes in the observables.
The resulting decomposition allows the principal directions to be ranked according to their effect on observables, thereby providing a clear characterization of the effective dimensionality of the parameter space accessible to future observations. 

The eigenvectors are ordered according to the magnitude of their corresponding eigenvalues. We denote by PC1 the principal direction associated with the largest eigenvalue of the PCA decomposition, and by PC2 the direction associated with the second-largest eigenvalue. For a parameter space of dimension $D=22$, the decomposition yields 22 orthogonal principal directions ranked by decreasing constraining power. 

\begin{figure*}[t]
  \centering
  \includegraphics[width=0.95\textwidth]{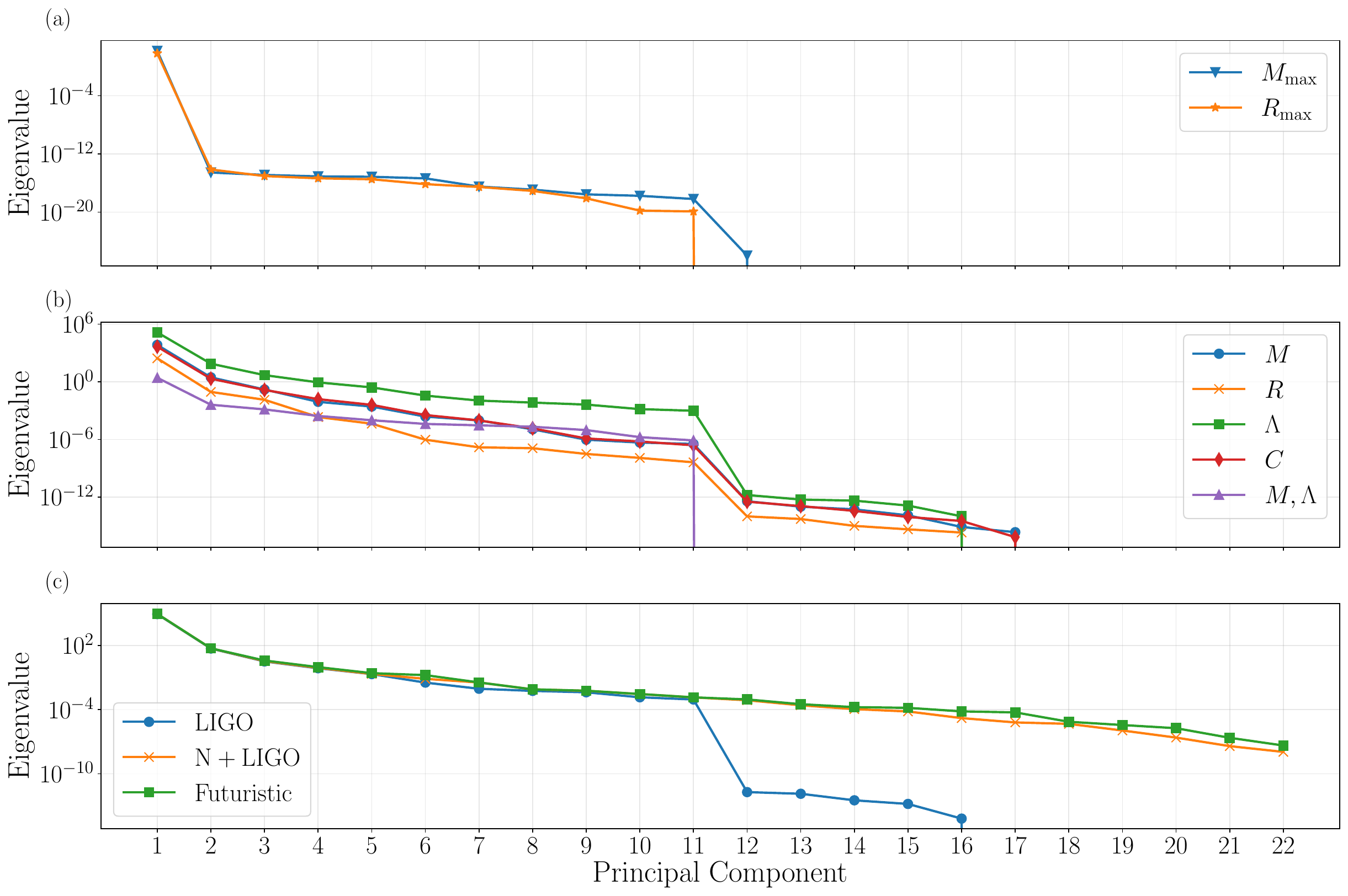}
\caption{
Eigenvalue spectra of the Fisher matrices for all observables and composite cases:
(a) endpoint observables ($M_{\max}$, $R_{\max}$),
(b) sequence observables ($M$, $R$, $\Lambda$, $C$, and $M,\Lambda$), and
(c) composite observational combinations (LIGO, N+LIGO, Futuristic).
}
  \label{fig:FIM_PCA_eigs}
\end{figure*}

For all cases considered in this work, the PCA exhibits a pronounced hierarchy: the first two eigenvalues dominate, while the remaining eigenvalues are significantly smaller in comparison, as shown in \Cref{fig:FIM_PCA_eigs}. This indicates that each observable is primarily sensitive to two independent parameter combinations, with PC1 representing the most strongly constrained direction and PC2 providing a subdominant contribution. Consequently, although the underlying parameter space is 22-dimensional, the effective dimensionality accessible to the observable is substantially reduced. 
\Cref{fig:FIM_PCA_eigs} displays the eigenvalue spectra in three panels: (a) endpoint observables ($M_{\max}$, $R_{\max}$), (b) bulk–sequence observables ($M$, $R$, $\Lambda$, $C$, and $M,\Lambda$), and (c) composite observational combinations (LIGO, N+LIGO, Futuristic). 

In panel~(a), a single dominant eigenvalue is followed by a sharp drop, indicating that the endpoint sensitivities are effectively one-dimensional; beyond the twelfth component, eigenvalues fall to the numerical noise floor. In panel~(b), two leading modes remain significant, with a cutoff near the sixteenth component (eleventh for the joint $M,\Lambda$ case), showing that the bulk observables share two main directions. Panel~(c) confirms this pattern for the composite cases: two principal directions capture nearly all information, with the LIGO combination retaining visible structure up to the sixteenth component. Overall, all three panels demonstrate that despite the formal dimensionality of the CMF parameter space, the astrophysical response is intrinsically low-rank dominated by one to two physically interpretable eigenmodes.

\Cref{tab:PC1_top11} and \Cref{tab:PC2_top11} summarize the eleven largest components (by absolute eigenvector weight) of the dominant and subdominant principal components, PC1 and PC2, for each observable and composite case. Two robust patterns emerge.

First, PC1 is consistently dominated by parameters associated with the scalar and chiral sectors of the CMF model. In particular, the vacuum dilaton field $\chi_0$, the scalar coupling $g_1^X$, and the leading scalar self–interaction parameter $k_0$ appear prominently across nearly all observables, including $M$, $R$, $\Lambda$, $C$, their combinations, and the endpoint quantities $M_{\max}$ and $R_{\max}$. This identifies PC1 as a coherent direction in parameter space that controls the leading correlated response of neutron-star observables. An exception occurs for the joint $(M,\Lambda)$ case, where the central energy density $\varepsilon_c$ and pion decay constant $f_\pi$ enter with comparable weight, reflecting the increased role of sampling location along the stellar sequence when cross-correlated observables are combined.

Second, PC2 captures subdominant and more heterogeneous effects. Its leading entries vary significantly between observables and typically involve parameters such as $\varepsilon_c$, $f_\pi$, $m_N$, and selected mass or coupling terms. Unlike PC1, no single sector or small subset of parameters systematically dominates PC2, indicating that it represents secondary directions associated with parameter degeneracies, sampling effects, or residual correlations rather than a universal physical control parameter.

These trends are consistent with the eigenvalue spectra shown in \Cref{fig:FIM_PCA_eigs}. In all cases, PC1 is separated by several orders of magnitude from higher components, while eigenvalues beyond PC2 rapidly decrease by several orders of magnitude and eventually reach the numerical noise floor. This sharp spectral decay indicates that additional principal components contribute negligibly compared to numerical noise, confirming that the effective sensitivity of neutron-star observables within the CMF parameter space is well captured by a low-dimensional subspace dominated by PC1 and, to a lesser extent, PC2.

For completeness, the full numerical principal-component coefficients for PC1 and PC2, including all subdominant components beyond the leading eleven entries shown here, are provided in Appendix~\ref{app:pca_tables}.
\begin{table*}[t]
\centering
\setlength{\tabcolsep}{1.4pt}
\begin{tabular}{l@{\hspace{2.5pt}}|@{}c@{}c@{}c@{}c@{}c@{}c@{}c@{}c@{}c@{}c@{}c}
\toprule
\multirow{2}{*}{\hspace{0.25cm}\textbf{Case}} 
& \multicolumn{11}{c}{\textbf{Top eleven parameters in PC1}} \\
\cmidrule(lr){2-12}
& 1 & 2 & 3 & 4 & 5 & 6 & 7 & 8 & 9 & 10 & 11 \\
\midrule
$M$
& \pcacellhex{2E86C1}{\chi_0}
& \pcavsep\pcacellhexn{5499C7}{g_{1}^{X}}
& \pcacellhex{5499C7}{k_0}
& \pcavsep\pcacellhexn{A9CCE3}{f_{\pi}}
& \pcacellhex{A9CCE3}{m_{N}}
& \pcacellhexn{A9CCE3}{m_{K}}
& \pcacellhexn{A9CCE3}{k_3}
& \pcavsep\pcacellhex{D4E6F8}{g_{N\omega}}
& \pcacellhex{D4E6F8}{k_2}
& \pcacellhexn{D4E6F8}{m_{\omega}}
& \pcacellhexn{D4E6F8}{k_1} \\
$R$
& \pcacellhex{2E86C1}{\chi_0}
& \pcavsep\pcacellhexn{5499C7}{g_{1}^{X}}
& \pcacellhex{5499C7}{k_0}
& \pcavsep\pcacellhexn{A9CCE3}{f_{\pi}}
& \pcacellhex{A9CCE3}{m_{N}}
& \pcacellhexn{A9CCE3}{m_{K}}
& \pcacellhexn{A9CCE3}{k_3}
& \pcavsep\pcacellhex{D4E6F8}{k_2}
& \pcacellhex{D4E6F8}{g_{N\omega}}
& \pcacellhexn{D4E6F8}{m_{\omega}}
& \pcacellhexn{D4E6F8}{k_1} \\
$\Lambda$
& \pcacellhex{2E86C1}{\chi_0}
& \pcavsep\pcacellhexn{5499C7}{g_{1}^{X}}
& \pcacellhex{5499C7}{k_0}
& \pcavsep\pcacellhexn{A9CCE3}{f_{\pi}}
& \pcacellhex{A9CCE3}{m_{N}}
& \pcacellhexn{A9CCE3}{m_{K}}
& \pcacellhexn{A9CCE3}{k_3}
& \pcavsep\pcacellhex{D4E6F8}{g_{N\omega}}
& \pcacellhex{D4E6F8}{k_2}
& \pcacellhexn{D4E6F8}{m_{\omega}}
& \pcacellhexn{D4E6F8}{k_1} \\
$C$
& \pcacellhex{2E86C1}{\chi_0}
& \pcavsep\pcahsep\pcacellhexn{5499C7}{g_{1}^{X}}
& \pcahsep\pcacellhex{5499C7}{k_0}
& \pcavsep\pcahsep\pcacellhexn{A9CCE3}{f_{\pi}}
& \pcahsep\pcacellhex{A9CCE3}{m_{N}}
& \pcahsep\pcacellhexn{A9CCE3}{m_{K}}
& \pcahsep\pcacellhexn{A9CCE3}{k_3}
& \pcavsep\pcacellhex{D4E6F8}{g_{N\omega}}
& \pcacellhex{D4E6F8}{k_2}
& \pcahsep\pcacellhexn{D4E6F8}{m_{\omega}}
& \pcahsep\pcacellhexn{D4E6F8}{k_1} \\
$M,\Lambda$
& \pcacellhexn{2E86C1}{\varepsilon_c}
& \pcavsep\pcahsep\pcacellhexn{A9CCE3}{f_{\pi}}
& \pcahsep\pcacellhex{A9CCE3}{m_{N}}
& \pcavsep\pcahsep\pcacellhexn{D4E6F8}{\chi_0}
& \pcahsep\pcacellhexn{D4E6F8}{g_{N\omega}}
& \pcahsep\pcacellhexn{D4E6F8}{g_{1}^{X}}
& \pcahsep\pcacellhexn{D4E6F8}{k_2}
& \pcacellhex{D4E6F8}{k_3}
& \pcacellhex{D4E6F8}{m_{K}}
& \pcavsep\pcahsep\pcacellhex{EAF2FB}{g_{N\phi}}
& \pcahsep\pcacellhex{EAF2FB}{k_1} \\
$M_{\max}$
& \pcacellhex{2E86C1}{\chi_0}
& \pcavsep\pcacellhexn{5499C7}{g_{1}^{X}}
& \pcahsep\pcacellhex{5499C7}{k_0}
& \pcavsep\pcahsep\pcacellhexn{A9CCE3}{f_{\pi}}
& \pcacellhex{A9CCE3}{m_{N}}
& \pcacellhexn{A9CCE3}{m_{K}}
& \pcahsep\pcacellhex{A9CCE3}{k_3}
& \pcavsep\pcacellhex{D4E6F8}{k_2}
& \pcacellhex{D4E6F8}{g_{N\omega}}
& \pcacellhexn{D4E6F8}{m_{\omega}}
& \pcacellhexn{D4E6F8}{k_1} \\
$R_{\max}$
& \pcacellhex{2E86C1}{\chi_0}
& \pcavsep\pcacellhexn{5499C7}{g_{1}^{X}}
& \pcavsep\pcahsep\pcacellhex{7FB3D5}{k_0}
& \pcahsep\pcacellhexn{7FB3D5}{f_{\pi}}
& \pcavsep\pcacellhex{A9CCE3}{m_{N}}
& \pcacellhexn{A9CCE3}{m_{K}}
& \pcavsep\pcahsep\pcacellhex{D4E6F8}{k_3}
& \pcacellhex{D4E6F8}{g_{N\omega}}
& \pcacellhex{D4E6F8}{k_2}
& \pcacellhexn{D4E6F8}{m_{\omega}}
& \pcacellhexn{D4E6F8}{k_1} \\
LIGO
& \pcacellhex{2E86C1}{\chi_0}
& \pcavsep\pcacellhexn{5499C7}{g_{1}^{X}}
& \pcacellhex{5499C7}{k_0}
& \pcavsep\pcacellhexn{A9CCE3}{f_{\pi}}
& \pcacellhex{A9CCE3}{m_{N}}
& \pcacellhexn{A9CCE3}{m_{K}}
& \pcacellhexn{A9CCE3}{k_3}
& \pcavsep\pcacellhex{D4E6F8}{g_{N\omega}}
& \pcacellhex{D4E6F8}{k_2}
& \pcacellhexn{D4E6F8}{m_{\omega}}
& \pcacellhexn{D4E6F8}{k_1} \\
N+LIGO
& \pcacellhex{2E86C1}{\chi_0}
& \pcavsep\pcacellhexn{5499C7}{g_{1}^{X}}
& \pcacellhex{5499C7}{k_0}
& \pcavsep\pcacellhexn{A9CCE3}{f_{\pi}}
& \pcacellhex{A9CCE3}{m_{N}}
& \pcacellhexn{A9CCE3}{m_{K}}
& \pcacellhexn{A9CCE3}{k_3}
& \pcavsep\pcacellhex{D4E6F8}{g_{N\omega}}
& \pcacellhex{D4E6F8}{k_2}
& \pcacellhexn{D4E6F8}{m_{\omega}}
& \pcacellhexn{D4E6F8}{k_1} \\
Futuristic
& \pcacellhex{2E86C1}{\chi_0}
& \pcavsep\pcacellhexn{5499C7}{g_{1}^{X}}
& \pcacellhex{5499C7}{k_0}
& \pcavsep\pcacellhexn{A9CCE3}{f_{\pi}}
& \pcacellhex{A9CCE3}{m_{N}}
& \pcacellhexn{A9CCE3}{m_{K}}
& \pcacellhexn{A9CCE3}{k_3}
& \pcavsep\pcacellhex{D4E6F8}{g_{N\omega}}
& \pcacellhex{D4E6F8}{k_2}
& \pcacellhexn{D4E6F8}{m_{\omega}}
& \pcacellhexn{D4E6F8}{k_1} \\
\bottomrule
\end{tabular}
\vspace{2pt}
\begin{center}
{\footnotesize Absolute eigenvector weight:}\\[2pt]
\pcacolorbar
\end{center}
\small
\caption{
Leading eleven parameter contributions (ranked by decreasing absolute eigenvector weight) to the first principal component (PC1) of the sensitivity matrix for each observable and composite case. PC1 is nearly universal across all observables. Vertical black lines mark changes in the order of magnitude of the ranked weights. Font color encodes the sign of the weight, with black denoting positive entries and magenta denoting negative entries. PC1 has dominant support from the scalar sector $\{\chi_0, g_1^X, k_0\}$, with subleading contributions from $\{k_1,k_2,k_3\}$, which identifies it as a coherent direction controlling the overall pressure scale of the EoS.
}
\label{tab:PC1_top11}
\end{table*}
\begin{table*}[t]
\centering
\setlength{\tabcolsep}{1.4pt}
\begin{tabular}{l@{\hspace{2.5pt}}|@{}c@{}c@{}c@{}c@{}c@{}c@{}c@{}c@{}c@{}c@{}c}
\toprule
\multirow{2}{*}{\hspace{0.25cm}\textbf{Case}} 
& \multicolumn{11}{c}{\textbf{Top elevel parameters in PC2}} \\
\cmidrule(lr){2-12}
& 1 & 2 & 3 & 4 & 5 & 6 & 7 & 8 & 9 & 10 & 11 \\
\midrule
$M$
& \pcacellhex{2E86C1}{\varepsilon_c}
& \pcavsep\pcacellhexn{7FB3D5}{f_{\pi}}
& \pcacellhex{7FB3D5}{m_{N}}
& \pcavsep\pcacellhex{A9CCE3}{f_{K}}
& \pcacellhexn{A9CCE3}{g_{1}^{X}}
& \pcahsep\pcacellhexn{A9CCE3}{k_0}
& \pcavsep\pcacellhexn{D4E6F8}{k_2}
& \pcacellhex{D4E6F8}{m_{K}}
& \pcacellhex{D4E6F8}{\epsilon}
& \pcacellhex{D4E6F8}{k_3}
& \pcacellhex{D4E6F8}{k_1} \\
$R$
& \pcahsep\pcacellhexn{2E86C1}{\varepsilon_c}
& \pcavsep\pcahsep\pcacellhex{7FB3D5}{f_{\pi}}
& \pcacellhexn{7FB3D5}{m_{N}}
& \pcavsep\pcahsep\pcacellhexn{A9CCE3}{f_{K}}
& \pcacellhex{A9CCE3}{k_0}
& \pcavsep\pcahsep\pcacellhexn{D4E6F8}{m_{K}}
& \pcahsep\pcacellhexn{D4E6F8}{k_3}
& \pcahsep\pcacellhexn{D4E6F8}{\epsilon}
& \pcahsep\pcacellhex{D4E6F8}{g_{1}^{X}}
& \pcahsep\pcacellhex{D4E6F8}{k_2}
& \pcacellhexn{D4E6F8}{k_1} \\
$\Lambda$
& \pcacellhexn{5499C7}{f_{\pi}}
& \pcacellhex{5499C7}{m_{N}}
& \pcavsep\pcacellhex{7FB3D5}{\varepsilon_c}
& \pcacellhex{7FB3D5}{f_{K}}
& \pcavsep\pcacellhexn{A9CCE3}{g_{1}^{X}}
& \pcacellhexn{A9CCE3}{k_0}
& \pcacellhexn{A9CCE3}{k_2}
& \pcacellhex{A9CCE3}{\epsilon}
& \pcacellhex{A9CCE3}{m_{K}}
& \pcacellhex{A9CCE3}{k_3}
& \pcavsep\pcacellhex{D4E6F8}{k_1} \\
$C$
& \pcacellhexn{5499C7}{f_{\pi}}
& \pcahsep\pcacellhex{5499C7}{m_{N}}
& \pcavsep\pcacellhex{7FB3D5}{\varepsilon_c}
& \pcacellhex{7FB3D5}{f_{K}}
& \pcavsep\pcacellhexn{A9CCE3}{g_{1}^{X}}
& \pcacellhexn{A9CCE3}{k_0}
& \pcacellhexn{A9CCE3}{k_2}
& \pcacellhex{A9CCE3}{m_{K}}
& \pcacellhex{A9CCE3}{\epsilon}
& \pcahsep\pcacellhex{A9CCE3}{k_3}
& \pcavsep\pcacellhex{D4E6F8}{k_1} \\
$M,\Lambda$
& \pcahsep\pcacellhex{5499C7}{k_2}
& \pcavsep\pcahsep\pcacellhex{7FB3D5}{k_3}
& \pcacellhexn{7FB3D5}{m_{\omega}}
& \pcacellhexn{7FB3D5}{\chi_0}
& \pcavsep\pcacellhex{A9CCE3}{m_{K}}
& \pcacellhexn{A9CCE3}{k_1}
& \pcahsep\pcacellhex{A9CCE3}{g_{N\omega}}
& \pcahsep\pcacellhex{A9CCE3}{k_0}
& \pcahsep\pcacellhex{A9CCE3}{m_{N}}
& \pcavsep\pcacellhexn{D4E6F8}{f_{K}}
& \pcacellhexn{D4E6F8}{m_{\phi}} \\
$M_{\max}$
& \pcacellhex{2E86C1}{k_3}
& \pcavsep\pcacellhexn{5499C7}{k_1}
& \pcavsep\pcahsep\pcacellhexn{7FB3D5}{g_{1}^{X}}
& \pcahsep\pcacellhexn{7FB3D5}{f_{K}}
& \pcavsep\pcahsep\pcacellhexn{A9CCE3}{m_{N}}
& \pcahsep\pcacellhex{A9CCE3}{f_{\pi}}
& \pcavsep\pcacellhex{D4E6F8}{k_2}
& \pcacellhexn{D4E6F8}{g_{N\omega}}
& \pcahsep\pcacellhex{D4E6F8}{\epsilon}
& \pcahsep\pcacellhexn{D4E6F8}{m_{\omega}}
& \pcahsep\pcacellhexn{D4E6F8}{\chi_0} \\
$R_{\max}$
& \pcahsep\pcacellhex{2E86C1}{g_{1}^{X}}
& \pcavsep\pcacellhex{5499C7}{\chi_0}
& \pcavsep\pcahsep\pcacellhex{A9CCE3}{k_3}
& \pcavsep\pcahsep\pcacellhex{D4E6F8}{k_0}
& \pcahsep\pcacellhexn{D4E6F8}{k_1}
& \pcahsep\pcacellhexn{D4E6F8}{f_{\pi}}
& \pcahsep\pcacellhex{D4E6F8}{m_{N}}
& \pcahsep\pcacellhexn{D4E6F8}{k_2}
& \pcavsep\pcahsep\pcacellhex{EAF2FB}{g_{N\omega}}
& \pcahsep\pcacellhexn{EAF2FB}{m_0}
& \pcahsep\pcacellhex{EAF2FB}{\varepsilon_c} \\
LIGO
& \pcacellhexn{5499C7}{f_{\pi}}
& \pcacellhex{5499C7}{m_{N}}
& \pcavsep\pcacellhex{7FB3D5}{f_{K}}
& \pcavsep\pcacellhex{A9CCE3}{\varepsilon_c}
& \pcacellhexn{A9CCE3}{k_0}
& \pcacellhexn{A9CCE3}{g_{1}^{X}}
& \pcacellhexn{A9CCE3}{k_2}
& \pcacellhex{A9CCE3}{\epsilon}
& \pcacellhex{A9CCE3}{m_{K}}
& \pcacellhex{A9CCE3}{k_3}
& \pcavsep\pcacellhex{D4E6F8}{k_1} \\
N+LIGO
& \pcacellhexn{5499C7}{f_{\pi}}
& \pcacellhex{5499C7}{m_{N}}
& \pcavsep\pcacellhex{7FB3D5}{f_{K}}
& \pcavsep\pcacellhex{A9CCE3}{\varepsilon_c}
& \pcacellhexn{A9CCE3}{k_0}
& \pcacellhexn{A9CCE3}{g_{1}^{X}}
& \pcacellhex{A9CCE3}{\epsilon}
& \pcacellhexn{A9CCE3}{k_2}
& \pcacellhex{A9CCE3}{m_{K}}
& \pcacellhex{A9CCE3}{k_3}
& \pcavsep\pcacellhex{D4E6F8}{k_1} \\
Futuristic
& \pcacellhexn{5499C7}{f_{\pi}}
& \pcacellhex{5499C7}{m_{N}}
& \pcavsep\pcacellhex{7FB3D5}{f_{K}}
& \pcavsep\pcacellhex{A9CCE3}{\varepsilon_c}
& \pcacellhexn{A9CCE3}{k_0}
& \pcacellhexn{A9CCE3}{g_{1}^{X}}
& \pcacellhex{A9CCE3}{\epsilon}
& \pcacellhexn{A9CCE3}{k_2}
& \pcacellhex{A9CCE3}{m_{K}}
& \pcacellhex{A9CCE3}{k_3}
& \pcavsep\pcacellhex{D4E6F8}{k_1} \\
\bottomrule
\end{tabular}
\vspace{2pt}
\begin{center}
{\footnotesize Absolute eigenvector weight:}\\[2pt]
\pcacolorbar
\end{center}
\small
\caption{
Leading eleven parameter contributions (ranked by decreasing absolute eigenvector weight) to the second principal component (PC2) of the sensitivity matrix for each observable and composite case. PC2 exhibits a more heterogeneous and observable-dependent structure than PC1. Its dominant entries frequently include the central energy density $\varepsilon_c$, decay constants, and selected scalar interactions, indicating that PC2 encodes variations in how stiffness develops with density rather than the overall pressure scale of the equation of state.
}
\label{tab:PC2_top11}
\end{table*}
%

\subsection{Consistency Across Observables}
A key outcome of the PCA analysis for the sensitivity matrices is the remarkable consistency of parameter sensitivities across different neutron-star observables and observational combinations. Despite probing distinct aspects of neutron-star structure, the responses of $M$, $R$, $\Lambda$, and $C$ are governed by a common low-dimensional structure in the CMF parameter space (see again \Cref{tab:PC1_top11}).

Across all single-observable cases and composite scenarios (LIGO, N+LIGO, and Futuristic), the dominant principal component (PC1) is aligned with a coherent direction primarily supported by the scalar and chiral sectors of the model. Parameters, such as the vacuum dilaton field $\chi_0$ and the scalar self-interaction coefficients $k_i$, consistently appear with the largest eigenvector weights, with additional but subleading contributions from quantities like $f_\pi$ and $m_N$. This dominant direction captures the dominant stiffness mode that coherently modifies pressure across densities of macroscopic observables under microscopic parameter changes. Its prominence is already visible in the relative-response figures (\Cref{fig:relative_response_Z1}–\Cref{fig:relative_response_Z2}), where the steepest local slopes cluster around the same subset of parameters.

The subdominant principal component (PC2, see again \Cref{tab:PC2_top11}) exhibits a more heterogeneous structure. Its leading contributions vary between observables and typically involve combinations of the central energy density $\varepsilon_c$, decay constants, and nucleon mass terms. Unlike PC1, PC2 does not correspond to a universal control direction but instead reflects secondary effects associated with sampling location along the stellar sequence. Its relative importance increases modestly when compactness or high-density endpoint observables are included, indicating a heightened sensitivity to how the underlying EoS evolves with density rather than to its overall response. PC2 encodes variations in how stiffness develops with density rather than the overall stiffness scale itself.

Importantly, combining gravitational-wave and X-ray information does not introduce new dominant directions in parameter space. Instead, joint constraints (e.g., N+LIGO) primarily reweight the existing principal components, enhancing sensitivity to decay constants and mass parameters without rotating the leading eigenvectors. This indicates that different observational channels probe the same underlying response subspace, albeit with different projections.

Finally, the eigenvalue spectra shown in \Cref{fig:FIM_PCA_eigs} demonstrate a rapid decay beyond the first one to two principal components for all cases considered. Higher-order components contribute negligibly compared to numerical uncertainties, confirming that, despite the nominal dimensionality of the \texttt{CMF++} parameter set, the local response of neutron-star observables around the fiducial model is effectively low-rank, with only $\sim 1$–$3$ relevant degrees of freedom.
%

\section{Discussion and Outlook}
\label{sec:discussion}

In this work we presented a systematic sensitivity analysis of neutron-star observables within the Chiral Mean Field (CMF) model using a Fisher-information-inspired framework. By combining the optimized \texttt{CMF++} implementation within the MUSES cyberinfrastructure with a derivative-based sensitivity formalism, we quantified how variations of microscopic CMF parameters (while ensuring reasonable saturation properties) propagate to macroscopic neutron-star observables, including the stellar mass $M$, radius $R$, tidal deformability $\Lambda$, and compactness $C$. Evaluating derivatives along sequences of stellar configurations and constructing global sensitivity matrices allowed us to identify the dominant directions in parameter space that control the response of neutron-star properties.

The motivation for such an analysis arises from the broader challenge of describing dense matter at high baryon density and low temperature, as realized in the cores of neutron stars. Key aspects of this regime remain uncertain, including the relevant particle content, the structure of the strong interaction at supranuclear densities, and the possible appearance of exotic phases. Realistic effective models of QCD, such as the CMF model, incorporate these ingredients in a thermodynamically-consistent framework but necessarily introduce a relatively large number of parameters. Traditionally, these parameters are calibrated sequentially: first to symmetric nuclear matter properties at saturation density, then to symmetry-energy observables, and finally to neutron-star properties. However, fitting parameters in stages neglects correlations between sectors of the model that may significantly influence the resulting equation of state. A probabilistic exploration of the full parameter space therefore, represents a more consistent strategy, provided that the computational tools exist to generate large numbers of EoSs efficiently. The present work constitutes a first step toward that goal.

Our results show that, in the vicinity of the fiducial CMF C4 parameter set used as a reference in this study, the macroscopic response of neutron-star observables to microscopic parameters is highly structured and effectively low-dimensional. Across all observables considered ($M$, $R$, $\Lambda$, and $C$), the dominant sensitivity direction is consistently aligned with a scalar-dominated sector of the CMF model, primarily spanned by the parameters $\{\chi_0, g_1^X, k_0\}$ with additional support from $\{k_1, k_2, k_3\}$ and subleading contributions from the pion decay constant $f_\pi$, the kaon mass $m_K$, and the nucleon vacuum mass $m_N$. Although strangeness degrees of freedom were not explicitly included in the present stellar calculations, the kaon mass remains an intrinsic parameter of the Lagrangian through explicit symmetry breaking and therefore contributes to the overall structure of the model. Together, these parameters regulate the pressure support of the EoS at supranuclear densities and consequently drive the leading variations in all stellar macroscopic properties.

A secondary sensitivity mode involves a more heterogeneous combination of parameters, including the central energy density $\varepsilon_c$, decay constants, and selected scalar couplings. 
In contrast to the dominant stiffness direction, this mode encodes variations in how the stiffness of the EoS develops with increasing density and reflects residual degeneracies along the stellar sequence. Its relative importance increases for observables probing the highest-density regime, such as configurations near the maximum mass. Beyond the first two principal components (occasionally accompanied by a weak third direction), the eigenvalue spectra of the sensitivity matrices decay rapidly toward the numerical noise floor. This behavior indicates that, locally around the fiducial point, the CMF parameter space effectively behaves as a two-dimensional response manifold.

It is important to emphasize that the Fisher-like analysis employed here is intrinsically local: the sensitivities are computed around a specific fiducial parameter set and therefore describe the linear response of observables to small parameter variations. Indeed, as illustrated in several examples throughout this work, the model exhibits noticeable nonlinear behavior once parameters are displaced sufficiently far from the fiducial point. Consequently, the low-dimensional structure identified here should be interpreted as a local property of the parameter space rather than a global one. Extending the analysis to larger regions of parameter space will require either higher-order expansions or global sampling techniques capable of capturing these nonlinearities.

The dominance of a single stiffness-related direction has important implications for the interpretation of multi-messenger astrophysical observations. Within the local Fisher approximation, current constraints from NICER X-ray measurements and gravitational-wave observations by the LIGO/Virgo/KAGRA collaboration primarily probe a common pressure-support direction in parameter space rather than orthogonal combinations associated with isovector or symmetry-breaking effects. Improvements in measurement precision are therefore expected to tighten constraints mainly along this leading eigenmode, while leaving other parameter combinations comparatively weakly constrained. However, this situation may change once additional types of data are incorporated. In particular, heavy-ion collision experiments probe complementary density and temperature regimes and are sensitive to aspects of the nuclear interaction not directly constrained by neutron-star observations. Including such data in a joint analysis would likely introduce additional independent constraints and could reveal new directions in parameter space beyond the dominant astrophysical stiffness mode. Similarly, post-merger gravitational-wave signals, cooling observations, and measurements of neutron-star moments of inertia may provide sensitivity to orthogonal combinations of parameters.

Another important caveat concerns the different physical origins and experimental constraints of the various sectors of the CMF model. The scalar sector is largely calibrated using baryon and pseudoscalar meson vacuum masses and meson decay constants, quantities that are experimentally well determined. In contrast, the vector sector is primarily constrained through nuclear saturation properties and meson masses, which leave significantly more freedom in the choice of coupling schemes and coupling strengths that reproduce the same bulk properties. The present study only enforces approximate symmetric-matter saturation conditions as a basic consistency check and therefore does not fully account for the differing levels of experimental accuracy associated with the scalar and vector sectors. Future analyses that incorporate additional nuclear and experimental constraints, such as symmetry-energy parameters and their density derivatives, will be necessary to assess how these sector-dependent uncertainties propagate into neutron-star observables.

The identification of a low effective dimensionality in the local response of the CMF model also motivates more efficient strategies for exploring its parameter space. Rather than treating all parameters as independent degrees of freedom, future Bayesian analyses could exploit the principal directions identified here to construct reduced parameterizations, for example through principal-component or active-subspace priors. Such dimensional reduction has the potential to accelerate posterior inference, mitigate degeneracies, and enable the consistent combination of heterogeneous observational constraints.

Several natural extensions of this work are therefore within reach. On the modeling side, incorporating additional degrees of freedom such as hyperons, quarks, or alternative vector self-interaction schemes will clarify how the dominant sensitivity directions evolve as new microscopic channels modify the high-density EoS. On the statistical side, integrating the Fisher-derived sensitivity structure with full multi-messenger likelihoods will enable hierarchical Bayesian analyses that jointly constrain CMF parameters and neutron-star observables. More broadly, embedding these techniques within the \texttt{CMF++} and MUSES workflows opens the possibility of adaptive sampling strategies, surrogate-model training, and ultimately a predictive, data-driven bridge between chiral microphysics and neutron-star phenomenology.

\acknowledgements

The authors thank Mateus Reinke Pelicer and Rajesh Kumar for their insightful discussions and valuable feedback that helped refine the physical interpretation and workflow integration of the \texttt{CMF++} analysis. All authors of this paper are part of the MUSES collaboration, which is supported by the U.S. National Science Foundation (NSF) under Grants No.~OAC-2103680, OAC-2005572, and OAC-2004879. J.N.H. acknowledges additional support from the U.S. Department of Energy, Office of Science, Nuclear Physics program under Grant No.~DE-SC0023861. N.C.-C. acknowledges support from the Illinois Center for Advanced Studies of the Universe (ICASU) and the Center for AstroPhysical Surveys (CAPS) Fellowship at the University of Illinois Urbana-Champaign. V.D. acknowledges additional support from the U.S. Department of Energy, Office of Science, Nuclear Physics program under Grant No.~DE-SC0024700 and the National Science Foundation under No.~Grant NP3M PHY2116686. N.Y. also acknowledges
support from the Simons Foundation through Award No.
896696, the Simons Foundation International through
Award No. SFI-MPS-BH-00012593-01, the NSF through
Grants No. PHY-2207650 and PHY-25-12423, and NASA
through Grant No. 80NSSC22K0806.
Figures in this manuscript were produced using matplotlib~\cite{Hunter:2007,thomas_a_caswell_2023_7697899}.
%

\appendix
\bibliography{inspire,not_inspire}
\newpage
%

\appendix
\setcounter{table}{0}
\renewcommand{\thetable}{A\arabic{table}}
\setcounter{figure}{0}
\renewcommand{\thefigure}{\thesection\arabic{figure}}
%

\section{Data Generation Workflow}
\label{app:workflow} 
This appendix summarizes the computational workflow used to generate stellar observables from a given CMF parameter vector $\boldsymbol{\lambda}$.  The pipeline begins with the specification of the model parameters and proceeds through nuclear matter validation, construction of a unified EoS, TOV integration via QLIMR to obtain neutron-star observables.
\begin{figure}[h]
    \centering
    \includegraphics[
        width=0.4\textwidth,
        trim=0 2.5cm 0 2.2cm,
        clip
    ]{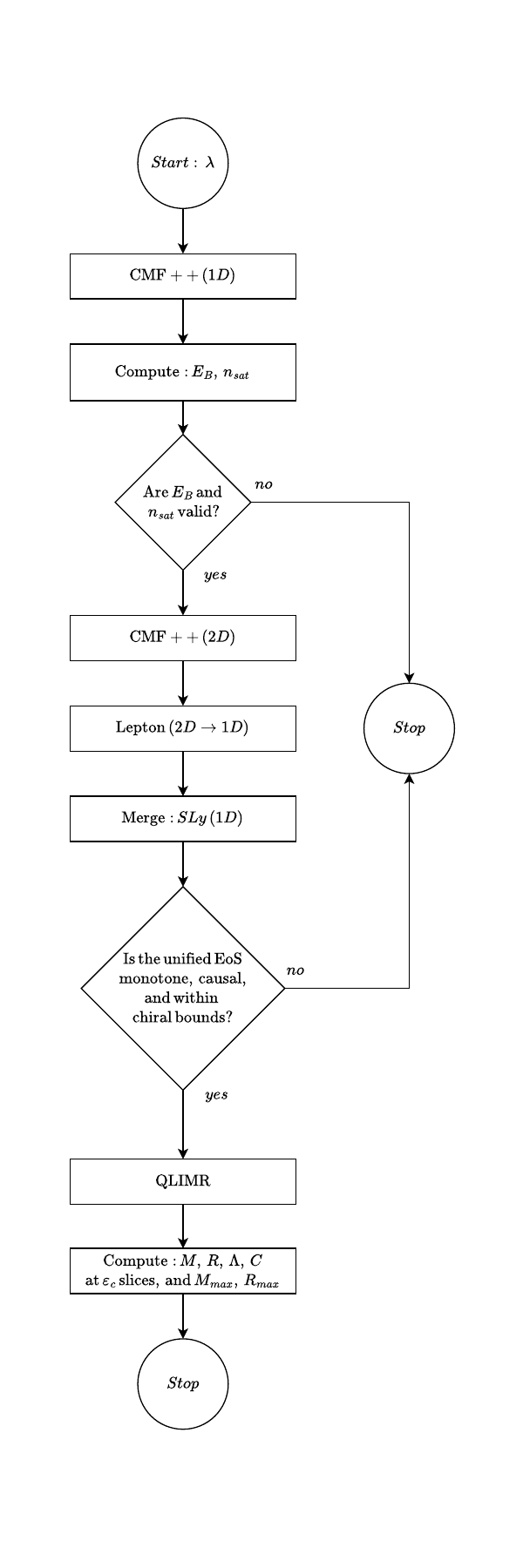}
    \caption{Workflow for generating neutron-star observables from a specific CMF configuration $\boldsymbol{\lambda}$.}
    \label{fig:data_workflow}
\end{figure}
For each parameter set $\boldsymbol{\lambda}$:
\begin{enumerate}
    \item \textbf{Nuclear matter initialization (1D).}  
    The CMF++ model is first solved in one dimension to compute the saturation properties, including the binding energy $E_B$ and saturation density $n_{\mathrm{sat}}$. It is calculated as a function of the baryon chemical potential $\mu_B$ for symmetric matter, zero strange and charge chemical potentials ($\mu_S=\mu_Q=0$).
    \item \textbf{Saturation filtering.}  
    Parameter sets that fail to satisfy empirical constraints on $E_B$ and $n_{\mathrm{sat}}$ are discarded.
    \item \textbf{Full CMF++ solution (2D).}  
    The thermodynamic quantities are computed in two dimensions (finite $\mu_B, \mu_Q$), after which charge neutrality and $\beta$-equilibrium are enforced via the Lepton module, reducing the system to a one-dimensional EoS.
    \item \textbf{Crust–core matching.}  
    The high-density CMF++ EoS is merged with the SLy crust EoS to construct a unified, thermodynamically consistent equation of state.
    \item \textbf{Physical consistency checks.}  
    The unified EoS is required to be monotonic, causal ($c_s^2 \le 1$), and within the prescribed chiral EFT bounds \(6 \le p(2n_{\rm sat}) \le 50~\mathrm{MeV/fm}^3\)~\cite{Huth:2021bsp}.
    \item \textbf{TOV integration.}  
    The QLIMR module solves the TOV equations, producing stellar structure sequences.
    \item \textbf{Observable extraction.}  
    Stellar observables are extracted at fixed central energy density slices $\varepsilon_c$, including 
    $M$, $R$, tidal deformability $\Lambda$, and compactness $C$, as well as global quantities such as $M_{\max}$ and $R_{\max}$.
\end{enumerate}
\setcounter{table}{0}
\renewcommand{\thetable}{B\arabic{table}}
\setcounter{figure}{0}
\renewcommand{\thefigure}{\thesection\arabic{figure}}
%

\section{Numerical derivative methods and convergence diagnostics}
\label{app:numerical_derivatives}
The derivatives $\partial O_\alpha / \partial \theta_j$ entering the sensitivity matrices were computed numerically by varying each CMF parameter $\lambda_j$ around its fiducial value $\lambda_j^*$ and re-evaluating the corresponding neutron-star observables $O_\alpha$. To ensure robustness against numerical noise and interpolation artifacts from the CMF-QLIMR workflow, a diverse collection of independent finite-difference and polynomial-based methods was implemented and cross-compared. 
All derivatives were evaluated using relative step sizes, rather than absolute changes. For each CMF parameter $\lambda_j$, the variation values were defined as
\[
\lambda_j = \lambda_j^* (1 \pm h),
\]
where $h$ was logarithmically spaced within the interval $[10^{-6},\,10^{-1}]$. This corresponds to symmetric step sizes up to a $10\%$ deviation from the fiducial parameter value, ensuring that the response of each observable is captured over several orders of magnitude while preserving the scale invariance. If the fiducial value is zero, relative perturbations are meaningless; in that case, an absolute logarithmic spacing from $10^{-6}$ up to the maximum value obtained from the global-bounds analysis was used. 
The derivative of each observable with respect to the stellar central energy density, $\partial O_\alpha / \partial \varepsilon_c$, was evaluated independently using a standard fourth-order central finite-difference stencil. This derivative captures the variation of observables along the equilibrium sequence of stellar configurations and is combined with the CMF-parameter derivatives in the total sensitivity analysis.
%

\subsection{Finite-difference stencils}
The simplest derivative approximations rely on discrete finite differences evaluated around the fiducial point. Let $O_+$ and $O_-$ denote the observable evaluated at $\lambda_j^* \pm h$, with $h$ the perturbation step. The following schemes were implemented:
\begin{description}
  \item[\textbf{Forward difference:}]
  \[
    \frac{\partial O}{\partial \lambda_j}
    \approx \frac{O_+ - O_0}{h}
    \quad \text{(first-order accuracy)}.
  \]
  \item[\textbf{Backward difference:}]
  \[
    \frac{\partial O}{\partial \lambda_j}
    \approx \frac{O_0 - O_-}{h}
    \quad \text{(first-order accuracy)}.
  \]
  \item[\textbf{Central 2-point difference:}]
  \[
    \frac{\partial O}{\partial \lambda_j}
    \approx \frac{O_+ - O_-}{2h}
    \quad \text{(second-order accuracy)}.
  \]
  \item[\textbf{Central 4-point difference:}]
  \[
    \frac{\partial O}{\partial \lambda_j}
    \approx
    \frac{-O_{+2h} + 8O_+ - 8O_- + O_{-2h}}{12h}
    \quad \text{(fourth-order accuracy)}.
  \]
\end{description}
These stencils provide a controlled hierarchy of accuracy and allow direct assessment of truncation errors.  
The fourth-order central scheme was typically adopted as the reference for convergence checks, as it provided the most reliable results.
%

\subsection{Polynomial and regression-based fits}
To complement the discrete finite-difference stencils and suppress local numerical noise, polynomial regression methods were employed to extract derivatives from multi-point fits of $O(\lambda_j)$ around $\lambda_j^*$. These regression-based techniques are less sensitive to random fluctuations and allow for a smoother estimation of local gradients, especially when the observable exhibits mild nonlinearity.
\begin{itemize}
  \item \textbf{Linear fits:} performed over 2-, 3-, 4-, and 5-point neighborhoods around $\lambda_j^*$, providing slope estimates equivalent to first-order derivatives.
  \item \textbf{Cubic fits:} constructed over 5-point symmetric sets ($\lambda_j^* \pm 2h, \pm h, \lambda_j^*$),  
  with analytical evaluation of the derivative of the fitted cubic polynomial at $\lambda_j^*$:
  \[
    \frac{\partial O}{\partial \lambda_j}\Big|_{\lambda_j^*}
    = 3a\,{\lambda_j^*}^2 + 2b\,\lambda_j^* + c,
  \]
  where $(a,b,c,d)$ are the cubic coefficients obtained from the least-squares fit.
\end{itemize}
These polynomial schemes complement finite differences by mitigating discretization noise and improving stability near nonlinear regions.
\setcounter{table}{0}
\renewcommand{\thetable}{C\arabic{table}}
\setcounter{figure}{0}
\renewcommand{\thefigure}{\thesection\arabic{figure}}
%

\section{Supplementary Sensitivity Matrices}
\label{app:extra_matrices}

In this appendix we present the remaining sensitivity matrices that complement the results discussed in the main text. 
The primary sensitivity matrices for the stellar mass were analyzed in \Cref{sec:result_matrices}, where we identified the dominant scalar-driven stiffness direction governing the response of neutron-star observables to variations of the CMF parameters. 
Here we provide the analogous matrices for the remaining observables discussed in the Results section, including the stellar radius, tidal deformability, compactness, and several combined or derived observables. 
These figures confirm that the hierarchical structure identified in the main text persists across all observables considered.

All matrices shown here are constructed using the definition given in \Cref{eq:fisher_relative} and are sorted by decreasing diagonal magnitude, following the same convention used for the matrices shown in the main text. 
Light (dark) colors indicate large (small) values of $|S_{ij}|$ on a logarithmic scale, while hatched cells denote negative correlations.

Across all observables, the matrices exhibit a remarkably stable hierarchical structure. 
The dominant parameter directions are consistently governed by the nonlinear scalar sector 
$\{\chi_0,\, g_1^X,\, k_0,\, k_1,\, k_2,\, k_3\}$, 
with subleading but coherent contributions from the isoscalar–vector sector 
$(m_\omega,\, g_{N\omega})$. 
Isovector, hidden-strangeness, and octet scalar couplings remain comparatively weakly constrained. 
As a result, the additional matrices presented below do not introduce qualitatively new dominant parameter directions; instead, they mainly rescale the relative weights of the same scalar-driven stiffness combinations identified in the main text.

\subsection{Radius and Tidal Deformability}

\begin{figure*}[t]
  \centering
  \includegraphics[width=0.48\textwidth]{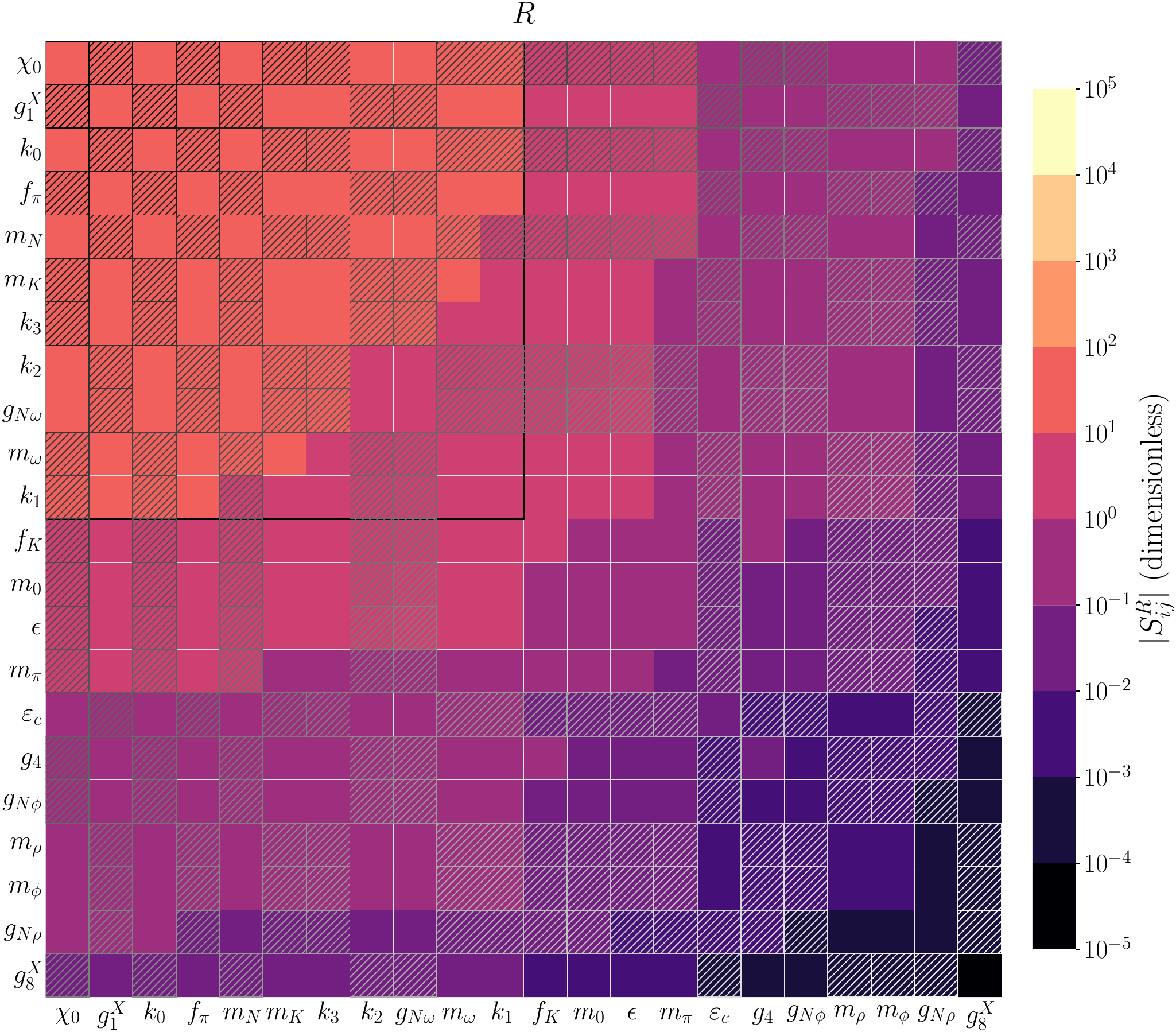}
  \hfill
  \includegraphics[width=0.48\textwidth]{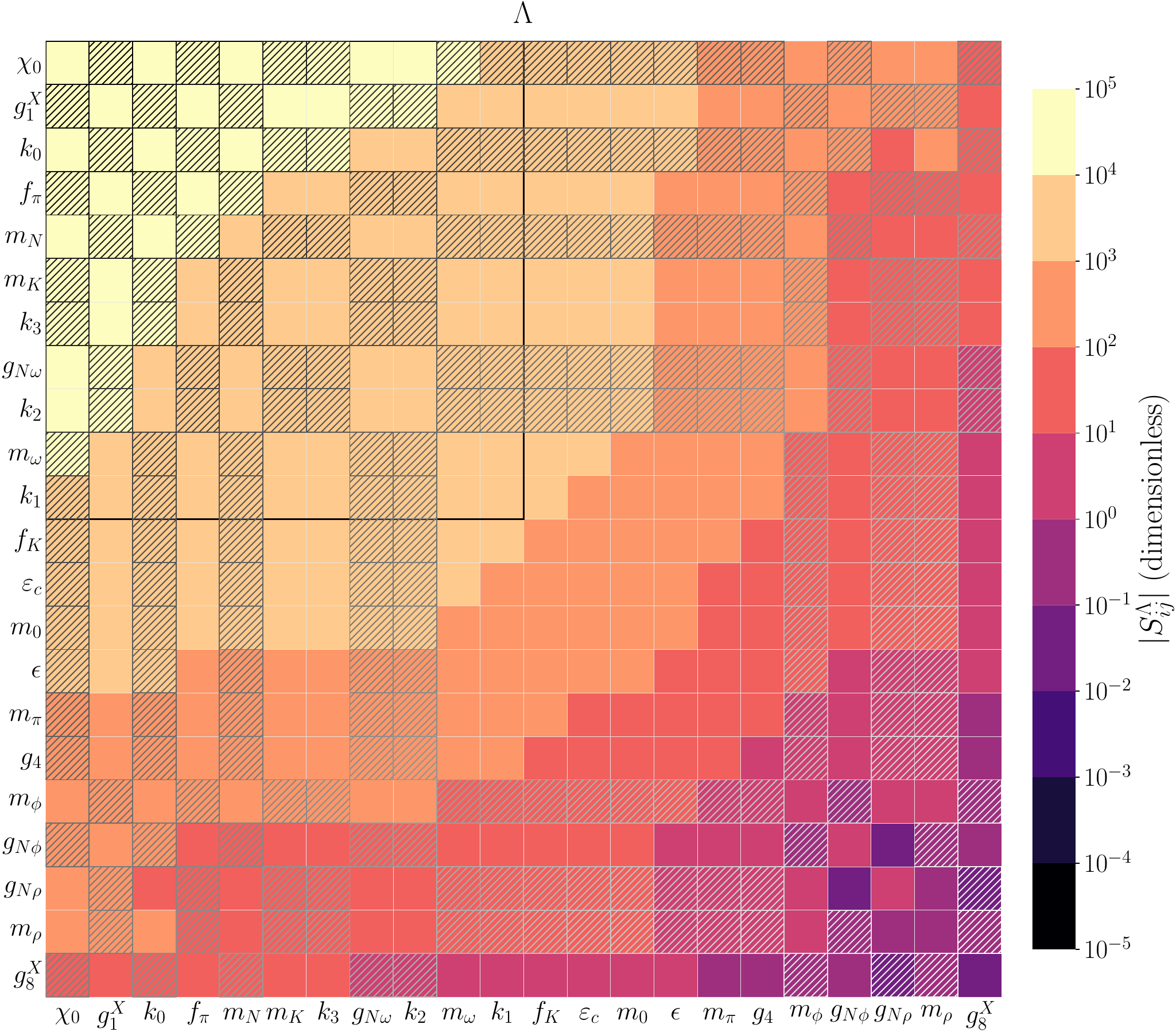}
  \caption{
  Left: sensitivity matrix $S^{R}$ (stellar radius). 
  Right: sensitivity matrix $S^{\Lambda}$ (tidal deformability). 
  Both matrices reproduce the scalar-dominated block structure observed for the mass observable in the main text. 
  The leading directions are controlled by nonlinear scalar self-interactions and their associated baryon couplings, while vector and isovector sectors contribute only at subleading order.
  }
  \label{fig:appendix_R_Lambda}
\end{figure*}

These observables probe the integrated pressure profile of the star and therefore provide complementary constraints on the EoS stiffness. 
The full sensitivity matrices for these observables are shown in \Cref{fig:appendix_R_Lambda}.

As expected, both matrices closely resemble the structure found for the mass observable. 
Variations of the scalar potential dominate the response, while vector and isovector couplings contribute at a much weaker level.

\subsection{Compactness and Mixed $M$,$\Lambda$ Sensitivity}

\begin{figure*}[t]
  \centering
  \includegraphics[width=0.48\textwidth]{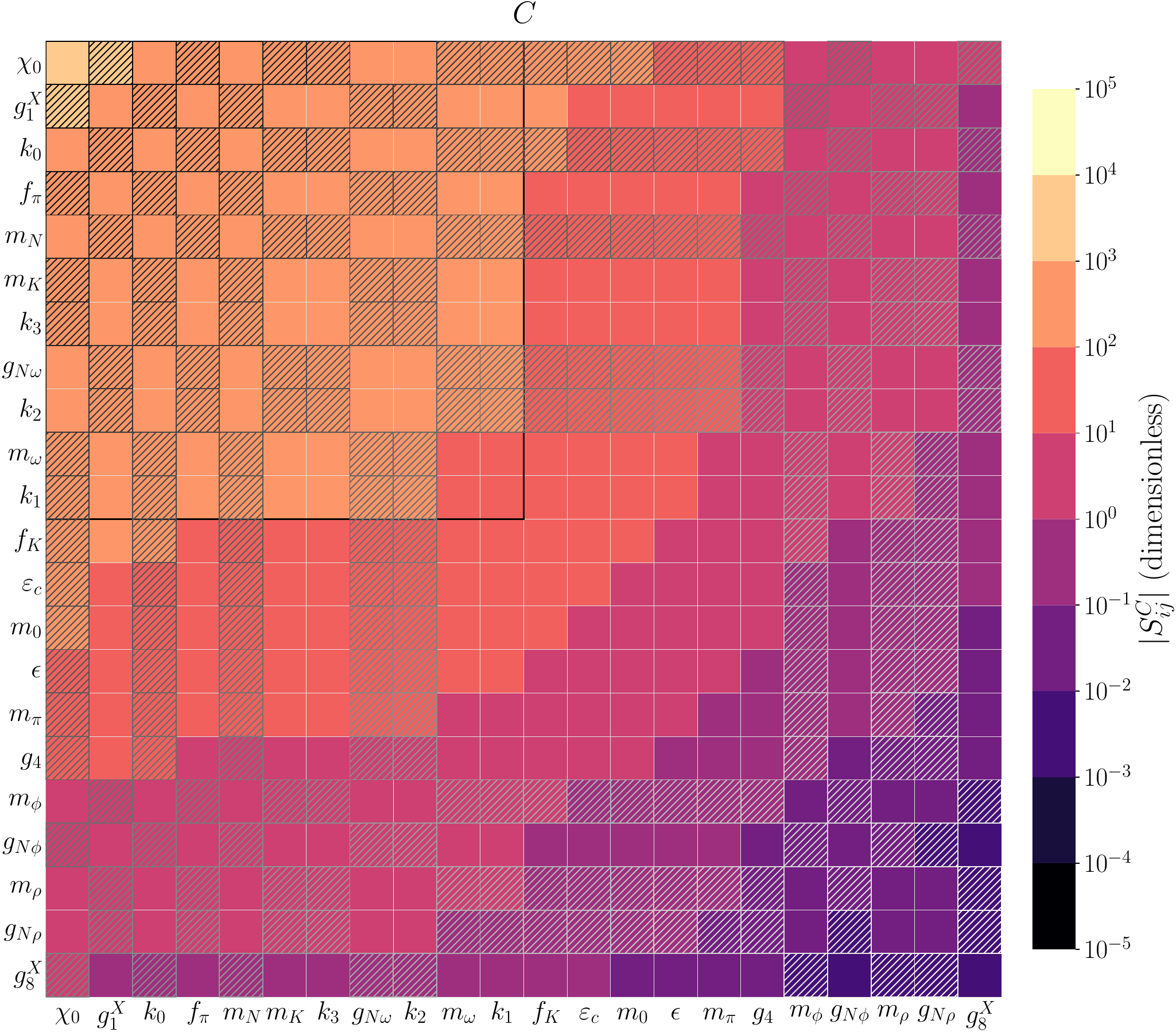}
  \hfill
  \includegraphics[width=0.48\textwidth]{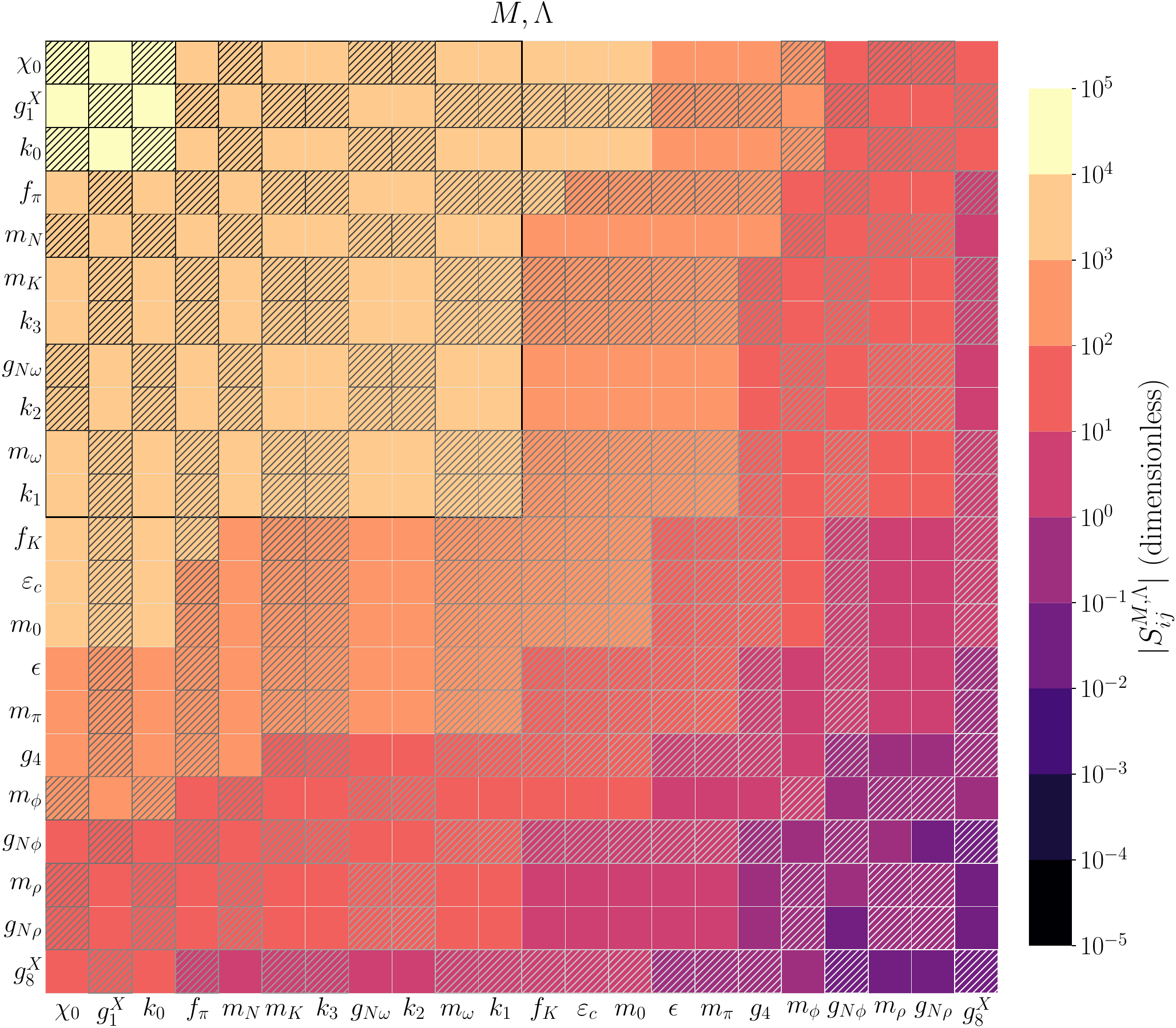}
  \caption{
  Left: sensitivity matrix $S^{C}$ (compactness). 
  Right: mixed sensitivity matrix $S^{M,\Lambda}$. 
  Both matrices retain the hierarchical ordering and block structure observed in $S^{M}$ and $S^{\Lambda}$ in the main text. 
  The dominant parameter combinations remain unchanged, confirming that mass and tidal observables probe essentially the same stiffness subspace.
  }
  \label{fig:appendix_C_MLambda}
\end{figure*}

Compactness and mixed observables were introduced in \Cref{sec:results}, where we examined how combinations of mass and radius-related quantities respond to changes in the microscopic parameters. 
The corresponding sensitivity matrices are displayed in \Cref{fig:appendix_C_MLambda}.

Compactness, being defined as a ratio of mass and radius, inherits the same scalar-driven structure. 
Similarly, the mixed matrix $S^{M,\Lambda}$ does not introduce new dominant parameter directions: its largest entries align with the scalar self-interaction block that controls both $M$ and $\Lambda$ independently. 
This confirms that current multimessenger observables predominantly constrain a common stiffness axis in parameter space.

\subsection{Maximum Mass and Maximum Radius}

\begin{figure*}[t]
  \centering
  \includegraphics[width=0.48\textwidth]{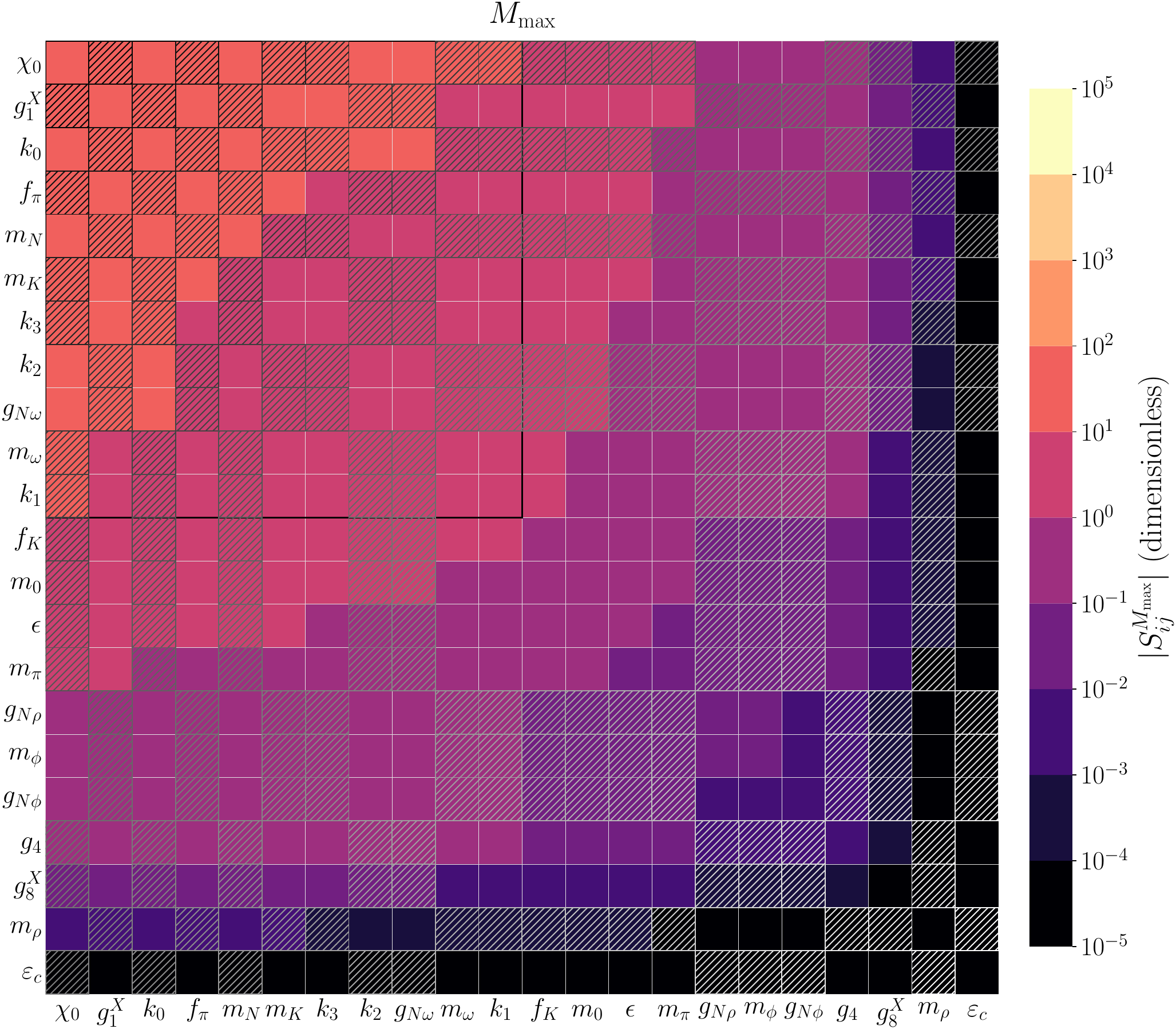}
  \hfill
  \includegraphics[width=0.48\textwidth]{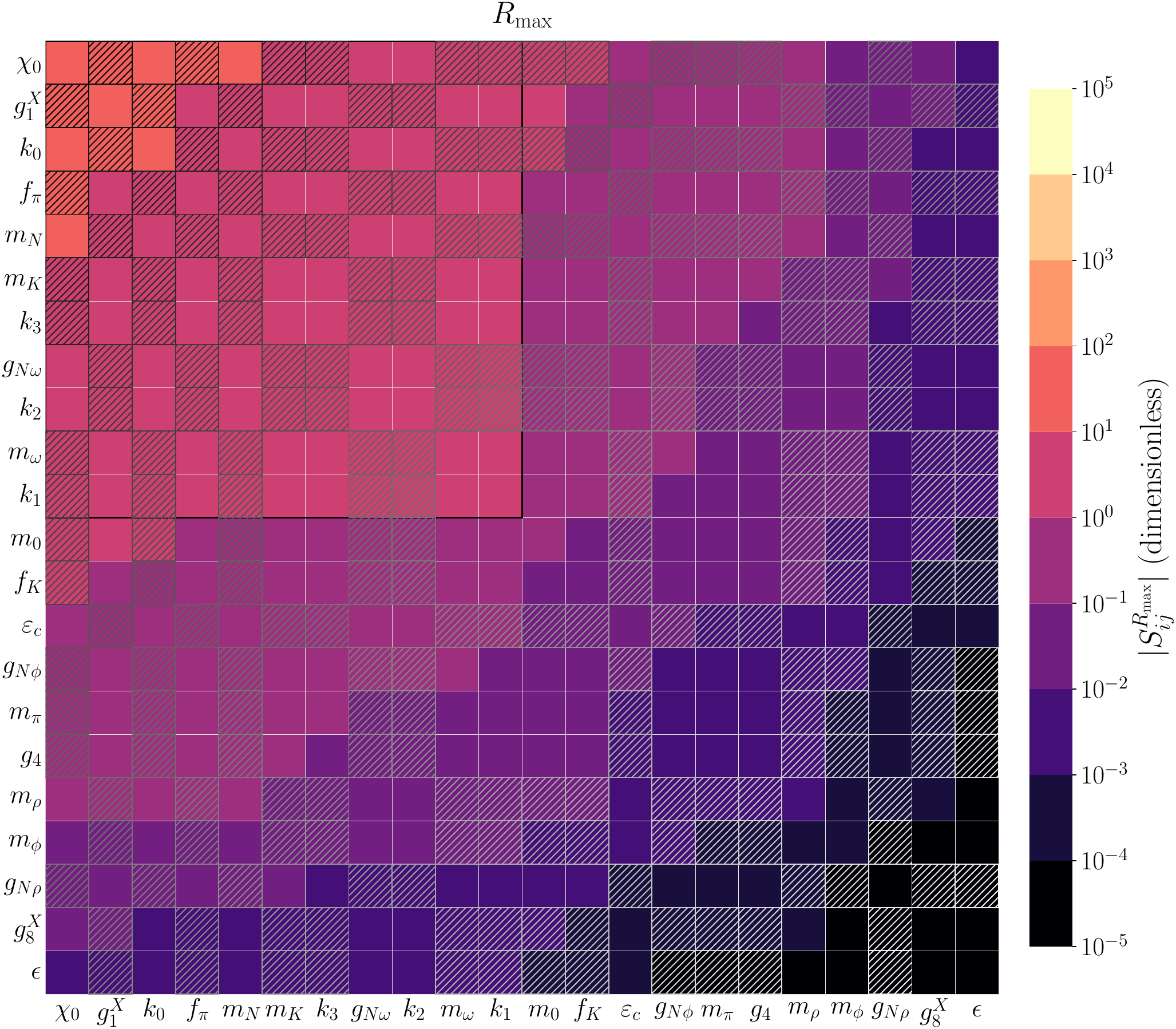}
  \caption{
  Left: sensitivity matrix $S^{M_{\max}}$. 
  Right: sensitivity matrix $S^{R_{\max}}$. 
  Maximum-mass observables preserve the same scalar-dominated hierarchy found in the main text, with modestly enhanced sensitivity to isoscalar–vector repulsion at the highest densities.
  }
  \label{fig:appendix_Mmax_Rmax}
\end{figure*}

Observables associated with the maximum-mass configuration are used to probe the highest-density region of the EoS. 
The corresponding sensitivity matrices are shown in \Cref{fig:appendix_Mmax_Rmax}.

Because these observables probe the high-density region of the EoS, the contribution of the vector sector, particularly the coupling $g_{N\omega}$, is slightly enhanced relative to canonical mass or radius observables. 
Nevertheless, the overall scalar hierarchy remains intact, and no orthogonal parameter combinations emerge.

\subsection{Combined and Projected Observables}

Finally, we show the sensitivity matrices corresponding to combined observational constraints and projected future measurements. 
These observables were discussed in \Cref{sec:results}, where we explored how the joint inclusion of multiple observational channels modifies the overall sensitivity structure.

\begin{figure*}[t]
  \centering
  \includegraphics[width=0.48\textwidth]{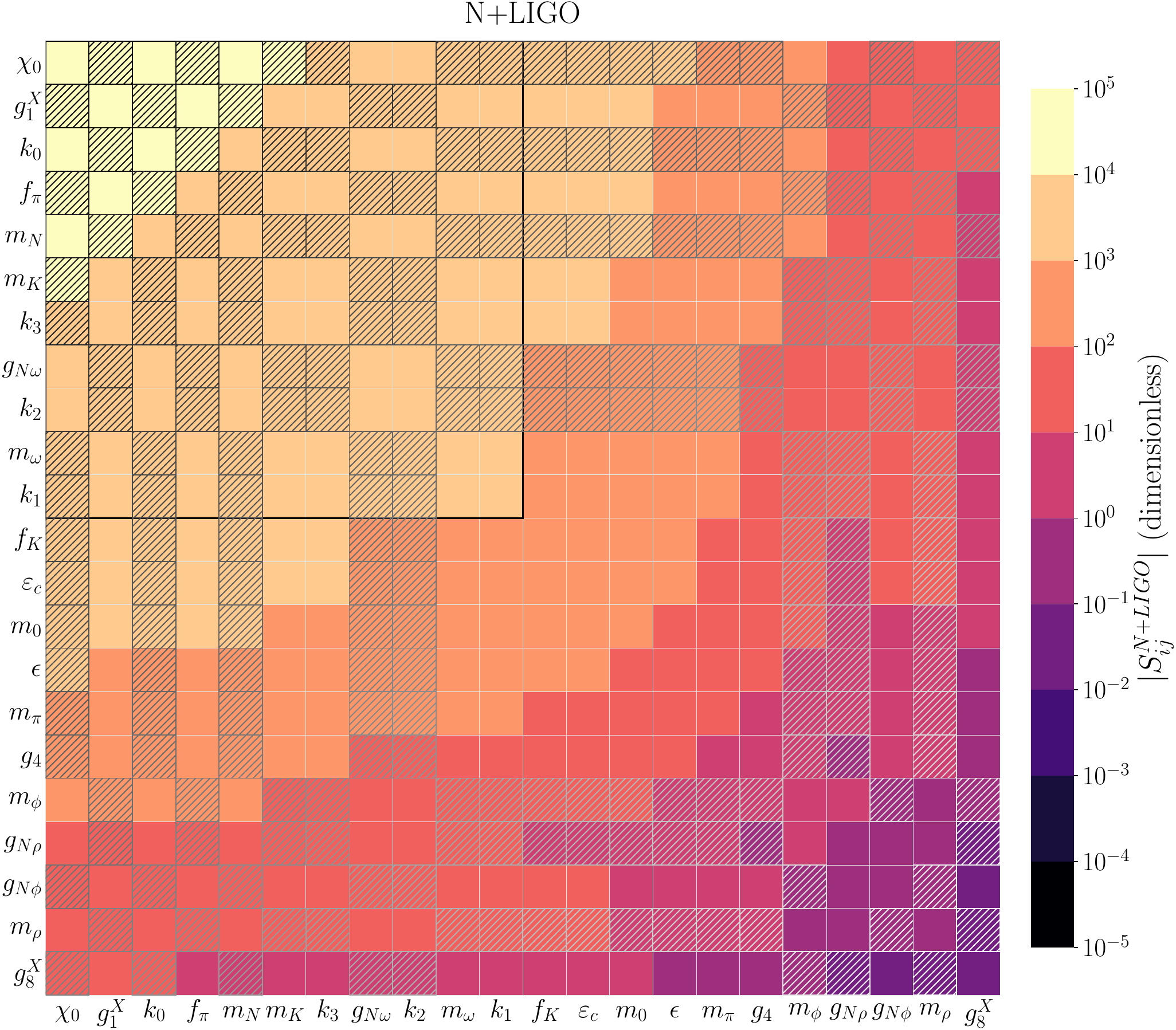}
  \hfill
  \includegraphics[width=0.48\textwidth]{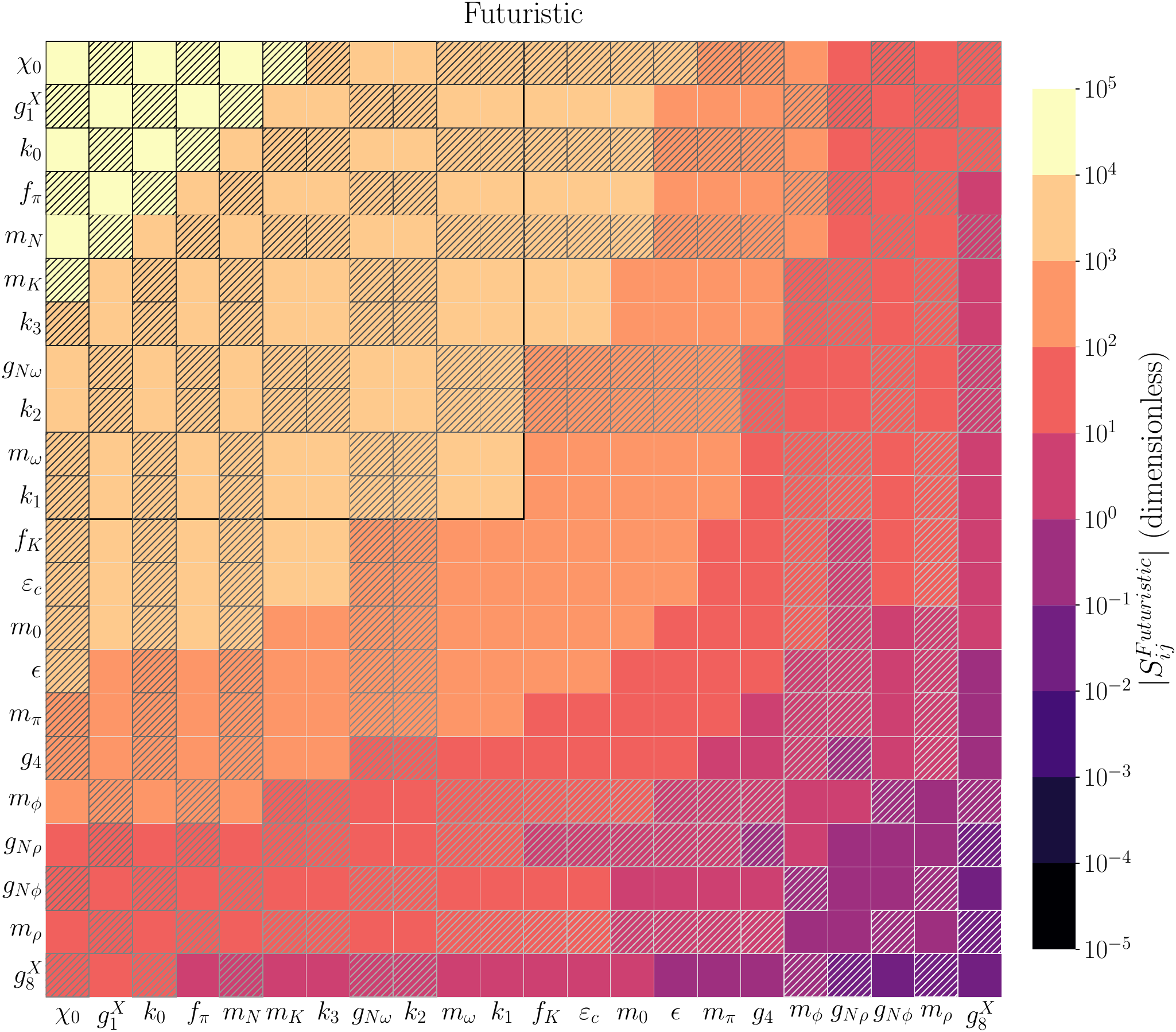}
  \caption{
  Left: combined $S^{\mathrm{N+LIGO}}$ matrix. 
  Right: projected futuristic-observables matrix. 
  In both cases the dominant scalar block remains unchanged. 
  Additional observational input primarily sharpens existing parameter directions rather than introducing new independent constraints.
  }
  \label{fig:appendix_combined}
\end{figure*}

Even when combining multiple observational channels, the information content remains concentrated along the same nonlinear scalar directions identified throughout this work. 
Future precision measurements therefore mainly reduce uncertainties along these established stiffness directions rather than uncovering new orthogonal parameter combinations.

In summary, all supplementary matrices reinforce a central conclusion of this study: neutron-star observables predominantly constrain a low-dimensional subspace of the CMF parameter space governed by nonlinear scalar self-interactions. 
Vector, isovector, and hidden-strangeness couplings remain comparatively weakly probed. 
The persistence of this structure across all observables indicates that both current and near-future multimessenger measurements access essentially the same stiffness manifold of dense QCD matter.
\setcounter{table}{0}
\renewcommand{\thetable}{D\arabic{table}}
\setcounter{figure}{0}
\renewcommand{\thefigure}{\thesection\arabic{figure}}
%

\section{PCA coefficient tables}
\label{app:pca_tables}

In this appendix we provide the full numerical coefficients of the principal directions discussed in \Cref{sec:results_pca}. 
\Cref{tab:pca_coeffs_pc1} lists the normalized eigenvector components associated with the first principal component (PC1) for each observable and composite case, while \Cref{tab:pca_coeffs_pc2} reports the corresponding coefficients for the second principal component (PC2). 
These tables complement the ranking summaries presented in the main text by showing both the magnitude and the sign of each parameter contribution, thereby making explicit which linear combinations of CMF parameters define the dominant and subdominant sensitivity directions.

The coefficients are normalized eigenvector components of the corresponding sensitivity matrices, so their absolute values quantify the relative importance of each parameter within a given principal direction, while their signs indicate whether a parameter contributes in phase or out of phase with the other components of that eigenmode. 
The horizontal separator highlights the subset of eleven parameters with the largest absolute contributions to PC1 for each case, with the exception of the mixed $(M,\Lambda)$ case discussed in the main text, for which the dominant ordering differs slightly. 
Parameters below the separator are subdominant and mainly contribute small corrections to the leading low-dimensional structure identified in \Cref{sec:results_pca}.

\begin{table*}[t!]
\centering
\small
\caption{Normalized eigenvector coefficients for the first principal component (PC1).}
\label{tab:pca_coeffs_pc1}
\begin{tabular}{c|@{}c@{}c@{}c@{}c@{}c@{}c@{}c@{}c@{}c@{}c}
\toprule
\textbf{Parameter} & \textbf{$M$} & \textbf{$R$} & \textbf{$\Lambda$} & \textbf{$C$} & \textbf{$M,\Lambda$} & \textbf{$M_{\max}$} & \textbf{$R_{\max}$} &  \textbf{LIGO} & \textbf{N+LIGO} & \textbf{Futuristic} \\ 
\midrule
$\chi_0$ & \pcacellhextable{2E86C1}{+0.527} & \pcacellhextable{2E86C1}{+0.530} & \pcacellhextable{2E86C1}{+0.525} & \pcacellhextable{2E86C1}{+0.526} & \pcacellhextablen{D4E6F8}{-0.182} & \pcacellhextable{2E86C1}{+0.520} & \pcacellhextable{2E86C1}{+0.528} & \pcacellhextable{2E86C1}{+0.525} & \pcacellhextable{2E86C1}{+0.525} & \pcacellhextable{2E86C1}{+0.525} \\
$g_{1}^{X}$ & \pcacellhextablen{5499C7}{-0.436} & \pcacellhextablen{5499C7}{-0.432} & \pcacellhextablen{5499C7}{-0.437} & \pcacellhextablen{5499C7}{-0.436} & \pcacellhextablen{D4E6F8}{-0.141} & \pcacellhextablen{5499C7}{-0.442} & \pcacellhextablen{5499C7}{-0.448} & \pcacellhextablen{5499C7}{-0.438} & \pcacellhextablen{5499C7}{-0.438} & \pcacellhextablen{5499C7}{-0.438} \\
$k_0$ & \pcacellhextable{5499C7}{+0.415} & \pcacellhextable{5499C7}{+0.418} & \pcacellhextable{5499C7}{+0.413} & \pcacellhextable{5499C7}{+0.414} & \pcacellhextable{EAF2FB}{+0.006} & \pcacellhextable{5499C7}{+0.418} & \pcacellhextable{7FB3D5}{+0.390} & \pcacellhextable{5499C7}{+0.413} & \pcacellhextable{5499C7}{+0.413} & \pcacellhextable{5499C7}{+0.413} \\
$f_{\pi}$ & \pcacellhextablen{A9CCE3}{-0.289} & \pcacellhextablen{A9CCE3}{-0.281} & \pcacellhextablen{A9CCE3}{-0.292} & \pcacellhextablen{A9CCE3}{-0.290} & \pcacellhextablen{A9CCE3}{-0.244} & \pcacellhextablen{A9CCE3}{-0.295} & \pcacellhextablen{7FB3D5}{-0.315} & \pcacellhextablen{A9CCE3}{-0.293} & \pcacellhextablen{A9CCE3}{-0.293} & \pcacellhextablen{A9CCE3}{-0.293} \\
$m_{N}$ & \pcacellhextable{A9CCE3}{+0.264} & \pcacellhextable{A9CCE3}{+0.256} & \pcacellhextable{A9CCE3}{+0.267} & \pcacellhextable{A9CCE3}{+0.265} & \pcacellhextable{A9CCE3}{+0.225} & \pcacellhextable{A9CCE3}{+0.269} & \pcacellhextable{A9CCE3}{+0.290} & \pcacellhextable{A9CCE3}{+0.268} & \pcacellhextable{A9CCE3}{+0.268} & \pcacellhextable{A9CCE3}{+0.268} \\
$m_{K}$ & \pcacellhextablen{A9CCE3}{-0.240} & \pcacellhextablen{A9CCE3}{-0.245} & \pcacellhextablen{A9CCE3}{-0.238} & \pcacellhextablen{A9CCE3}{-0.239} & \pcacellhextable{D4E6F8}{+0.125} & \pcacellhextablen{A9CCE3}{-0.240} & \pcacellhextablen{A9CCE3}{-0.222} & \pcacellhextablen{A9CCE3}{-0.238} & \pcacellhextablen{A9CCE3}{-0.238} & \pcacellhextablen{A9CCE3}{-0.238} \\
$k_3$ & \pcacellhextablen{A9CCE3}{-0.212} & \pcacellhextablen{A9CCE3}{-0.217} & \pcacellhextablen{A9CCE3}{-0.210} & \pcacellhextablen{A9CCE3}{-0.211} & \pcacellhextable{D4E6F8}{+0.137} & \pcacellhextablen{A9CCE3}{-0.211} & \pcacellhextablen{D4E6F8}{-0.197} & \pcacellhextablen{A9CCE3}{-0.210} & \pcacellhextablen{A9CCE3}{-0.210} & \pcacellhextablen{A9CCE3}{-0.210} \\
$g_{N\omega}$ & \pcacellhextable{D4E6F8}{+0.169} & \pcacellhextable{D4E6F8}{+0.170} & \pcacellhextable{D4E6F8}{+0.168} & \pcacellhextable{D4E6F8}{+0.169} & \pcacellhextablen{D4E6F8}{-0.176} & \pcacellhextable{D4E6F8}{+0.161} & \pcacellhextable{D4E6F8}{+0.179} & \pcacellhextable{D4E6F8}{+0.168} & \pcacellhextable{D4E6F8}{+0.168} & \pcacellhextable{D4E6F8}{+0.168} \\
$k_2$ & \pcacellhextable{D4E6F8}{+0.168} & \pcacellhextable{D4E6F8}{+0.172} & \pcacellhextable{D4E6F8}{+0.167} & \pcacellhextable{D4E6F8}{+0.168} & \pcacellhextablen{D4E6F8}{-0.138} & \pcacellhextable{D4E6F8}{+0.165} & \pcacellhextable{D4E6F8}{+0.154} & \pcacellhextable{D4E6F8}{+0.166} & \pcacellhextable{D4E6F8}{+0.166} & \pcacellhextable{D4E6F8}{+0.166} \\
$m_{\omega}$ & \pcacellhextablen{D4E6F8}{-0.149} & \pcacellhextablen{D4E6F8}{-0.149} & \pcacellhextablen{D4E6F8}{-0.149} & \pcacellhextablen{D4E6F8}{-0.149} & \pcacellhextable{EAF2FB}{+0.057} & \pcacellhextablen{D4E6F8}{-0.147} & \pcacellhextablen{D4E6F8}{-0.152} & \pcacellhextablen{D4E6F8}{-0.149} & \pcacellhextablen{D4E6F8}{-0.149} & \pcacellhextablen{D4E6F8}{-0.149} \\
$k_1$ & \pcacellhextablen{D4E6F8}{-0.134} & \pcacellhextablen{D4E6F8}{-0.136} & \pcacellhextablen{D4E6F8}{-0.133} & \pcacellhextablen{D4E6F8}{-0.133} & \pcacellhextable{EAF2FB}{+0.090} & \pcacellhextablen{D4E6F8}{-0.132} & \pcacellhextablen{D4E6F8}{-0.123} & \pcacellhextablen{D4E6F8}{-0.132} & \pcacellhextablen{D4E6F8}{-0.132} & \pcacellhextablen{D4E6F8}{-0.132} \\
\hline
$f_{K}$ & \pcacellhextablen{EAF2FB}{-0.058} & \pcacellhextablen{EAF2FB}{-0.062} & \pcacellhextablen{EAF2FB}{-0.056} & \pcacellhextablen{EAF2FB}{-0.057} & \pcacellhextable{EAF2FB}{+0.077} & \pcacellhextablen{EAF2FB}{-0.058} & \pcacellhextablen{EAF2FB}{-0.028} & \pcacellhextablen{EAF2FB}{-0.055} & \pcacellhextablen{EAF2FB}{-0.055} & \pcacellhextablen{EAF2FB}{-0.055} \\
$m_0$ & \pcacellhextablen{EAF2FB}{-0.043} & \pcacellhextablen{EAF2FB}{-0.043} & \pcacellhextablen{EAF2FB}{-0.044} & \pcacellhextablen{EAF2FB}{-0.044} & \pcacellhextablen{EAF2FB}{-0.005} & \pcacellhextablen{EAF2FB}{-0.044} & \pcacellhextablen{EAF2FB}{-0.045} & \pcacellhextablen{EAF2FB}{-0.044} & \pcacellhextablen{EAF2FB}{-0.044} & \pcacellhextablen{EAF2FB}{-0.044} \\
$\varepsilon_c$ & \pcacellhextablen{EAF2FB}{-0.039} & \pcacellhextable{EAF2FB}{+0.004} & \pcacellhextablen{EAF2FB}{-0.052} & \pcacellhextablen{EAF2FB}{-0.050} & \pcacellhextablen{2E86C1}{-0.845} & \pcacellhextablen{EAF2FB}{-0.000} & \pcacellhextable{EAF2FB}{+0.019} & \pcacellhextablen{EAF2FB}{-0.055} & \pcacellhextablen{EAF2FB}{-0.055} & \pcacellhextablen{EAF2FB}{-0.055} \\
$\epsilon$ & \pcacellhextablen{EAF2FB}{-0.025} & \pcacellhextablen{EAF2FB}{-0.028} & \pcacellhextablen{EAF2FB}{-0.023} & \pcacellhextablen{EAF2FB}{-0.024} & \pcacellhextable{EAF2FB}{+0.031} & \pcacellhextablen{EAF2FB}{-0.028} & \pcacellhextable{EAF2FB}{+0.000} & \pcacellhextablen{EAF2FB}{-0.023} & \pcacellhextablen{EAF2FB}{-0.023} & \pcacellhextablen{EAF2FB}{-0.023} \\
$m_{\pi}$ & \pcacellhextablen{EAF2FB}{-0.014} & \pcacellhextablen{EAF2FB}{-0.014} & \pcacellhextablen{EAF2FB}{-0.014} & \pcacellhextablen{EAF2FB}{-0.014} & \pcacellhextable{EAF2FB}{+0.001} & \pcacellhextablen{EAF2FB}{-0.015} & \pcacellhextablen{EAF2FB}{-0.008} & \pcacellhextablen{EAF2FB}{-0.013} & \pcacellhextablen{EAF2FB}{-0.014} & \pcacellhextablen{EAF2FB}{-0.014} \\
$g_{4}$ & \pcacellhextablen{EAF2FB}{-0.006} & \pcacellhextablen{EAF2FB}{-0.006} & \pcacellhextablen{EAF2FB}{-0.006} & \pcacellhextablen{EAF2FB}{-0.006} & \pcacellhextable{EAF2FB}{+0.018} & \pcacellhextablen{EAF2FB}{-0.006} & \pcacellhextablen{EAF2FB}{-0.007} & \pcacellhextablen{EAF2FB}{-0.006} & \pcacellhextablen{EAF2FB}{-0.006} & \pcacellhextablen{EAF2FB}{-0.006} \\
$m_{\phi}$ & \pcacellhextable{EAF2FB}{+0.004} & \pcacellhextable{EAF2FB}{+0.003} & \pcacellhextable{EAF2FB}{+0.004} & \pcacellhextable{EAF2FB}{+0.004} & \pcacellhextable{EAF2FB}{+0.067} & \pcacellhextable{EAF2FB}{+0.009} & \pcacellhextable{EAF2FB}{+0.002} & \pcacellhextable{EAF2FB}{+0.005} & \pcacellhextable{EAF2FB}{+0.005} & \pcacellhextable{EAF2FB}{+0.005} \\
$m_{\rho}$ & \pcacellhextable{EAF2FB}{+0.002} & \pcacellhextable{EAF2FB}{+0.003} & \pcacellhextable{EAF2FB}{+0.002} & \pcacellhextable{EAF2FB}{+0.002} & \pcacellhextablen{EAF2FB}{-0.074} & \pcacellhextable{EAF2FB}{+0.000} & \pcacellhextable{EAF2FB}{+0.006} & \pcacellhextable{EAF2FB}{+0.002} & \pcacellhextable{EAF2FB}{+0.002} & \pcacellhextable{EAF2FB}{+0.002} \\
$g_{N\phi}$ & \pcacellhextablen{EAF2FB}{-0.002} & \pcacellhextablen{EAF2FB}{-0.004} & \pcacellhextablen{EAF2FB}{-0.002} & \pcacellhextablen{EAF2FB}{-0.002} & \pcacellhextable{EAF2FB}{+0.099} & \pcacellhextable{EAF2FB}{+0.007} & \pcacellhextablen{EAF2FB}{-0.010} & \pcacellhextablen{EAF2FB}{-0.002} & \pcacellhextablen{EAF2FB}{-0.002} & \pcacellhextablen{EAF2FB}{-0.002} \\
$g_{N\rho}$ & \pcacellhextable{EAF2FB}{+0.001} & \pcacellhextable{EAF2FB}{+0.001} & \pcacellhextable{EAF2FB}{+0.002} & \pcacellhextable{EAF2FB}{+0.001} & \pcacellhextable{EAF2FB}{+0.019} & \pcacellhextable{EAF2FB}{+0.009} & \pcacellhextablen{EAF2FB}{-0.001} & \pcacellhextable{EAF2FB}{+0.002} & \pcacellhextable{EAF2FB}{+0.002} & \pcacellhextable{EAF2FB}{+0.002} \\
$g_{8}^{X}$ & \pcacellhextablen{EAF2FB}{-0.000} & \pcacellhextablen{EAF2FB}{-0.000} & \pcacellhextablen{EAF2FB}{-0.000} & \pcacellhextablen{EAF2FB}{-0.000} & \pcacellhextablen{EAF2FB}{-0.001} & \pcacellhextablen{EAF2FB}{-0.000} & \pcacellhextable{EAF2FB}{+0.000} & \pcacellhextablen{EAF2FB}{-0.000} & \pcacellhextablen{EAF2FB}{-0.000} & \pcacellhextablen{EAF2FB}{-0.000} \\
\bottomrule
\end{tabular}
\vspace{4pt}
\vspace{4pt}
\centering
{\footnotesize Absolute eigenvector weight:}\\[2pt]
\pcacolorbar
\end{table*}

\begin{table*}[t]
\centering
\small
\caption{Normalized eigenvector coefficients for the first principal component (PC2).}
\label{tab:pca_coeffs_pc2}
\begin{tabular}{c|@{}c@{}c@{}c@{}c@{}c@{}c@{}c@{}c@{}c@{}c}
\toprule
\textbf{Parameter} & \textbf{$M$} & \textbf{$R$} & \textbf{$\Lambda$} & \textbf{$C$} & \textbf{$M,\Lambda$} & \textbf{$M_{\max}$} & \textbf{$R_{\max}$} & \textbf{LIGO} & \textbf{N+LIGO} & \textbf{Futuristic} \\ 
\midrule
$\chi_0$ & \pcacellhextablen{EAF2FB}{-0.087} & \pcacellhextable{EAF2FB}{+0.049} & \pcacellhextablen{EAF2FB}{-0.095} & \pcacellhextablen{D4E6F8}{-0.100} & \pcacellhextablen{7FB3D5}{-0.340} & \pcacellhextablen{D4E6F8}{-0.103} & \pcacellhextable{5499C7}{+0.431} & \pcacellhextablen{EAF2FB}{-0.094} & \pcacellhextablen{EAF2FB}{-0.095} & \pcacellhextablen{EAF2FB}{-0.094} \\
$g_{1}^{X}$ & \pcacellhextablen{A9CCE3}{-0.237} & \pcacellhextable{D4E6F8}{+0.159} & \pcacellhextablen{A9CCE3}{-0.265} & \pcacellhextablen{A9CCE3}{-0.266} & \pcacellhextable{D4E6F8}{+0.106} & \pcacellhextablen{7FB3D5}{-0.378} & \pcacellhextable{2E86C1}{+0.783} & \pcacellhextablen{A9CCE3}{-0.266} & \pcacellhextablen{A9CCE3}{-0.267} & \pcacellhextablen{A9CCE3}{-0.266} \\
$k_0$ & \pcacellhextablen{A9CCE3}{-0.202} & \pcacellhextable{A9CCE3}{+0.204} & \pcacellhextablen{A9CCE3}{-0.261} & \pcacellhextablen{A9CCE3}{-0.243} & \pcacellhextable{A9CCE3}{+0.236} & \pcacellhextable{EAF2FB}{+0.094} & \pcacellhextable{D4E6F8}{+0.184} & \pcacellhextablen{A9CCE3}{-0.273} & \pcacellhextablen{A9CCE3}{-0.272} & \pcacellhextablen{A9CCE3}{-0.271} \\
$f_{\pi}$ & \pcacellhextablen{7FB3D5}{-0.388} & \pcacellhextable{7FB3D5}{+0.332} & \pcacellhextablen{5499C7}{-0.443} & \pcacellhextablen{5499C7}{-0.436} & \pcacellhextablen{EAF2FB}{-0.016} & \pcacellhextable{A9CCE3}{+0.241} & \pcacellhextablen{D4E6F8}{-0.150} & \pcacellhextablen{5499C7}{-0.448} & \pcacellhextablen{5499C7}{-0.449} & \pcacellhextablen{5499C7}{-0.447} \\
$m_{N}$ & \pcacellhextable{7FB3D5}{+0.382} & \pcacellhextablen{7FB3D5}{-0.332} & \pcacellhextable{5499C7}{+0.431} & \pcacellhextable{5499C7}{+0.430} & \pcacellhextable{A9CCE3}{+0.234} & \pcacellhextablen{A9CCE3}{-0.277} & \pcacellhextable{D4E6F8}{+0.137} & \pcacellhextable{5499C7}{+0.435} & \pcacellhextable{5499C7}{+0.436} & \pcacellhextable{5499C7}{+0.434} \\
$m_{K}$ & \pcacellhextable{D4E6F8}{+0.193} & \pcacellhextablen{D4E6F8}{-0.182} & \pcacellhextable{A9CCE3}{+0.228} & \pcacellhextable{A9CCE3}{+0.223} & \pcacellhextable{A9CCE3}{+0.263} & \pcacellhextablen{EAF2FB}{-0.010} & \pcacellhextablen{EAF2FB}{-0.031} & \pcacellhextable{A9CCE3}{+0.234} & \pcacellhextable{A9CCE3}{+0.233} & \pcacellhextable{A9CCE3}{+0.233} \\
$k_3$ & \pcacellhextable{D4E6F8}{+0.179} & \pcacellhextablen{D4E6F8}{-0.181} & \pcacellhextable{A9CCE3}{+0.208} & \pcacellhextable{A9CCE3}{+0.204} & \pcacellhextable{7FB3D5}{+0.400} & \pcacellhextable{2E86C1}{+0.529} & \pcacellhextable{A9CCE3}{+0.273} & \pcacellhextable{A9CCE3}{+0.212} & \pcacellhextable{A9CCE3}{+0.212} & \pcacellhextable{A9CCE3}{+0.212} \\
$g_{N\omega}$ & \pcacellhextablen{EAF2FB}{-0.041} & \pcacellhextablen{EAF2FB}{-0.008} & \pcacellhextablen{EAF2FB}{-0.025} & \pcacellhextablen{EAF2FB}{-0.038} & \pcacellhextable{A9CCE3}{+0.240} & \pcacellhextablen{D4E6F8}{-0.147} & \pcacellhextable{EAF2FB}{+0.072} & \pcacellhextablen{EAF2FB}{-0.019} & \pcacellhextablen{EAF2FB}{-0.020} & \pcacellhextablen{EAF2FB}{-0.019} \\
$k_2$ & \pcacellhextablen{D4E6F8}{-0.199} & \pcacellhextable{D4E6F8}{+0.149} & \pcacellhextablen{A9CCE3}{-0.235} & \pcacellhextablen{A9CCE3}{-0.228} & \pcacellhextable{5499C7}{+0.418} & \pcacellhextable{D4E6F8}{+0.156} & \pcacellhextablen{D4E6F8}{-0.127} & \pcacellhextablen{A9CCE3}{-0.240} & \pcacellhextablen{A9CCE3}{-0.239} & \pcacellhextablen{A9CCE3}{-0.239} \\
$m_{\omega}$ & \pcacellhextablen{EAF2FB}{-0.002} & \pcacellhextable{EAF2FB}{+0.015} & \pcacellhextablen{EAF2FB}{-0.007} & \pcacellhextablen{EAF2FB}{-0.005} & \pcacellhextablen{7FB3D5}{-0.378} & \pcacellhextablen{D4E6F8}{-0.131} & \pcacellhextable{EAF2FB}{+0.027} & \pcacellhextablen{EAF2FB}{-0.008} & \pcacellhextablen{EAF2FB}{-0.008} & \pcacellhextablen{EAF2FB}{-0.008} \\
$k_1$ & \pcacellhextable{D4E6F8}{+0.135} & \pcacellhextablen{D4E6F8}{-0.105} & \pcacellhextable{D4E6F8}{+0.161} & \pcacellhextable{D4E6F8}{+0.156} & \pcacellhextablen{A9CCE3}{-0.241} & \pcacellhextablen{5499C7}{-0.471} & \pcacellhextablen{D4E6F8}{-0.152} & \pcacellhextable{D4E6F8}{+0.165} & \pcacellhextable{D4E6F8}{+0.165} & \pcacellhextable{D4E6F8}{+0.164} \\
$f_{K}$ & \pcacellhextable{A9CCE3}{+0.268} & \pcacellhextablen{A9CCE3}{-0.216} & \pcacellhextable{7FB3D5}{+0.335} & \pcacellhextable{7FB3D5}{+0.315} & \pcacellhextablen{D4E6F8}{-0.199} & \pcacellhextablen{7FB3D5}{-0.333} & \pcacellhextablen{EAF2FB}{-0.007} & \pcacellhextable{7FB3D5}{+0.347} & \pcacellhextable{7FB3D5}{+0.346} & \pcacellhextable{7FB3D5}{+0.346} \\
$m_0$ & \pcacellhextablen{EAF2FB}{-0.026} & \pcacellhextable{EAF2FB}{+0.019} & \pcacellhextablen{EAF2FB}{-0.032} & \pcacellhextablen{EAF2FB}{-0.030} & \pcacellhextablen{EAF2FB}{-0.006} & \pcacellhextablen{EAF2FB}{-0.075} & \pcacellhextablen{EAF2FB}{-0.052} & \pcacellhextablen{EAF2FB}{-0.033} & \pcacellhextablen{EAF2FB}{-0.033} & \pcacellhextablen{EAF2FB}{-0.033} \\
$\varepsilon_c$ & \pcacellhextable{2E86C1}{+0.594} & \pcacellhextablen{2E86C1}{-0.727} & \pcacellhextable{7FB3D5}{+0.336} & \pcacellhextable{7FB3D5}{+0.395} & \pcacellhextable{EAF2FB}{+0.036} & \pcacellhextablen{EAF2FB}{-0.000} & \pcacellhextable{EAF2FB}{+0.037} & \pcacellhextable{A9CCE3}{+0.279} & \pcacellhextable{A9CCE3}{+0.281} & \pcacellhextable{A9CCE3}{+0.289} \\
$\epsilon$ & \pcacellhextable{D4E6F8}{+0.185} & \pcacellhextablen{D4E6F8}{-0.165} & \pcacellhextable{A9CCE3}{+0.231} & \pcacellhextable{A9CCE3}{+0.218} & \pcacellhextable{EAF2FB}{+0.036} & \pcacellhextable{D4E6F8}{+0.140} & \pcacellhextablen{EAF2FB}{-0.000} & \pcacellhextable{A9CCE3}{+0.240} & \pcacellhextable{A9CCE3}{+0.240} & \pcacellhextable{A9CCE3}{+0.240} \\
$m_{\pi}$ & \pcacellhextable{EAF2FB}{+0.037} & \pcacellhextablen{EAF2FB}{-0.040} & \pcacellhextable{EAF2FB}{+0.048} & \pcacellhextable{EAF2FB}{+0.043} & \pcacellhextablen{EAF2FB}{-0.068} & \pcacellhextablen{EAF2FB}{-0.010} & \pcacellhextable{EAF2FB}{+0.003} & \pcacellhextable{EAF2FB}{+0.051} & \pcacellhextable{EAF2FB}{+0.050} & \pcacellhextable{EAF2FB}{+0.050} \\
$g_{4}$ & \pcacellhextablen{EAF2FB}{-0.002} & \pcacellhextable{EAF2FB}{+0.000} & \pcacellhextablen{EAF2FB}{-0.003} & \pcacellhextablen{EAF2FB}{-0.003} & \pcacellhextablen{EAF2FB}{-0.064} & \pcacellhextable{EAF2FB}{+0.031} & \pcacellhextablen{EAF2FB}{-0.001} & \pcacellhextablen{EAF2FB}{-0.003} & \pcacellhextablen{EAF2FB}{-0.003} & \pcacellhextablen{EAF2FB}{-0.003} \\
$m_{\phi}$ & \pcacellhextable{EAF2FB}{+0.053} & \pcacellhextablen{EAF2FB}{-0.009} & \pcacellhextable{EAF2FB}{+0.058} & \pcacellhextable{EAF2FB}{+0.059} & \pcacellhextablen{D4E6F8}{-0.151} & \pcacellhextablen{EAF2FB}{-0.009} & \pcacellhextablen{EAF2FB}{-0.004} & \pcacellhextable{EAF2FB}{+0.057} & \pcacellhextable{EAF2FB}{+0.057} & \pcacellhextable{EAF2FB}{+0.057} \\
$m_{\rho}$ & \pcacellhextablen{EAF2FB}{-0.013} & \pcacellhextable{EAF2FB}{+0.013} & \pcacellhextable{EAF2FB}{+0.001} & \pcacellhextablen{EAF2FB}{-0.010} & \pcacellhextablen{D4E6F8}{-0.127} & \pcacellhextable{EAF2FB}{+0.000} & \pcacellhextablen{EAF2FB}{-0.006} & \pcacellhextable{EAF2FB}{+0.006} & \pcacellhextable{EAF2FB}{+0.005} & \pcacellhextable{EAF2FB}{+0.005} \\
$g_{N\phi}$ & \pcacellhextable{EAF2FB}{+0.073} & \pcacellhextablen{EAF2FB}{-0.003} & \pcacellhextable{EAF2FB}{+0.069} & \pcacellhextable{EAF2FB}{+0.079} & \pcacellhextablen{EAF2FB}{-0.009} & \pcacellhextablen{EAF2FB}{-0.011} & \pcacellhextable{EAF2FB}{+0.003} & \pcacellhextable{EAF2FB}{+0.066} & \pcacellhextable{EAF2FB}{+0.066} & \pcacellhextable{EAF2FB}{+0.065} \\
$g_{N\rho}$ & \pcacellhextable{EAF2FB}{+0.029} & \pcacellhextable{EAF2FB}{+0.073} & \pcacellhextable{EAF2FB}{+0.053} & \pcacellhextable{EAF2FB}{+0.041} & \pcacellhextablen{D4E6F8}{-0.104} & \pcacellhextable{EAF2FB}{+0.015} & \pcacellhextablen{EAF2FB}{-0.001} & \pcacellhextable{EAF2FB}{+0.056} & \pcacellhextable{EAF2FB}{+0.055} & \pcacellhextable{EAF2FB}{+0.055} \\
$g_{8}^{X}$ & \pcacellhextable{EAF2FB}{+0.004} & \pcacellhextable{EAF2FB}{+0.000} & \pcacellhextable{EAF2FB}{+0.006} & \pcacellhextable{EAF2FB}{+0.005} & \pcacellhextable{EAF2FB}{+0.005} & \pcacellhextable{EAF2FB}{+0.003} & \pcacellhextablen{EAF2FB}{-0.001} & \pcacellhextable{EAF2FB}{+0.006} & \pcacellhextable{EAF2FB}{+0.006} & \pcacellhextable{EAF2FB}{+0.006} \\
\bottomrule
\end{tabular}
\vspace{4pt}
\vspace{4pt}
\centering
{\footnotesize Absolute eigenvector weight:}\\[2pt]
\pcacolorbar
\end{table*}
\end{document}